\documentclass[12pt,a4paper]{article}
\pdfoutput=1
\usepackage[utf8]{inputenc}
\usepackage[english]{babel}
\usepackage{amsmath}
\usepackage{bigints}
\usepackage{subfig}
\usepackage{amsfonts}
\usepackage{amssymb}
\usepackage{cite}
\usepackage{graphicx}
\usepackage{mathrsfs} 
\usepackage{hyperref} 
\usepackage{slashed} 
\newcommand{\be}{\begin{equation}}
\newcommand{\ee}{\end{equation}}
\newcommand{\bea}{\begin{eqnarray}}
\newcommand{\eea}{\end{eqnarray}}

\newcommand{\gbl}{g_{\rm BL}}

\newcommand{\zbl}{Z_{\rm BL}}
\newcommand{\mzbl}{M_{\zbl}}
\newcommand{\mdm}{M_{\psi_1}}
\newcommand{\dm}{\psi_1}
\newcommand{\ubl}{{\rm U}(1)_{\rm B-L}}

\catcode`@=12
\def\la{\mathrel{\mathchoice {\vcenter{\offinterlineskip\halign{\hfil
$\displaystyle##$\hfil\cr<\cr\sim\cr}}}
{\vcenter{\offinterlineskip\halign{\hfil$\textstyle##$\hfil\cr<\cr\sim\cr}}}
{\vcenter{\offinterlineskip\halign{\hfil$\scriptstyle##$\hfil\cr<\cr\sim\cr}}}
{\vcenter{\offinterlineskip\halign{\hfil$\scriptscriptstyle##$\hfil\cr<\cr\sim
\cr}}}}}
\def\ga{\mathrel{\mathchoice {\vcenter{\offinterlineskip\halign{\hfil
$\displaystyle##$\hfil\cr>\cr\sim\cr}}}
{\vcenter{\offinterlineskip\halign{\hfil$\textstyle##$\hfil\cr>\cr\sim\cr}}}
{\vcenter{\offinterlineskip\halign{\hfil$\scriptstyle##$\hfil\cr>\cr\sim\cr}}}
{\vcenter{\offinterlineskip\halign{\hfil$\scriptscriptstyle##$\hfil\cr>\cr\sim
\cr}}}}}
\begin{document}
\thispagestyle{empty}
\begin{center}
{\Large \bf
{Calculation of Momentum Distribution Function of a Non-thermal
Fermionic Dark Matter}}\\
\vspace{0.25cm}
\begin{center}
{{\bf Anirban Biswas} \footnote[1]{\makebox[1.cm]
{Email:} anirbanbiswas@hri.res.in},
{\bf Aritra Gupta}\footnote[2]{\makebox[1.cm]
{Email:} aritra@hri.res.in}}\\
\vspace{0.5cm}
{\it Harish-Chandra Research Institute, Chhatnag Road,
Jhunsi, Allahabad 211 019, INDIA}
\end{center}
\vspace{1cm}
{\bf ABSTRACT} \\
\end{center}
The most widely studied scenario in dark matter phenomenology is
the thermal WIMP scenario. Inspite of numerous efforts
to detect WIMP, till now we have no direct evidence for it. 
A possible explanation for this non-observation of dark matter
could be because of its very feeble interaction strength and hence,
failing to thermalise with the rest of the cosmic soup. In other words,
the dark matter might be of non-thermal origin where the relic density
is obtained by the so-called freeze-in mechanism. Furthermore,
if this non-thermal dark matter is itself produced substantially
from the decay of another non-thermal mother particle, then their
distribution functions may differ in both size and shape from
the usual equilibrium distribution function. In this work,
we have studied such a non-thermal (fermionic) dark matter
scenario in the light of a {\it new} type of $\ubl$ model.
The $\ubl$ model is interesting, since, besides being anomaly
free, it can give rise to neutrino mass by Type II see-saw
mechanism. Moreover, as we will show, it can accommodate a
non-thermal fermionic dark matter as well. Starting from the
collision terms, we have calculated the momentum distribution
function for the dark matter by solving a coupled system of
Boltzmann equations. We then used it to calculate the final
relic abundance, as well as other relevant physical quantities.
We have also compared our result with that obtained from solving
the usual Boltzmann (or rate) equations directly in terms of
comoving number density, $Y$. Our findings suggest that the
latter approximation is valid only in cases where the system
under study is close to equilibrium, and hence should be used
with caution.
\vskip 1cm
\newpage
\section{Introduction}
\label{intro}
With the discovery of the flat rotation curves a few decades back
\cite{astro-ph/0010594}, there was no doubt about the presence of
dark matter in the Universe. From the recent observations of Planck
\cite{1502.01589}, the existence of this mysterious dark matter
has become even more certain. Their results also indicate
a huge triumph of the $\Lambda$CDM cosmology where theory
and experiments match to a great degree of accuracy.
But unfortunately, all these predictions about the existence
of dark matter were through indirect methods, mostly exploiting
the gravitational interaction of the dark matter
(for e.g. the detection of dark matter through the observation
of flat rotation curves, or through gravitational lensing).
Nothing much can be said about the \textit{particle} nature of
the dark matter, for e.g. whether it is a scalar or a fermion,
what type of interaction it has with the Standard Model (SM)
particles (if any at all), what is the strength of these type
of interactions etc. For example dark matter mass can vary
from $10^{-15}-10^{15}$ GeV, while its scattering cross
section to SM particles ranges from $10^{-76}-10^{-41}$ cm$^{2}$
\cite{ibarra-talk}. This has led to many scientifically motivated
speculations about its nature. A very interesting scenario
is the WIMP (Weakly Interacting Massive Particles), where dark
matter particles interact with the weak interaction
strength and is simultaneously able to satisfy the relic
density constraint. These are thermal relics, which remained
in the Universe as dark matter particles after the
process of \textit{thermal freeze-out}
\cite{Srednicki:1988ce, Gondolo:1990dk}.
From the direct detection point of view, we also have
very little success. Non-observations from the recent
direct detection experiments \cite{1608.07648, 1612.01223},
have put a strong upper bound on the dark matter-nucleon interaction
cross section. Future experiments \cite{1512.07501, 1606.07001}
will make this limit even stronger. With the increasing
sensitivity of these direct detection experiments, the dark
matter nucleon cross section can become as low as
the neutrino-nucleon elastic scattering cross section
\cite{1310.8327}. So in a few years time, we are going to
be in the era where we can not distinguish a dark matter
signal from that of a neutrino. Only possible way of
distinguishing the two will then be directional
searches \cite{1602.03781}. Another alternative idea to
explore is to go beyond the thermal scenario, and assume
that the dark matter is even more weakly interacting
than the WIMPs (hence explaining the null result in
the direct detection experiments) and consequently has
never been able to attain thermal equilibrium. The relic density
is obtained through \textit{freeze-in} scenario \cite{Hall:2009bx}.
Many models that explains the relic density through
this type of mechanism have been studied in detail in Refs.
\cite{Hall:2009bx, Yaguna:2011qn, Molinaro:2014lfa,
Biswas:2015sva, Merle:2015oja, Shakya:2015xnx, Biswas:2016bfo,
Konig:2016dzg}. Earlier works however had already focussed
on the production mechanism of lighter states from decays of
heavier parent particles in the early Universe
\cite{Kawasaki:1992kg,Kaplinghat:2005sy}. For example,
sterile neutrino production from the decay of heavier particles
have been discussed in Refs. \cite{Shaposhnikov:2006xi,
Kusenko:2006rh,Petraki:2007gq}. Approximate analytical solutions
for non-thermal dark matter production from decays can be
found in Ref. \cite{Bezrukov:2014qda}.

In this work we studied the viability of a non-thermal
dark matter candidate within the framework of a
recently proposed model called the \textit{new} $\ubl$ model
\cite{1607.04029}. It is {\it new} in the sense that this
model unlike the usual $\ubl$ model has no right handed neutrinos.
Four chiral fermions are however introduced for anomaly
cancellations. These chiral fermions linearly combine together
in mass basis to give rise to two Dirac fermions namely
$\dm$ and $\psi_2$. Two extra scalars were introduced
in order to give masses to these fermions.
The charge assignment of the new particles
under the $\ubl$ gauge group is consequently different
from the usual model. Since the model is symmetric under
a local gauge group (i.e. $\ubl$), it naturally has an
extra gauge boson ($\zbl$), which gets mass after this
extra gauge symmetry is broken spontaneously. We will take
the lightest of the two Dirac fermions ($\psi_1$)
as our non-thermal dark matter candidate.

Another novel feature of this new model
(as was already noted in \cite{1607.04029}) is that it
can explain the generation of neutrino mass through
a Type II see-saw mechanism upon the introduction
of a new scalar triplet ($\Delta$) with suitable {\rm B-L} charge.

The study assuming the lightest Dirac fermion to
be a thermal dark matter has already been done in Ref. \cite{1607.04029}.
We see from their analysis that the relic density constraint
is actually satisfied within very small regions.
It is satisfied either when $M_{DM} \sim M_{\zbl}/2$
(i.e. near the resonance) or when dark matter mass is
$\sim 4$ TeV. But a priori there is no reason for
the dark matter mass to be $\sim M_{\zbl}/2$
as there is no symmetry in the Lagrangian, which can relate
the masses of dark matter ($\dm$) and $\zbl$ in the above mentioned way.
This naturally motivates one to study the implications
of a non-thermal dark matter candidate within this frame work.
Imposing the non-thermality condition implies that unlike in
the thermal case, the dark matter particles are so feebly
interacting that they never attain thermal equilibrium.
An approximate mathematical statement in this regard will
be $\frac{n_{eq}\langle \sigma v \rangle}{H} < 1$,
which means that the interaction rate for scattering of
dark matter particles is less than the expansion rate of
the Universe and hence the particles fail
to scatter with other particles within the thermal plasma
and so remains out of the thermal soup. We have shown
in this work that, this model can indeed accommodate
a non-thermal dark matter candidate with correct relic density.
We have solved a coupled set of Boltzmann equations
to find the momentum distribution function for
the dark matter particles. Knowledge of the non-equilibrium
momentum distribution function (unlike in the usual
scenarios where only the comoving number density ($Y=n/s$)
is solved for) will allow us to calculate all the relevant
quantities of interest like the relic density (from freeze in),
constraints from structure formation, bounds from relativistic
degrees of freedom etc. It is well known that, if the particles
under consideration are produced from a non-thermal source
(e.g. from the decay of an out of equilibrium mother particle)
then solving the usual Boltzmann equations in terms of $Y$ is
only an approximate method to find the comoving number density.
This formalism will provide roughly the correct result as long as
we do not move far from equilibrium. In light of this, we have
also discussed and compared results from our exact calculations
with that obtained from the above mentioned approximate method.

An important difference with the earlier work
is that unlike in \cite{1607.04029}, here all the (three)
mixing angles between the three scalars (i.e.
SM Higgs, and the two non-standard Higgs) are
taken into account and we have found that in some cases,
two of them significantly control the final DM
abundance.

The rest of the work is divided as follows:
In Section \ref{model} we have elaborately discussed
about the {\it new} $\ubl$ model. Section \ref{sec:FIMP}
deals with the FIMP scenario and also
with the coupled Boltzmann equations
needed to solve the non-thermal momentum distribution
function of DM.
The results that we have found by solving the coupled
Boltzmann equations are presented in Section \ref{res}.
In Section \ref{constraints} we discuss
about the relevant theoretical as well as experimental
constraints on this non-thermal dark matter scenario. 
Finally the conclusion is given Section \ref{conclu}.
The detailed derivations all the collision terms
as well as the relevant vertex factors and decay widths
are given in the Appendix.
\section{A \textit{new} U(1)$_{\rm B-L}$ extension of Standard Model}
\label{model}
We consider a \textit{new} $\ubl$ extension of the Standard Model of particle
physics. The model has been proposed in Ref. \cite{1607.04029}. This model
does not contain any sterile neutrino like the \textit{minimal} $\ubl$ model
\cite{0812.4313} which is usually studied in the literature. The gauge group
however is the same i.e. ${\rm SU}(2)_{\rm L}\times {\rm U}(1)_{\rm Y}\times
\ubl$. But in the absence of the three sterile neutrinos we need
some extra chiral fermions for the cancellation of axial
vector anomaly \cite{Adler:1969gk} and mixed
gravitational-gauge anomaly \cite{Delbourgo:1972xb}.
Hence four chiral fermions namely $\zeta$, $\eta_L$,
${\chi_1}_R$ and ${\chi_2}_R$ with suitable ${\rm B-L}$
charges are introduced. In order to generate Dirac type
mass terms for these chiral fermions in a gauge invariant
manner we need two distinct scalar fields ($\phi_1$, $\phi_2$)
with different ${\rm B-L}$ charges. All the fields and their
corresponding charges under ${\rm SU}(2)_{\rm L}$,
${\rm U}(1)_{\rm Y}$ and $\ubl$ gauge groups are given
in Table \ref{tabcharge}. The presence of a new gauge symmetry
($\ubl$) also introduces its corresponding gauge boson
($\zbl$) to the particle spectrum and $\zbl$ becomes
massive whenever the proposed ${\rm B-L}$ symmetry
is broken spontaneously by the VEVs of scalar fields.
\begin{table}[h!]
\begin{center}
\vskip 0.5cm
\begin{tabular} {|c|c|c|c|c|c|c|}
\hline
&{\bf Field} & $\boldsymbol{{\rm SU(2)_{L}}}$ &
$\boldsymbol{{\rm U(1)_{Y}}} $&
$\boldsymbol{\ubl}$& {\bf VEV}\\
&& {\bf charge} & {\bf charge} & {\bf charge}&\\
\hline
\hline
&$l_{\rm L}\equiv (\nu_{\rm L}\,\,\,\, e_{\rm L})^{\rm T}$&
2& -$\frac{1}{2}$& -1 & \\ \cline{2-5}
&$Q_{\rm L}\equiv (u_{\rm L}\,\,\,\, d_{\rm L})^{\rm T}$&
2& $\frac{1}{6}$& $\frac{1}{3}$ & \\ \cline{2-5}
{\bf SM Fermions}&$e_{\rm R}$& 1& -$1$& -1 & 0\\ \cline{2-5}
&$u_{\rm R}$& 1& $\frac{2}{3}$& $\frac{1}{3}$ & \\ \cline{2-5}
&$d_{\rm R}$& 1& -$\frac{1}{3}$& $\frac{1}{3}$ & \\ \cline{2-5}
\hline
\hline
&$\xi_{\rm L}$& 1& 0& $\frac{4}{3}$ & \\ \cline{2-5}
{\bf BSM Fermions}&$\eta_{\rm L}$& 1& 0& $\frac{1}{3}$ & 0\\ \cline{2-5}
&${\chi_1}_{\rm R}$& 1& 0& $-\frac{2}{3}$ & \\ \cline{2-5}
&${\chi_2}_{\rm R}$& 1& 0& $-\frac{2}{3}$ & \\ \cline{2-5}
\hline
\hline
&{$H$} & 2 &$\frac{1}{2}$& 0& $v$\\ \cline{2-6}
{\bf Scalars}&{$\phi_1$} & 1 &$0$& $1$& ${v_1}$\\ \cline{2-6}
&{$\phi_2$} & 1 &$0$& $2$& ${v_2}$\\ \cline{2-6}
&{$\Delta$} & 3 &$1$& $-2$& ${v_{t}}$\\
\hline
\hline
\end{tabular}
\end{center}
\caption{${\rm SU}(2)_{\rm L}$, ${\rm U}(1)_{\rm Y}$ and
$\ubl$ charges and corresponding VEVs of all the fields
involved in the present model.}
\label{tabcharge}
\end{table} 

The gauge invariant Lagrangian for these new fields is given by:
\begin{eqnarray}
\mathscr{L}_{\rm BL} 
&=& i \, \overline{\eta_{\rm L}}\,\gamma_{\mu}\,D^\mu_{\eta}\,\eta_{\rm L} 
+i \, \overline{\xi_{\rm L}}\,\gamma_{\mu}\,D^\mu_{\xi}\,\xi_{\rm L} 
+ i \sum_{i = 1}^2 \, \overline{{{\chi}_i}_{\rm R}}
\,\gamma_{\mu}\,D^\mu_{\chi_i}\,
{{\chi}_i}_{\rm R} - \frac{1}{4} F_{\zbl}^{\mu \nu}
{F_{\zbl}}_{\,\mu \nu}  
\nonumber \\
&& + \sum_{i = 1}^2(D^\mu_{\phi_i} \phi_i)^\dagger
({D_{\phi_i}}_{\mu} \phi_i)
-\sum_{i = 1}^2 \left({y_{\xi}}_i\, \overline{\xi_{\rm L}}
\,{\chi_i}_{\rm R}\,\phi_2 + {y_{\eta}}_i\,
\overline{\eta_{\rm L}}\,{\chi_i}_{\rm R}\, \phi_1+ h.c.\,
\right) 
\nonumber \\
&& -\,V( H,\,\phi_1,\,\phi_2 ) + \mathscr{L}_{\Delta} \,,
\label{lagrang}
\end{eqnarray}
where ${D_{\psi}}_{\mu}$ is the covariant derivative for the field $\psi$
($\psi=\eta_{\rm L},\,\xi_{\rm L},\,{\chi_i}_{\rm R}$
and $\phi_i$). General expression of ${D_{\psi}}_{\mu}$
for a field $\psi$ with a ${\rm B-L}$ charge
$Q_{\rm B-L}(\psi)$ is given by
\begin{eqnarray*}
{D_{\psi}}_{\mu} = \left({\partial}_{\mu} +
i\,Q_{\rm BL}(\psi)\,g_{\rm BL}\,{\zbl}_\mu \gamma^\mu \right)\,.
\end{eqnarray*} 
Here $\gbl$ is the new gauge coupling corresponding to the
gauge group $\ubl$ while $F_{\zbl}^{\mu \nu}$ is the usual field tensor
of the new gauge boson $\zbl$. The Yukawa couplings of the chiral fermions
are denoted by ${y_{\xi}}_i$ and ${y_{\eta}}_i$. These chiral fermions
$\eta_{\rm L},\,\xi_{\rm L},\,{\chi_1}_{\rm R}$
and ${\chi_2}_{\rm R}$ in gauge basis do not represent
any physical fermionic field. In mass basis, they combine together
to give rise new physical states $\psi_1$ and $\psi_2$ with
masses $M_{\psi_1}$ and $M_{\psi_2}$ respectively. The
scalar potential including all possible gauge invariant as
well as renormalisable interaction terms among $H$, $\phi_1$
and $\phi_2$ is given by:
\begin{eqnarray}
V(H,\phi_1,\phi_2)&=& \mu^2_H  H^\dagger H + \lambda_H (H^\dagger H)^2 
+ \mu^2_1 \phi^\dagger_1 \phi_1 + \lambda_1 (\phi^\dagger_1 \phi_1)^2 
+\mu^2_2 \phi^\dagger_2 \phi_2 + \lambda_2 (\phi^\dagger_2 \phi_2)^2 
\nonumber \\ &&
+\rho_{1} (H^\dagger H) (\phi^\dagger_1 \phi_1)
+\rho_{2} (H^\dagger H) (\phi^\dagger_2 \phi_2)
+\lambda_{3} (\phi^\dagger_1 \phi_1) (\phi^\dagger_2 \phi_2)
\nonumber \\ && 
+\mu \left( \phi_2 \phi^{\dagger^2}_1+
\phi_2^\dagger \phi^{^2}_1 \right)\,,
\label{eq:V}
\end{eqnarray}
where $H$ is the usual Standard Model Higgs doublet,
while $\phi_1$ and $\phi_2$ are the new scalars which are
required to generate fermion masses in a gauge invariant way
after symmetry breaking. The $\ubl$ symmetry is assumed to
be broken spontaneously above the electroweak phase
transition scale. The scalar potential defined above
should be bounded from below. In other words, it should
have stable minima. The existence of a stable minimum of
the potential puts some conditions on the quartic couplings. These are known
as the vacuum stability condition, and are given by:
\begin{eqnarray}
\lambda_H,\,\, \lambda_1,\,\, \lambda_2 &\geq & 0\,,\nonumber \\
\rho_1 + \sqrt{\lambda_H \lambda_1} &\geq & 0\,, \nonumber \\
\rho_2 + \sqrt{\lambda_H \lambda_2} &\geq & 0\,, \nonumber \\ 
\lambda_3 + \sqrt{\lambda_1 \lambda_2} & \geq & 0 \,,
\label{vs1}
\end{eqnarray}
and 
\begin{eqnarray}
&&\sqrt{\lambda_H\,\lambda_1\,\lambda_2} + \rho_1\,\sqrt{\lambda_2}+
\rho_2\,\sqrt{\lambda_{1}} + \lambda_3\,\sqrt{\lambda_H} +
\nonumber \\
&& \sqrt{2(\rho_1 + \sqrt{\lambda_H \lambda_1})\,
(\rho_2 + \sqrt{\lambda_H \lambda_2})\,
(\lambda_3 + \sqrt{\lambda_1 \lambda_2})} \geq 0\,.
\label{vs2}
\end{eqnarray}
The neutral component of the Higgs doublet ($H^0$) and the other
two scalars acquire VEVs after symmetry breaking:
\begin{eqnarray}
H^0 &=&\frac{1}{\sqrt{2} }(v+\tilde{h})+  \frac{i}{\sqrt{2} } \tilde{G}
\, , \nonumber \\
 \phi_1 &=& \frac{1}{\sqrt{2} }(v_1+\tilde{h}_1)+ 
 \frac{i}{\sqrt{2} } \tilde{A}_1\,, \nonumber \\
 \phi_2 &=& \frac{1}{\sqrt{2} }(v_2+\tilde{h}_2)+ 
 \frac{i}{\sqrt{2} } \tilde{A}_2\, ,
\end{eqnarray}
where $v$, $v_1$ and $v_2$ are the respective VEVs, $\tilde{h}$,
$\tilde{h}_1$ and $\tilde{h}_2$ are the CP-even scalars,
while $\tilde{G}$, $\tilde{A}_1$ and $\tilde{A}_2$ are the CP-odd counterparts.
From the minimisation condition, i.e. equating the first order
derivative of the scalar potential $V(H,\,\phi_1,\,\phi_2)$ to zero with
respect to each of the scalars, we get the following equations:
\begin{eqnarray}
\mu^2_H &=& -\left(\lambda_H v^2 + \frac{ \rho_1}{2} v^2_1 + 
\frac{ \rho_2}{2} v^2_2 \right),  \nonumber \\
\mu^2_1 &=& -\left(\lambda_1 v^2_1 + \frac{ \rho_1}{2} v^2 + 
\frac{ \lambda_3}{2} v^2_2+\sqrt{2} v_2\,\mu \right)\,, \nonumber \\
\mu^2_2 &=& -\left(\lambda_2 v^2_2 + \frac{ \rho_2}{2} v^2 + 
\frac{ \lambda_3}{2} v^2_1 +\frac{1}{\sqrt{2}} \dfrac{v^2_1\,\mu}{v_2} \right). 
\label{extrema}
\end{eqnarray}

After the spontaneous breaking of all the gauge symmetries that
we have imposed on the model Lagrangian (Eq. (\ref{lagrang})), three CP-even
scalars ($\tilde{h}$, $\tilde{h}_1$, $\tilde{h}_2$) mix among themselves.
With respect to the basis states $\tilde{h}$-$\tilde{h}_1$-$\tilde{h}_2$
(gauge basis), the mass matrix of the CP-even scalars is given by
\begin{eqnarray}
\mathscr{M}^2_{\rm CP\,even}=
\begin{pmatrix}
2\lambda_Hv^2 & \rho_1v \,v_1 & \rho_2v \,v_2\\
\rho_1v\,v_1 & 2\lambda_1v_1^2 & (\lambda_3 v_2+\sqrt{2}\mu)\,v_1\\
\rho_2v \,v_2 & (\lambda_3v_2+\sqrt{2}\,\mu)\,v_1& (2\lambda_2v_2^2
-\frac{\mu\,v_1^2}{\sqrt{2}\,v_2})
\end{pmatrix}\,.
\label{scalarmix}
\end{eqnarray}
It should be noted that, while deriving the mass matrix, we have
used the conditions obtained from extremising the scalar potential
i.e. Eq. (\ref{extrema}). Now, in order to find the physical scalar
states and their respective masses we have to find a new basis states
($h_1$, $h_2$, $h_3$) with respect to which the above mass matrix
becomes diagonal. This new basis states are known as the mass basis.
As in this case, the CP-even scalars mass matrix is a real symmetric one
(assuming all the parameters in the Lagrangian are real),
the gauge basis and mass basis states must be
related by an orthogonal matrix which is the PMNS
matrix with zero complex phase. The three mixing angles are
$\theta_{12},\,\theta_{13},\,\theta_{23}$. So we have:
\begin{eqnarray*}
&&\mathscr{U}_{\rm PMNS}(\theta_{12},\,\theta_{23},
\,\theta_{13})=\nonumber\\&&
\begin{pmatrix}	
\cos\theta_{12}\cos\theta_{13} &\sin\theta_{12}\cos\theta_{13}&\sin\theta_{13}\\
-\sin\theta_{12}\cos\theta_{23}-\cos\theta_{12}\sin\theta_{23}
\sin\theta_{13}&\cos\theta_{12}\cos\theta_{23}-\sin\theta_{12}
\sin\theta_{23}\sin\theta_{13}&\sin\theta_{23}\cos\theta_{13}\\
\sin\theta_{12}\sin\theta_{23}-\cos\theta_{12}\cos\theta_{23}
\sin\theta_{13}&-\cos\theta_{12}\sin\theta_{23}-\sin\theta_{12}
\cos\theta_{23}\sin\theta_{13}&\cos\theta_{23}\cos\theta_{13}
\end{pmatrix}\,,
\end{eqnarray*}
and hence the gauge basis and the mass basis states are related by:
\begin{eqnarray}
\begin{pmatrix}
h_1\\
h_2\\
h_3
\end{pmatrix}=\mathscr{U}_{\rm PMNS}(\theta_{12},\,\theta_{23},\,\theta_{13})
\begin{pmatrix}
\tilde{h}\\
\tilde{h}_1\\
\tilde{h}_2 
\end{pmatrix} \,.
\end{eqnarray}

Like the the CP-even scalar sector, the CP odd sector also exhibits
mixing between the pseudo scalars. However in this case, only the pseudo
scalars ($\tilde{A}_1$, $\tilde{A}_2$) of the singlets $\phi_1$
and $\phi_2$ mix with each other. This is because the CP odd scalar
($\tilde{G}$) of the Higgs doublet $H$ does not mix with the
CP odd portion of the other two complex scalars ($\phi_1$ and $\phi_2$),
which are SU(2)$_{\rm L}$ singlets. This is due to
the fact that with a doublet and a complex singlet scalar
we cannot write a gauge invariant term in the Lagrangian and
also all the VEVs are assumed to be real and associated with
the CP even sector. Hence terms involving odd powers of $\tilde{G}$
is absent here. The CP odd scalars mixing matrix is thus given by:
\begin{eqnarray}
\mathscr{M}^2_{CP-odd}=
\sqrt{2}\,\begin{pmatrix}
-\,2\mu\,v_2 & \,\mu\,v_1\\
\mu\,v_1 & -\,\frac{\mu\,v_1^2}{2\,v_2}
\end{pmatrix}\,.
\end{eqnarray}
On diagonalisation we find that one of the eigenvalues
of the matrix is zero as expected and which corresponds
to a massless Goldstone mode. The mass of only physical
pseudo scalar is given by:
\begin{eqnarray}
M_A^2=-\frac{\mu\,v_2}{\sqrt{2}\,\beta^2}
\left(1+4\,\beta^2\right)\,,
\end{eqnarray}
where $\beta = \dfrac{v_2}{v_1}$, the ratio of VEVs of
$\phi_2$ and $\phi_1$. Since mass of this pseudo scalar is
always positive, the above equation implies that $\mu < 0$.
Also in terms of the mixing angle $\alpha$ between $A_1$ and
$A_2$, the expression of $M^2_A$ can also be written in the
following form
\begin{eqnarray}
M^2_A = - 2\sqrt{2}\dfrac{\mu\,v_2}{\sin^2 \alpha}
\end{eqnarray}
with mixing angle $\alpha=\tan^{-1} 2\,\beta$. 

The fermions in the present model also get masses after
the spontaneous breaking of the $\ubl$ gauge symmetry.
The masses of the fermions arise from the Yukawa
interaction terms appearing in Eq. (\ref{lagrang}), when
$\phi_1$ and $\phi_2$ get their VEVs. The Yukawa interaction terms
involving only chiral fermions in Eq.\,\,(\ref{lagrang}), can also
be written in the following matrix form
\begin{eqnarray}
\mathscr{L}_{\rm fermion-mass} &=&\left(\begin{array}{cc}
\overline{\xi_{\rm L}}& \overline{\eta_{\rm L}}\end{array}\right)
\mathscr{M}_{\rm fermion}
\left(\begin{array}{c}{\chi_1}_{\rm R}\\
{\chi_2}_{\rm R} \end{array}\right) + h.c.,
\end{eqnarray}
where
\begin{eqnarray}
\mathscr{M}_{\rm fermion} &=& 
\left( \begin{array}{cc}  {y_{\xi}}_1 v_2 &  {y_{\xi}}_2 v_2    \\
{y_{\eta}}_1v_1 & {y_{\eta}}_1 v_1
\end{array}\right)
\end{eqnarray}
is the mass matrix for the chiral fermions, which can
in general be diagonalised by a bi-unitary transformation.
From the expression of mass matrix, one can
notice that the $\mathscr{M}_{\rm fermion}$ is not a symmetric
matrix (Dirac type). Hence in the mass basis we have two
physical Dirac fermions ($\psi_1$ and $\psi_2$).
The mass and gauge basis states are related by:
\begin{eqnarray}
\begin{pmatrix}{\xi}_{\rm L}\\ {\eta}_{\rm L}\end{pmatrix}=
\mathscr{U}_{\rm L}\begin{pmatrix}
{\psi_2}_{\rm L}\\ {\psi_1}_{\rm L}\end{pmatrix},\qquad
\begin{pmatrix}{\chi_1}_{\rm R}\\ {\chi_2}_{\rm R}\end{pmatrix}
=\mathscr{U}_{\rm R}
\begin{pmatrix}{\psi_2}_{\rm R}\\ {\psi_1}_{\rm R}\end{pmatrix}.
\end{eqnarray}
Where $\mathscr{U}_{\rm L,\,R}$ are two unitary matrices
and for the case when all the Yukawa couplings (${y_{\xi}}_i$
and ${y_{\eta}}_i$) are real numbers, these matrices can
be the usual $2\times 2$ rotation matrix. Therefore,
for this case $\mathscr{U}_{\rm L,\,R}$ can be written as
\begin{eqnarray}
\mathscr{U}_{\rm L,\,R}
=\begin{pmatrix}
\cos\theta_{\rm L,\,R} &\sin\theta_{\rm L,\,R}\\
-\sin\theta_{\rm L,\,R} &\cos\theta_{\rm L,\,R}
\end{pmatrix}\,
\end{eqnarray}
with $\theta_{\rm L,\,R}$ are the respective mixing
angles for the left chiral and the right chiral states.
In the mass basis, the two physical fermionic states
are $\psi_1={\psi_1}_{\rm L}+{\psi_1}_{\rm R}$,
$\psi_2={\psi_2}_{\rm L}+{\psi_2}_{\rm R}$ and the lightest
one would automatically be stable, hence can serve
as a viable dark matter candidate. Without any
loss of generality, throughout the present work,
we assume the lightest fermion $\psi_1$ is our dark
matter candidate. 

The breaking of $\ubl$ symmetry, besides giving masses to
the fermions also makes the extra gauge boson $\zbl$ massive.
Its mass is given by :
\begin{eqnarray}
M^2_{\zbl} =\left(\dfrac{\gbl\,v_2}{\beta}\right)^2(1+4\beta^2)\,.
\label{zbl-mass}
\end{eqnarray}

The set of independent parameters relevant for our analysis are as follows: \\
$\theta_{12},\,\theta_{13},\,\theta_{23},\,\theta_{\rm L},
\,\theta_{\rm R},\,M_{h_2},\,M_{h_3},\,M_A,\,M_{\psi_1},
\,M_{\psi_2},\,\mzbl,\,\gbl$ and $\beta$. Other model parameters
can be written in terms of all these independent variables. In addition,
we have chosen $h_1$ as the SM-like Higgs boson which has recently
been discovered by ATLAS \cite{1207.7214}, CMS \cite{1207.7235} collaborations
of LHC at CERN and consequently we have kept fixed $M_{h_1}$
and $v$ at 125.5 GeV and 246 GeV respectively. 
The relevant vertex factors (in terms of the independent parameters)
that we will need in our further calculations of DM distribution
function as well as its comoving number density, are given in
the Appendix \ref{cc}.

As there are no right handed neutrinos in this new $\ubl$
model, which are usually present in $\ubl$ extended Standard
Model to cancel gauge anomaly, light active neutrinos remain
massless. We can overcome this situation
by using Type-II see-saw mechanism \cite{Magg:1980ut,
Lazarides:1980nt, Mohapatra:1980yp}
for which one has to introduce a
scalar field $\Delta$ which is
a triplet under SU$(2)_{\rm L}$. In Eq. (\ref{lagrang})
the term $\mathscr{L}_{\Delta}$ represents the Lagrangian
for the triplet $\Delta$ field. The $\Delta$
field also has a ${\rm B-L}$ charge -2, which
is required to write a gauge invariant Yukawa term
involving $\Delta$ and two lepton doublet ($l_{\rm L}$) via
$\mathscr{L}_{\Delta}\supset-{Y_{\nu}}_{{}_{\alpha\beta}}
{l^{\rm T}_{\alpha}}_{\rm L}\,C i\,\sigma_2\,\Delta\,
{l_{\beta}}_{\rm L}$, where ${l_{\alpha}}_{\rm L}$
is the usual left handed lepton doublet of flavour $\alpha$
while $C$ is the charge conjugation matrix. Therefore,
neutrinos become massive with ${m_{\nu}}_{ij} =
{Y_{\nu}}_{{}_{ij}}\,\dfrac{v_t}{\sqrt{2}}$,
when the neutral component of $\Delta$ acquires a VEV $v_t$.
However, the VEV of $\Delta$ field is related to
that of SM Higgs doublet through the relation
$v_t \sim \dfrac{\mu\,v^2}{\sqrt{2}\,M^2_{\Delta}}$ \cite{1105.1925}
(when $v>>v_t$, required for $\rho$ parameter to be
equal to 1). Here $M^2_{\Delta}$ is the coefficient
for the quadratic term (mass term) of
$\Delta$ in the $\mathscr{L}_{\Delta}$ ($\supset
-M^2_{\Delta} {\rm Tr}(\Delta^\dagger\Delta)$) while
$\mu$ is the coefficient of the trilinear term between
two Higgs doublets ($H$) and a $\Delta$. In our present case,
such a trilinear interaction term is although forbidden,
but can be generated from a term like $\lambda^{\prime} H^{\rm T} i
\sigma_2 \Delta^{\dagger} H\,\phi_2$ in a gauge
invariant manner, when $\phi_2$ gets its VEV. Therefore
in our case $\mu = \lambda^{\prime}\dfrac{v_2}{\sqrt{2}}$ and
consequently ${m_{\nu}}_{{}_{ij}} =
{Y_{\nu}}_{{}_{ij}}\,\dfrac{\lambda^{\prime}\,v_2\,v^2}
{2\sqrt{2}\,M^2_{\Delta}}$. Hence, in order to produce
neutrino masses $\sim \mathcal O(0.1)$ eV, we need $M_{\Delta}
\sim 10^8$ GeV for $Y_{\nu} \sim 10^{-1}$ and $\lambda^{\prime}\,v_2
\sim 1$ TeV (possible as we have assumed before
that the ${\rm B-L}$ symmetry breaking occurs well
above the EWPT). As a results the masses of the scalar fields
within the triplet $\Delta$ will be several orders of magnitude
higher than those of particles we are considering
in this work. Hence the effect of the formers will be
negligibly small at that epoch of the
Universe (Temperature $\leq 10$ TeV) where we
have done our analysis. 
\section{The FIMP paradigm}
\label{sec:FIMP}
Now we turn to the problem of investigating a non-thermal
fermionic dark matter candidate ($\psi_1$) within the framework
of this new $\ubl$ model. As already discussed before, since
the thermal scenario is only viable either near the resonance,
or near the high mass range where mass of dark matter
$\sim \mathcal{O}\,(4\, \rm TeV)$, hunt for a non-thermal
dark matter candidate is quite natural. In the usual
scenario (i.e. the thermal scenario), dark matter has
weak but sizeable interaction with other particles in
the thermal plasma. But as the Universe evolves it
freezes out and drops out of the thermal bath. Freeze-out
occurs because the rate of collision of DM particles
falls below the expansion rate after a certain time,
and the dark matter species retains its value of
comoving number density at the freeze out temperature.
But situations may be such that from the very beginning
dark matter particles are so very weakly interacting
with the particles in the thermal soup, that they never
enter thermal equilibrium in the first place.
So their initial number density is almost negligible.
But as the Universe evolves, these may begin to
be produced (mostly from the decays of) heavier
mother particle(s). In the case where the mother
particles are in thermal equilibrium, the production of
these non-thermal dark matter particles is most
significant at around $T_{\rm Universe} \sim M$,
where $M$ is the mass of the mother particle. So, starting
from a negligible initial number density, the number density
of the dark matter particles will increase and
may finally evolve to match the relic density constraint.
Moreover, there can be a situation when the mother particles are not
even in equilibrium. Then we will also have to solve
the momentum distribution function for the mother particle
as well. This, (as we will see here) leads to a coupled
set of Boltzmann equations. Since the initial number density
of these non-thermal dark matter is extremely small, inverse
reactions are often neglected while solving the Boltzmann
equations \cite{1404.2220, 1502.01011}. Since the dominant production mode
of a non-thermal dark matter is the decay of heavy
particles at the early epoch, the condition for non-thermality
is given by $\dfrac{\Gamma}{H}<1\biggm\lvert_{T\sim M}$ \cite{1305.6587},
where $H$ is the Hubble parameter, $M$ is the mass of the decaying
mother particle while $\Gamma$ is the corresponding decay width.
This gives an order of magnitude estimate (upper bound) of the coupling
strength needed for a species remains out of equilibrium in the
early Universe. Using the non-thermality criterion we find that
for a decaying particle of mass $\mathcal{O}$(TeV), the extra
gauge coupling $\gbl$ must be less than $10^{-7}$. 

Most of the earlier studies involving calculation of DM relic abundance have
attempted to solve the Boltzmann equation in terms of the comoving number density
$Y=\dfrac{n}{s}$ of the relic particle. But this approach
is valid as long as the decaying and the annihilating particles
(except one whose comoving number density is being solved)
are in thermal equilibrium or at least their distribution functions are
similar in shape to the equilibrium distribution function
and do not vary much from the latter. However this situation
is certainly not guaranteed here, since one of the decaying
particles ($\zbl$) is not in equilibrium. Thus 
in order to compute the DM relic density, first we need to calculate 
the momentum distribution function of $\zbl$ followed by that of $\dm$.
Hence, we have solved a set of coupled Boltzmann equations at the level of momentum
distribution functions for each of $\zbl$ and $\dm$
(other decaying particles are assumed to be in thermal
equilibrium) following Ref. \cite{1609.01289}.
Once we have the knowledge about both the distribution
functions, it is straight forward to calculate
the other physical quantities like comoving number density,
relic density etc. 
\subsection{Coupled Boltzmann equations and its solution}
\label{sec:BEQ}
The Boltzmann equation for the distribution function $f(p)$,
in its most general form can be written in terms of the
Liouville operator ($\hat{L}$) and the collision term ($\mathcal{C}$). 
Symbolically, it is written as:
\begin{eqnarray*}
\hat{L}\,f=\mathcal{C}[f]\,.
\end{eqnarray*}
For an isotropic and homogeneous Universe, using the FRW metric
we find that $\hat{L}=\frac{\partial}{\partial t}
-Hp\frac{\partial}{\partial p}$, where $p=\lvert \vec{p} \rvert$
is the absolute value of the particle's three momentum.
As in Ref. \cite{1609.01289} making the transformation of variable:
\begin{eqnarray}
r&=&\frac{M_{sc}}{T}\,,
\label{coord1}\\
\xi_p&=&\Bigg(\frac{g_s(T_0)}{g_s(T)}\Bigg)^{1/3}\frac{p}{T}\,,
\label{coord2}
\end{eqnarray}
where $M_{sc}$ and $T_0$ are some reference mass scale and
temperature respectively, we find that the Liouville operator takes the following
form:
\begin{eqnarray}
\hat{L}&=& r H \Bigg(1+\frac{T\,g_s^\prime}{3\,g_s}\Bigg)^{-1}
\frac{\partial}{\partial r}\,,
\label{liouville}
\end{eqnarray}
where $g_s(T)$ is the effective number of degrees of freedom
related to the entropy density of the Universe while $g_s^\prime$
denotes differentiation of $g_s$ with respect to temperature $T$.
The form of the function $g_s\,(T)$ is taken from Ref. \cite{Gondolo:1990dk}
(Fig. 1). The bulk of the contribution to the effective degrees
of freedom ($g_s$) comes from the relativistic (SM) particles
in equilibrium with the thermal soup\,\,\footnote{The contribution
of BSM particles to $g_s\,(T)$ however will not affect the
results presented in this work because all of our BSM particles
including the dark matter ($\psi_1$) are either out of equilibrium
from the thermal bath or their masses are such, that they have
become non-relativistic by the time the dark matter production
starts dominating.}. The reference mass scale $M_{sc}$ is taken
to be the mass of the Standard Model Higgs boson ($M_{h_1}$)
throughout the rest of the work.

The main production channels for the non-thermal dark matter
$\psi_1$ are from the decays of $h_1,\,h_2$ and $\zbl$.
All of these BSM particles have been assumed to have mass
of $\sim \mathcal{O}$(TeV). Among the three decaying particles,
$\zbl$ is itself very feebly interacting (due to very low value
of $\gbl$) and remains outside the thermal soup. The BSM scalar $h_2$
can be in thermal equilibrium, as it can interact with the SM
particles through its mixing with $h_1$, which need not be too
small even in the non-thermal scenario. In whole of the analysis
that will follow, (for simplicity) we have assumed that
the CP odd scalar $A$, the extra fermion $\psi_1$ and one
of the three CP even scalars (say $h_3$) are much heavier
than rest of the particles and hence they have negligible
abundance during the epoch of interest here
(due to exponential Boltzmann suppression). So the production
of dark matter particles from these very heavy states can
safely be neglected since there are almost no particles
left in the thermal bath to produce $\psi_1$. So, $\psi_1$
is partly produced from the decay of $h_1$ and $h_2$ which are in
thermal equilibrium, and consequently the usual equilibrium
Boltzmann distribution function has been assumed for them.
$\psi_1$ is also produced from the decay of $\zbl$
which is out of equilibrium, and hence we have to solve 
for its non-equilibrium distribution function separately.
Hence we have to solve two coupled Boltzmann equations.
From the first one we calculate the non-equilibrium momentum
distribution function of $\zbl$. This solution is then
used in the second equation to find the final non-equilibrium
momentum distribution function of $\psi_1$. The scattering
terms contribute very little in the freeze-in scenario
and hence left out in rest of the analysis
\cite{Biswas:2016bfo,1502.01011}. The coupled set of
Boltzmann equations necessary for calculating the
momentum distribution function of $\psi_1$ are as follows:
\begin{eqnarray}
\,\,\hat{L}\,f_{\zbl}&=&\mathcal{C}^{h_2 \rightarrow \zbl\zbl}
+\mathcal{C}^{\zbl\rightarrow all}\,,
\label{fzbl}\\
\hat{L}\,f_{\psi_1}&=&\sum_{s=h_1,\,h_2}
\mathcal{C}^{S\rightarrow\overline{\psi_1} \psi_1}
+\mathcal{C}^{\zbl \rightarrow \overline{\psi_1} \psi_1}\,.
\label{fpsi}
\end{eqnarray} 
Here $\mathcal{C}^{A \rightarrow B B}$s are the collision
terms corresponding to the interaction depicted in the
superscript. Before proceeding further, let us pause here to discuss
a small subtlety. We know that the SM particles gain their
masses after electroweak phase transition (EWPT) which occurs
when the temperature of the Universe is $T_{\rm EWPT} \sim 153$ GeV
\cite{hep-ph/0702143}. So while evolving the Boltzmann equations,
as written above, from a initial temperature
$T_{\rm in}$ ($>T_{\rm EWPT}$) we have to bear in mind that
when $T_{\rm Universe} > T_{\rm EWPT}$, the decay of SM Higgs boson
($h_1$) is not allowed kinetically. This is because $h_1$
is not massive during that epoch and hence cannot decay.
Its decay will be an important part when the Universe
cools down below $T_{\rm EWPT}$. On the other hand,
the BSM scalar $h_2$ can however always decay
since it gets its mass from the spontaneous breaking
of the new $\ubl$ symmetry which is assumed to occur
at a much higher temperature than $T_{\rm EWPT}$. 

As discussed earlier, the simplistic form of the
Liouville operator in Eq. (\ref{liouville}) can be used only
when we are in a specially chosen coordinate system defined by
$\xi_p$ and $r$. The final solution of the momentum
distribution function will thus, in general be a function
of both $r \equiv\frac{M_{sc}}{T}$ and $\xi_p$ defined
in Eqs. ((\ref{coord1})--(\ref{coord2})). For example,
$f_{\zbl}=f_{\zbl}(\xi_p,r)$. For our convenience,
let us further define: 
\begin{eqnarray} 
\Bigg(\dfrac{g_s(T)}{g_s(T_0)}\Bigg)^{1/3} =
\Bigg(\dfrac{g_s(M_{sc}/r)}{g_s(M_{sc}/r_0)}\Bigg)^{1/3}
\equiv \mathcal{B}(r)
\label{Br}
\end{eqnarray}
where, $T_0$ (and the corresponding $r_0$) is some reference
temperature, which we take to be equal to the initial temperature
$T_{\rm in}=10$ TeV.
The collision terms corresponding to Eq. (\ref{fzbl})
are as follows:
\begin{eqnarray}
\mathcal{C}^{h_2 \rightarrow \zbl\zbl} &=&
\dfrac{r}{8\pi M_{sc}}\dfrac{\mathcal{B}^{-1}(r)}
{\xi_p \sqrt{\xi_p^2\mathcal{B}(r)^2+
\left(\dfrac{M_{\zbl}\,r}{M_{sc}}\right)^2}}
\dfrac{g_{h_2\zbl\zbl}^2}{6}\left(2+
\dfrac{(M_{h_2}^2-2M_{\zbl}^2)^2}
{4M_{\zbl}^4}\right) \nonumber \\
&&\times \left(e^{-\sqrt{\left(\xi_{k}^{\rm min}\right)^2
\mathcal{B}(r)^2+\left(\frac{M_{h_2}\,r}{M_{sc}}\right)^2}}
\,-\,e^{-\sqrt{\left(\xi_{k}^{\rm max}\right)^2
\mathcal{B}(r)^2+\left(\frac{M_{h_2}\,r}{M_{sc}}\right)^2}}
\right)\,,
\label{c1}
\\
\mathcal{C}^{\zbl\rightarrow all} &=&
-\dfrac{\Gamma_{\zbl \rightarrow all}\,M_{\zbl}\,r}
{M_{sc}\sqrt{\xi_p^2\,\mathcal{B}(r)^2
+\left(\frac{M_{\zbl}\,r}{M_{sc}}\right)^2}} f_{\zbl}(\xi_p,r)\,.
\label{c2}
\end{eqnarray}
Here, in the above two equations $\xi_k \equiv
\dfrac{1}{\mathcal{B}(r)}\,\dfrac{k}{T}$ is the variable
corresponding to the three momentum $k$ of
the decaying particle (i.e. $h_2$). It is integrated
over from $\xi_k^{min}$ to $\xi_k^{max}$ where each of
these are functions of $\xi_p$ and $r$ (and also of
masses of the particles involved in the corresponding process).
$M_{sc}$, as already mentioned, is some reference mass scale,
which we take to be equal to $M_{h_1}$. The quantity
$\Gamma_{\zbl \rightarrow all}$ is the total decay width of
$\zbl$. Explicit expression of the total
decay width as well as the detailed derivation of the collision term
$\mathcal{C}^{h_2 \rightarrow \zbl\zbl}$ are given
in the Appendix (\ref{width}, \ref{app:h2-zblzbl}).
Further, $g_{h_2\zbl\zbl}$ is the vertex factor of
an interaction vertex containing fields $h_2\,\zbl\,\zbl$
and its expression in terms of chosen set of independent
parameters is also given in the Appendix \ref{cc}.
The detailed derivation of other collision term
$\mathcal{C}^{\zbl\rightarrow all}$
is also given in Appendix \ref{app:zbl-all}.

The collision terms appearing in Eq. (\ref{fpsi})
can similarly be written as:
\begin{eqnarray}
\mathcal{C}^{s \rightarrow \overline{\psi_1} \psi_1} &=&
\dfrac{r}{8\pi M_{sc}}\dfrac{\mathcal{B}^{-1}(r)}
{\xi_p \,\sqrt{\xi_p^2\mathcal{B}(r)^2+
\left(\dfrac{M_{\dm}\,r}{M_{sc}}\right)^2}}
\,g_{s\overline{\psi}_1\dm}^2\left(M_{s}^2-4M_{\dm}^2\right) \nonumber \\
&&\times \left(e^{-\sqrt{\left(\widehat{\xi_{k}}^{\rm min}\right)^2
\mathcal{B}(r)^2+\left(\frac{M_{s}\,r}{M_{sc}}\right)^2}}\,
-\,e^{-\sqrt{\left(\widehat{\xi_{k}}^{\rm max}\right)^2
\mathcal{B}(r)^2+\left(\frac{M_{s}\,r}{M_{sc}}\right)^2}}
\right)\,,
\label{spsi}
\end{eqnarray}
\begin{eqnarray}
\mathcal{C}^{\zbl \rightarrow \overline{\psi_1}\psi_1}&=&
\dfrac{r}{4\,\pi\,M_{sc}}\dfrac{\mathcal{B}(r)}
{\xi_p\,\sqrt{\xi_p^2\,\mathcal{B}(r)^2+\left(\frac{M_{\psi_1}\,r}
{M_{sc}}\right)^2}} \times \left(M_{\zbl}^2
\left(a_{\dm}^2+b_{\dm}^2\right)+2 M_{\dm}^2
\left(a_{\dm}^2-2\,b_{\dm}^2\right)\right)\nonumber \\
&&~~~~~\times \bigintss_{\widetilde{\xi_k}^{\rm min}}
^{\widetilde{\xi_k}^{\rm max}}\dfrac{\xi_k \, f_{\zbl}
(\xi_k,\,r)\,d\xi_k}{\sqrt{\xi_k^2\,\mathcal{B}(r)^2
+\left(\frac{M_{\zbl}\,r}{M_{sc}}
\right)^2}}\,,
\label{zblpsi}
\end{eqnarray}
where the superscript $s$ is a generic symbol
denoting the decay of $\psi_1$ from any of the scalars $h_1,\,h_2$.
As we mentioned above, the expressions
of the coupling $g_{s\overline{\psi_1}\psi_1}$ in terms of the
independent parameters are given in the Appendix \ref{cc}.
The value of the function $f_{\zbl}$ in Eq. (\ref{zblpsi})
is obtained by solving the first Boltzmann equation,
i.e. Eq. (\ref{fzbl}). The derivation of these collision
terms are also roughly sketched in Appendix \ref{app:spsi1psi1}
and \ref{app:zblpsi1psi1}.
\section{Results}
\label{res}
Having developed the structure of the coupled set of Boltzmann
equations that we will use to find the momentum distribution
functions of $\zbl$ and $\psi_1$, we can now proceed further
to solve them numerically. For our numerical calculation
we have always taken $\dfrac{M_{h_1}}{2}\leq \mzbl
\leq \dfrac{M_{h_2}}{2}$, so that the extra gauge boson
can be produced from the decay of $h_2$ only. Introduction
of another decay mode only complicates the numerics
while giving rise to no extra interesting features.
The present section can be broadly categorised in
two parts, i) $\beta = 1$ and ii) $\beta \ll 1$,
depending on the relative contributions of
different decay modes in the final relic abundance of
$\psi_1$. For definiteness, we have chosen
$\beta = 10^{-3}$ as a representative value in the
$\beta \ll 1$ case. All of our arguments and
discussions in this section are with respect to
two benchmarks, one corresponding to $\beta =1$
and the other corresponding to $\beta = 10^{-3}$.

Once the momentum distribution function for $\dm$
is calculated (solving Eq. (\ref{fzbl}) and Eq. (\ref{fpsi})),
it is then easy to calculate other quantities of physical
importance. The first order moment of the distribution function
for e.g. gives an idea about the number density of the concerned
particle, i.e. $n \sim \bigintssss d^3p\, f(p)$ or
in terms of $\xi_p$, it is given by:
\begin{eqnarray} 
n(r) &=& \dfrac{g\, T^3}{2\pi^2} \,
\mathcal{B}(r)^3 \bigintssss d\xi_p\,\xi_p^2\, f_{\psi_1}(\xi_p)\,,
\label{n}
\end{eqnarray}
where $g$ is the internal degree of freedom of the
particle under consideration and $\mathcal{B}(r)$ is
defined in Eq. (\ref{Br}). Other symbols have their
usual meaning. Our primary quantity of interest
in the rest of this section is the comoving
number density $Y=\dfrac{n}{s}$, where $s$
is the entropy density of the Universe, given by:
\begin{eqnarray}
s &=& \dfrac{2\pi^2}{45} \, g_s(T)\,T^3\,.
\label{s}
\end{eqnarray}
Here $T$ is the temperature
and $g_s(T)$ is degrees of freedom corresponding to
the entropy density $s$ of the Universe. The relic abundance
of our dark matter $\psi_1$ is simply related to the
comoving number density $Y$ by \cite{Edsjo:1997bg}:
\begin{eqnarray}
\Omega_{\psi_1}h^2 &=& 2.755\times 10^8
\bigg(\dfrac{M_{\psi_1}}{\rm GeV}\bigg)
\,Y_{\psi_1}(T_{\rm Now})\,, 
\label{relic}
\end{eqnarray}
where $T_{\rm Now}$ is the temperature of
the Universe at the present epoch. In the present scenario,
the temperature $T$ can be easily calculated
if $r \left(\equiv \dfrac{M_{sc}}{T}\right)$ is known.

The values of different independent parameters in
our benchmark scenarios have been tabulated in Table
\ref{tab1} (left) for $\beta =1$ and Table \ref{tab1}
(right) for $\beta = 10^{-3}$. The two benchmarks are so
chosen such that the final $Y_{\dm}$ calculated using
these parameters give the correct relic density when
plugged in Eq. (\ref{relic}). As we will see later,
in the $\beta =1$ scenario, if we fix the scalars mixing
angles to values of $\mathcal{O}(0.1)$ rad or
less \footnote{to satisfy the bounds on the signal strength of
SM Higgs boson \cite{1606.02266}.}, the contributions
arising from the scalar decay channels to
the total comoving number density ($Y_{\dm}$) become
quite low. Almost the whole of $\dm$ is produced
from the decay of $\zbl$. The percentage contribution
of the scalar decay modes to $Y_{\dm}$, in this case,
is thus not much sensitive to the values of
the mixing angles ($\theta$s $\leq 0.1$ rad).
This can be easily understood from the expressions
of $g_{h_1\overline{\psi_1} \dm}$ and $g_{h_2\overline{\psi_1} \dm}$
given in the Appendix\,\,\,\ref{cc}. 
The situation is however different when $\beta=10^{-3}$.
For our chosen benchmark, values of the dark matter--scalar
couplings now become sizeable and also sensitive to
$\theta_{13}$ ($h_1\overline{\psi_1}\dm$ coupling)
and $\theta_{23}$ ($h_2\overline{\psi_1}\dm$ coupling).
The benchmark in this case is chosen in such a
way so that we can have equal contributions
to the final comoving number density of
the dark matter ($Y_{\dm}$) from
$h_1$, $h_2$ and $\zbl$ decays. For definiteness,
the value of the arbitrary mass scale $M_{sc}$ has been
fixed at the Standard Model Higgs mass.
In passing, let us comment on the values of the couplings
related to the scalar sector (generically denoted by $\lambda$, say).
For our chosen set of independent parameters, the values of the scalar
coupling constants can be solved uniquely by using Eq. (\ref{scalarmix}).
We have checked that these values are also very small and
in general (for the chosen mass hierarchy between $M_{\zbl}$
and $M_{h_2}$), $\lambda \ga \mathcal{O}(\gbl^2)$. However,
we have verified that these scalar sector couplings
satisfy the theoretical constraints arising from the vacuum
stability conditions (see Eq. ((\ref{vs1})--(\ref{vs2}))).
The aforementioned mass hierarchy along with the fact that
$\gbl$ is very small (due to non-thermality) leads to a
corresponding hierarchy in the scalar sector couplings
(e.g. $\lambda_1 \sim 10^{-20}$ and $\rho_1 \sim 10^{-10}$).
At this point, we should however be careful, so that the radiative
corrections to the couplings are small enough to make our choice
feasible. For example, the most dominant contribution to the one
loop correction of $\lambda_1$ is through the SM Higgs boson and
it is $\propto \rho_1^2\,\int_{0}^{1}\,dx\,{\rm ln}
\left(\frac{\Lambda^2}{M_{0}^2(x,p^2)}\right)$, where $\Lambda$
is some chosen cut-off scale and $M_{0}^2(x,p^2)=M_{h_1}^2-x(1-x)p^2$
with $p$ being the total incoming four momentum. With our chosen
set of parameters this correction indeed turns out to be $\la \lambda_1$.

\begin{table}[h!]
	\begin{center}
		\begin{tabular} {||c||c||}
			\hline
			\hline
			Input Parameters & Corresponding values \\
			\hline
			\hline
			$\mzbl$ & 1 TeV\\
			\hline
			$M_{h_2}$ & 5 TeV\\
			\hline
			$M_{\psi_1}$ & 10 GeV \\
			\hline
			$\gbl$ & $4.87\times 10^{-11}$ \\
			\hline
			$\theta_{12}$ & 0.1 rad\\
			\hline
			$\theta_{13}$ & 0.1 rad\\
			\hline
			$\theta_{23}$ & 0.1 rad\\
			\hline
			$\theta_{\rm L}=\theta_{\rm R}$ & $\pi/4$ rad\\
			\hline
			\hline
		\end{tabular}
		\quad
		\begin{tabular} {||c||c||}
			\hline
			\hline
			Input Parameters & Corresponding values \\
			\hline
			\hline
			$\mzbl$ & 1 TeV\\
			\hline
			$M_{h_2}$ & 5 TeV\\
			\hline
			$M_{\psi_1}$ & 10 GeV \\
			\hline
			$\gbl$ & $1.75\times 10^{-11}$ \\
			\hline
			$\theta_{12}$ & 0.1000 rad\\
			\hline
			$\theta_{13}$ & $9.58\times 10^{-3}$ rad\\
			\hline
			$\theta_{23}$ & $6.18\times 10^{-2}$ rad\\
			\hline
			$\theta_{\rm L}=\theta_{\rm R}$ & $\pi/4$ rad\\
			\hline
			\hline
		\end{tabular}
	\end{center}
	\caption{Values of different input parameters used
	in our analysis. Benchmark corresponding to $\beta =1$
	(left) and $\beta = 10^{-3}$ (right).}
	\label{tab1}
\end{table}

\begin{figure}[h!]
	\centering
	\includegraphics[height=8.5cm,width=6.5cm,angle=-90]
	{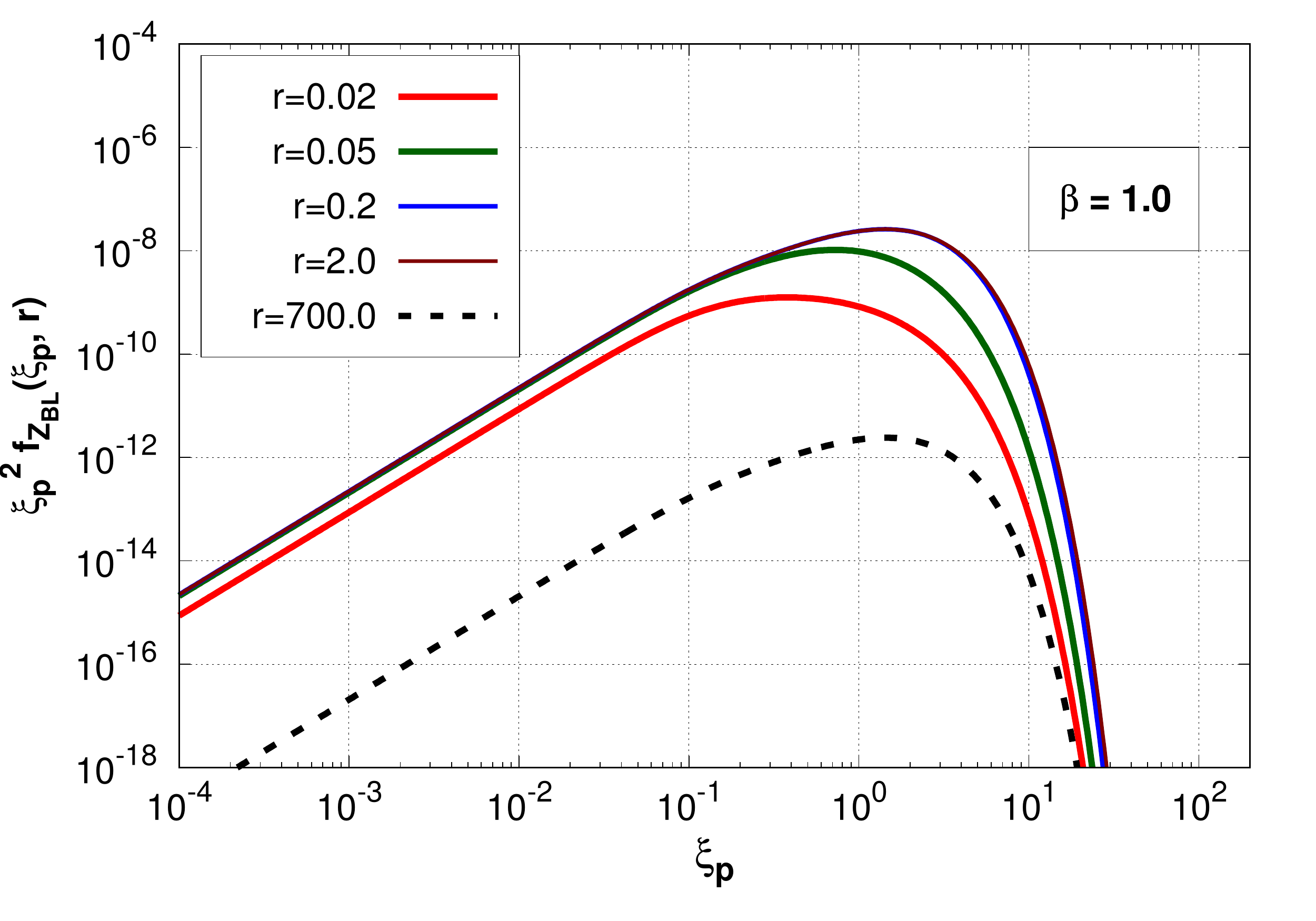}
	\includegraphics[height=8.5cm,width=6.5cm,angle=-90]
	{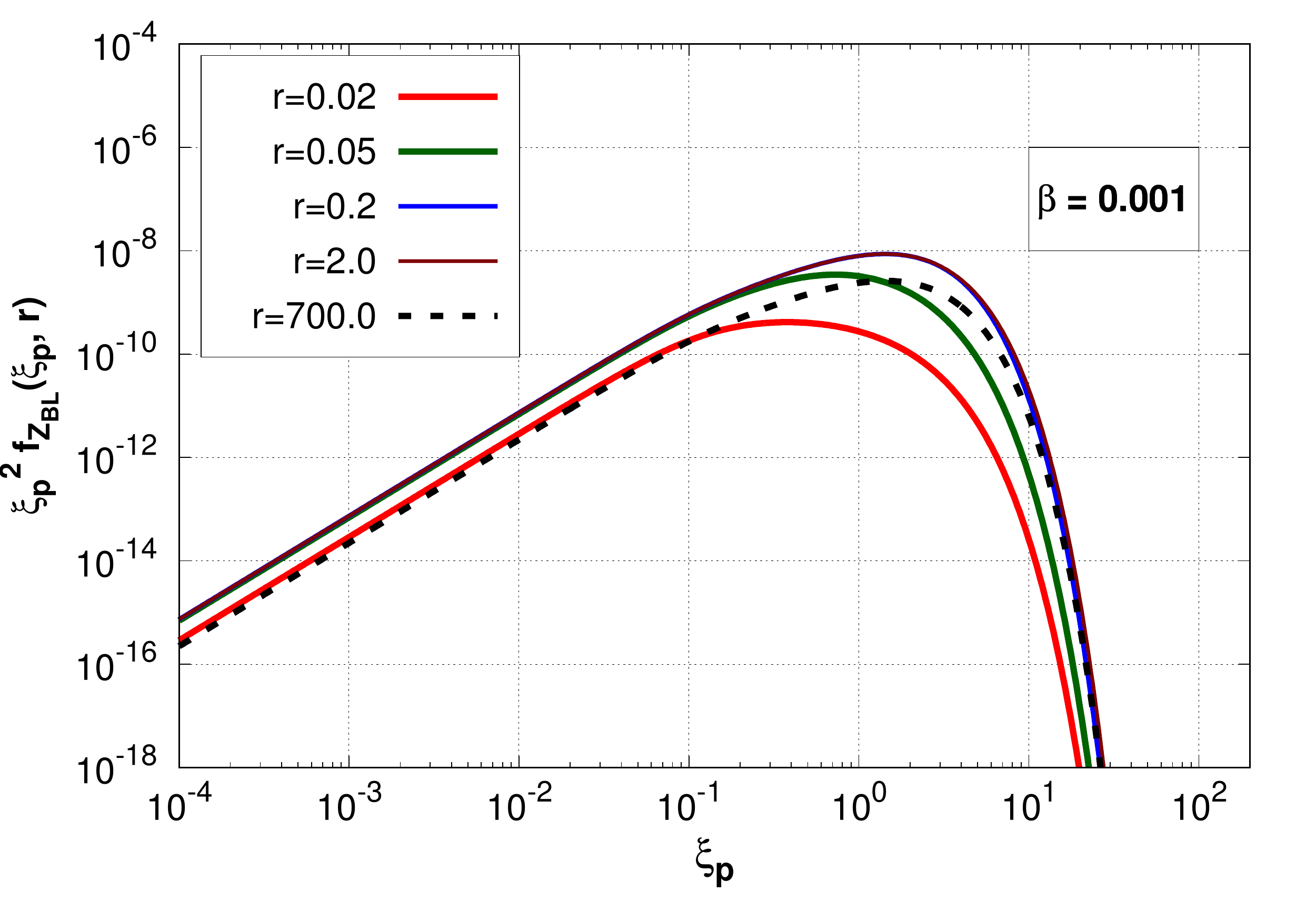}
	\caption{Non-thermal momentum distribution function
	$f_{\zbl}$ plotted as a function of the dimensionless
	variable $\xi_{p}$ for $\beta =1$ (left) and
	$\beta=10^{-3}$ (right). The curves are shown for
	different values of $r = \frac{M_{sc}}{T}$.}
	\label{beta1fzbl}
\end{figure}

Let us now try to solve the Boltzmann equations
(Eqs. ((\ref{fzbl})--(\ref{fpsi})))
numerically. The first step, of course, is to solve the
non-equilibrium momentum distribution function of $\zbl$.
Using Eq. (\ref{fzbl}) along with Eqs. ((\ref{c1}) and (\ref{c2})),
we solve for the non-thermal momentum distribution function of
$\zbl$ i.e. $f_{\zbl}$ as shown in Fig. \ref{beta1fzbl} (left)
for $\beta =1$. In the y-axis we have plotted $\xi_p^2 \, f_{\zbl}(\xi_p,r)$,
since area under this curve will readily give us an idea about
the number density of the particle species under consideration
(at a fixed temperature). Initially, at the onset, as $r$ increases
(i.e. the temperature of the Universe decreases), we expect
that more and more $\zbl$ will be produced from the decay
of $h_2$. In other words, the area under the curve should
increase. This is exactly what we see as we go from $r=0.02$ (red solid line)
to $r=0.05$ (green solid line) in the plot. Then, with further lowering of
temperature (increment in $r$), the process of depletion of
$\zbl$ through its decay starts to compete with the
production, and hence, no appreciable change in the number
density is expected. This is reflected in the curves
corresponding to $r=0.2$ (blue solid line) and $r=2.0$
(brown solid line). At a much lower
temperature, production of $\zbl$ almost ceases due the
Boltzmann suppression of $h_2$ abundance. So $\zbl$ gets
depleted through its decay, and number density is expected
to fall. This is observed in Fig. \ref{beta1fzbl} (left)
for the black dotted line corresponding to $r=700$.
Similar plot for the $\beta =10^{-3}$ case is also
shown in Fig. \ref{beta1fzbl} (right).

\begin{figure}[h!]
	\centering
	\includegraphics[height=8.5cm,width=6.5cm,angle=-90]
	{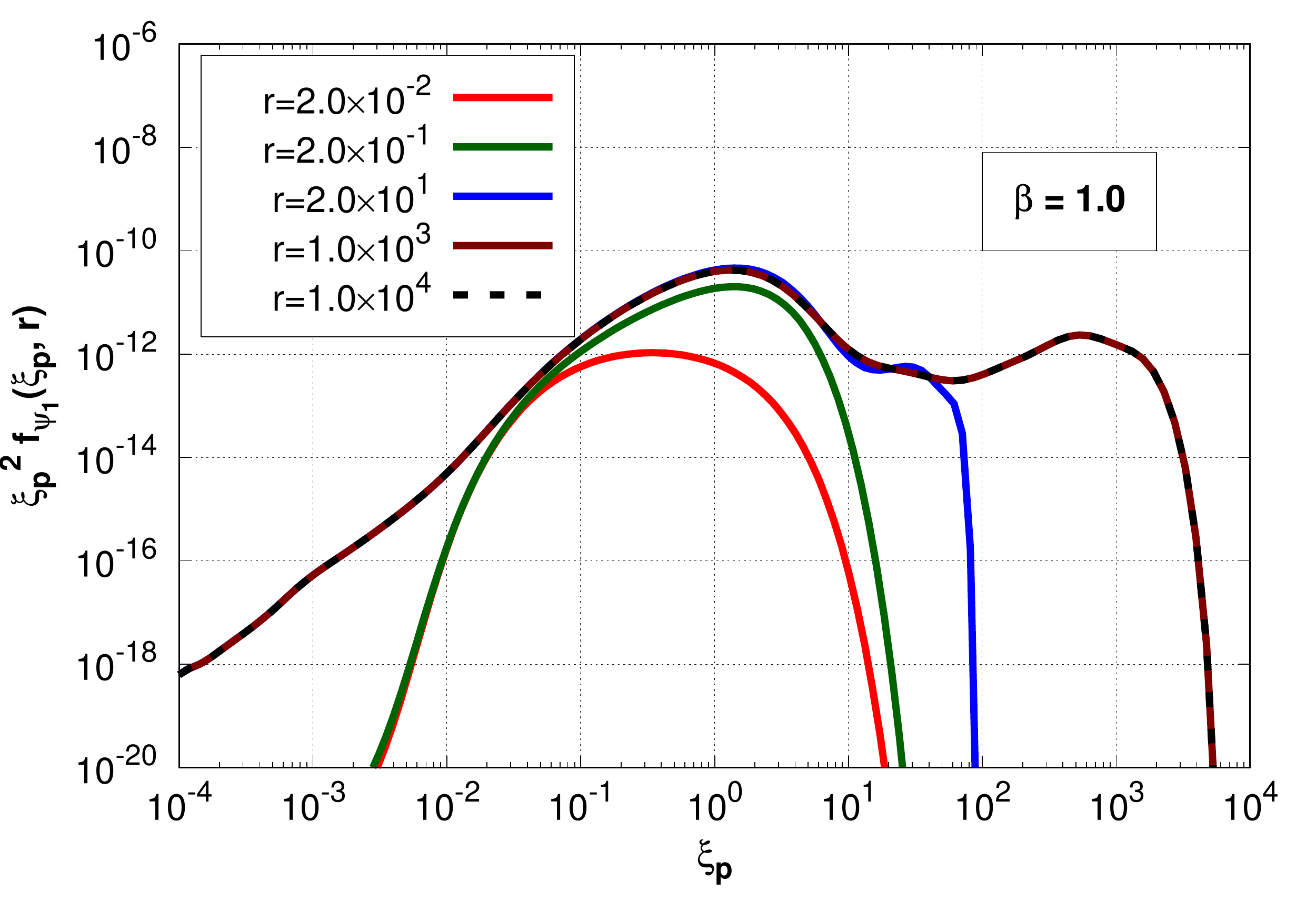}
	\includegraphics[height=8.5cm,width=6.5cm,angle=-90]
	{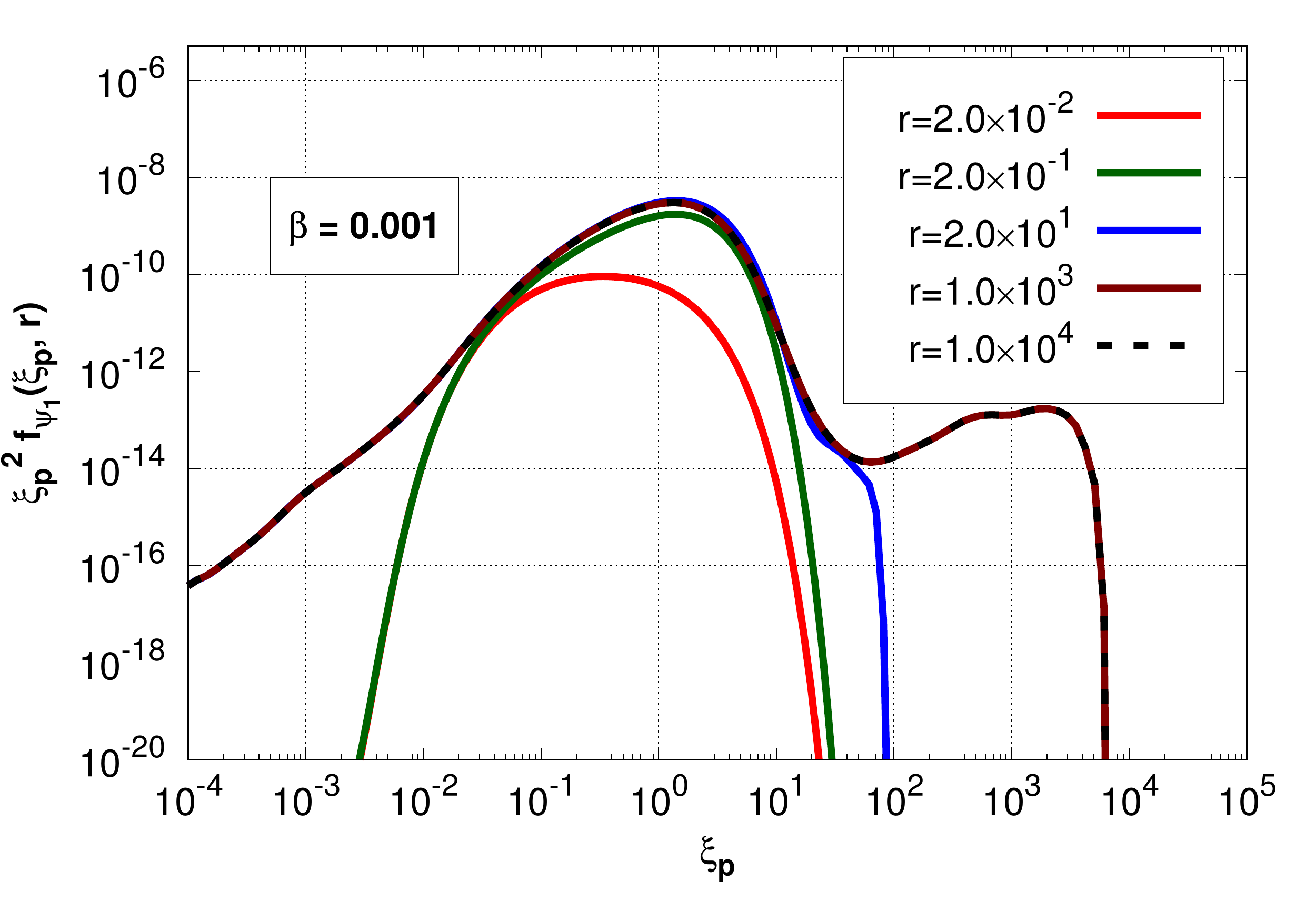}
	\caption{Non-thermal momentum distribution function
	$f_{\psi_1}$ plotted as a function of the dimensionless
	variable $\xi_{p}$ for $\beta=1$ (left) and $\beta=10^{-3}$
	(right). The curves are shown for different values of
	$r = \frac{M_{sc}}{T}$.}
	\label{beta1psi}
\end{figure}

With this, we now proceed to find the non-equilibrium distribution
function for our dark matter particle $\psi_1$ using Eq. (\ref{fpsi}).
It is shown in Fig. \ref{beta1psi} (left) for $\beta=1$.
Similar to the $\zbl$ case, here also with the decrease of
temperature more and more $\psi_1$ particles are produced
from the decays massive bosons such as $h_1$, $h_2$
and $\zbl$. Hence the area under the curves
increases as we go from $r=0.02$ to $r=1000$. With further
increase in $r$ we expect that the {\it rate}
of production of $\psi_1$ should decrease and
consequently the (comoving) number density will
cease to change, since for this high value of $r$
(low temperature) the number densities of all the
decaying bosons have become extremely dilute
and also $\psi_1$ itself is stable. This can be verified,
if we compare the curves corresponding to $r=10^3$ and
$r=10^4$. Similar features are also observed for
$\beta=10^{-3}$ case which has been shown in
Fig.\,\,\ref{beta1psi} (right).

\begin{figure}[h!]
	\centering
	\includegraphics[height=8.5cm,width=6.5cm,angle=-90]
	{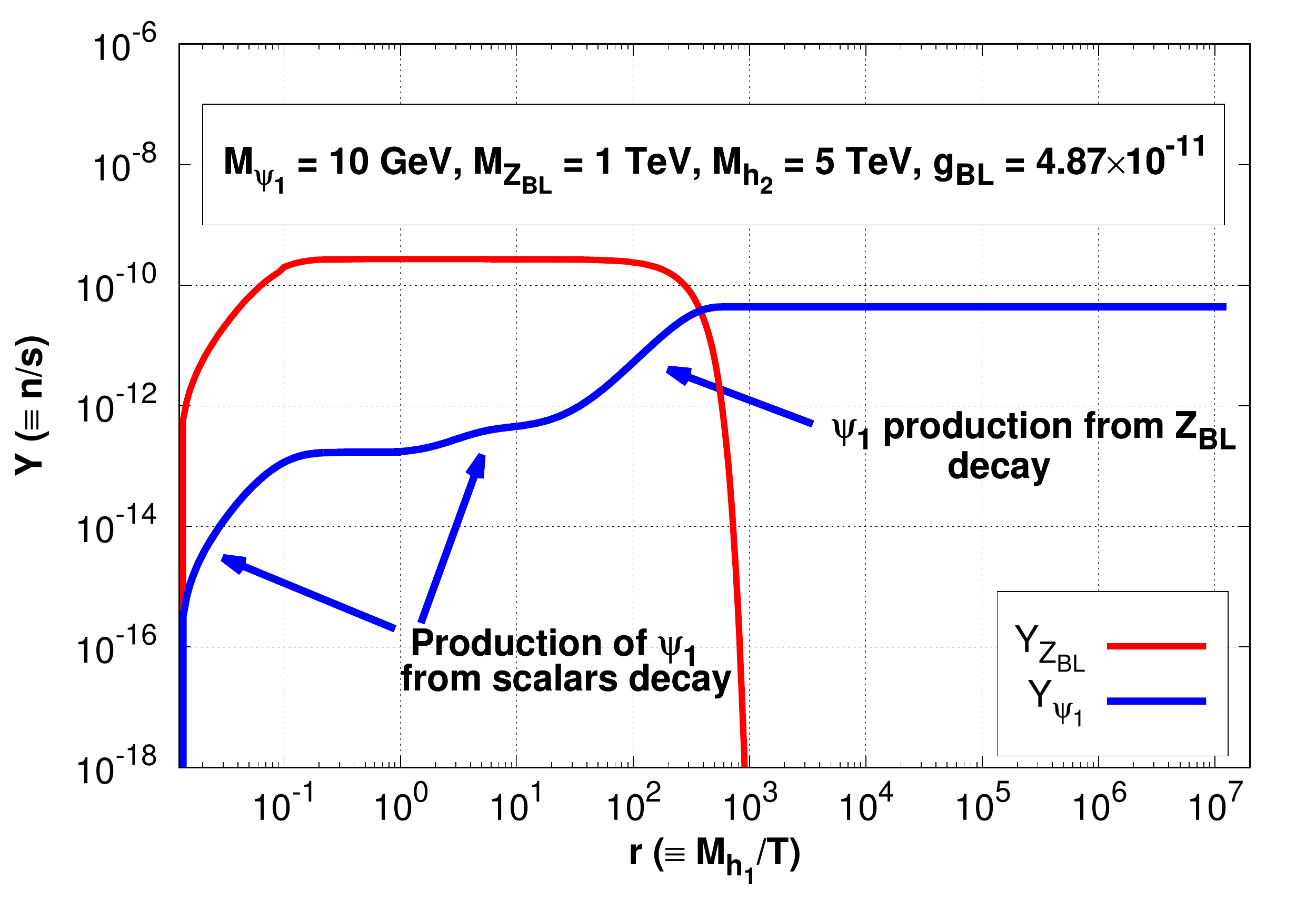}
	\includegraphics[height=8.5cm,width=6.5cm,angle=-90]
	{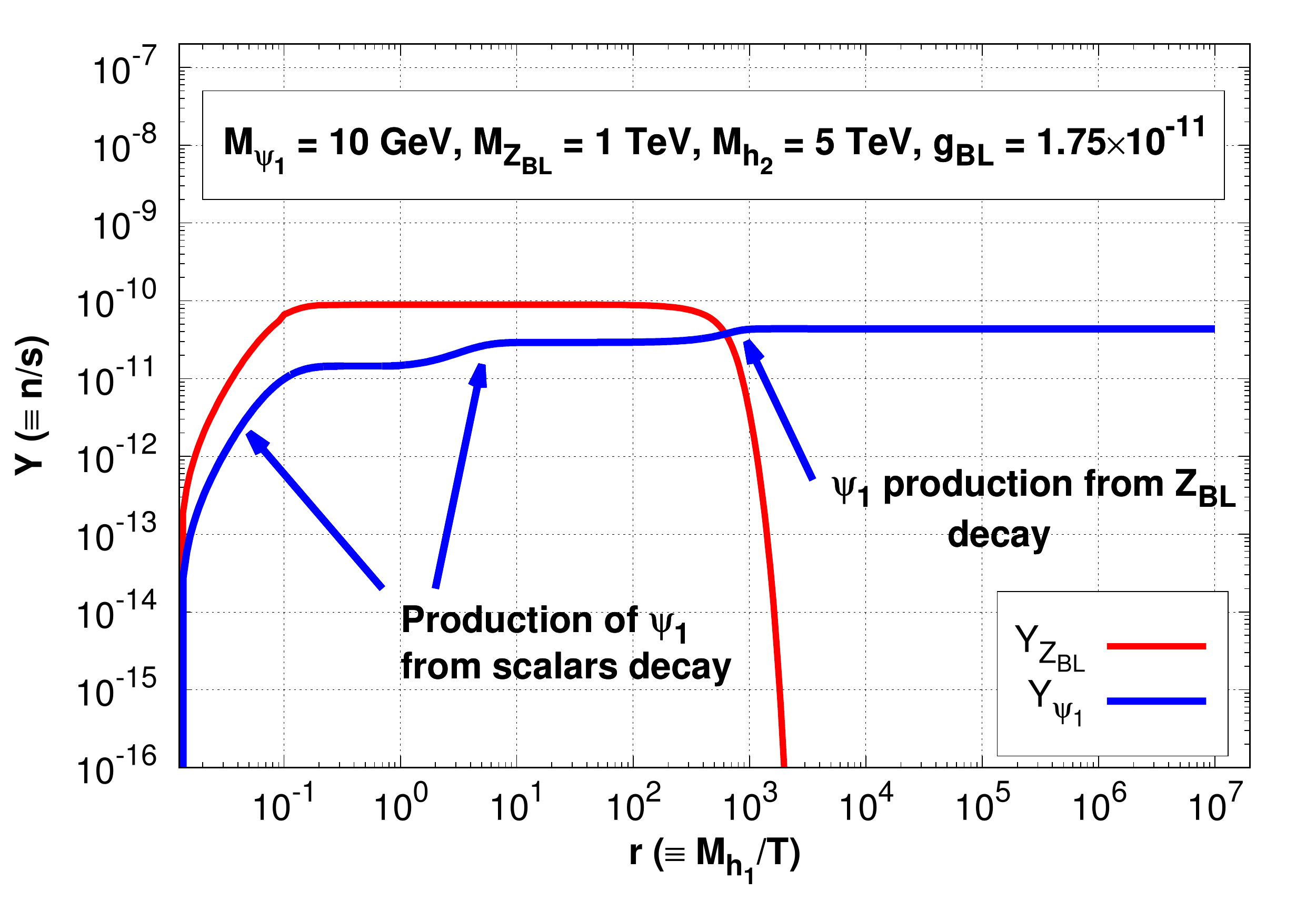}
	\caption{Variation of comoving number density
	of $\zbl$ and $\psi_1$ with respect to $r$.
	{\bf Left}: $\beta=1$ and {\bf Right}: $\beta=10^{-3}$.}
	\label{beta1y}
\end{figure}

All of the features that we have discussed so far with
respect to the momentum distribution functions are
reflected clearly if we plot the variation of the comoving
number density of $\zbl$ and $\psi_1$ with respect to $r$.
The comoving number density $Y$ is easily calculable by using
Eqs. ((\ref{n}) and (\ref{s})), once the momentum distribution
function of the corresponding species is known . We plot our
numerical results in both panels of Fig. \ref{beta1y}.

For the $\beta=1$ case, the scalars $h_1$ and $h_2$
contribute minimally to the comoving number density of
$\psi_1$. The bulk of the contribution comes from
$\zbl$. In the left panel of Fig. \ref{beta1y},
we find that the comoving number density of
$\zbl$ first rises with $r$. Initially, there is also a
similar rise in the number density of $\psi_1$ as well. However,
the rate of increment of $Y_{\dm}$ is small compared to
$Y_{\zbl}$ for $r\leq0.1$ since in this regime, the main production
channel of $\psi_1$ is the decay from BSM scalar $h_2$,
which is presently contributing very little to $Y_{\psi}$.
Then as $r$ increases, the number density of $\zbl$
flattens out due to the competing decay and production terms
while $Y_{\psi_1}$ rises slightly due to its
production from the decay of SM-like Higgs boson $h_1$.
With the further increase of $r$, the
$\zbl$ number density falls off as
the decay modes of $\zbl$ become
dominant over its production process (i.e. production from
the Boltzmann suppressed $h_2$).
Consequently, there is a sharp rise
in $Y_{\psi}$ as more and more $\psi_1$
starts producing dominantly from $\zbl$ decay.
Finally, for $r > 10^3$ there is practically
no $\zbl$ is left for decay to $\dm$,
and hence in absence of any sources
$Y_{\psi_1}$ freezes-in to a constant value.
For the other case i.e. when $\beta=10^{-3}$,
the situation is exactly same as with $\beta=1$
except in this case all the production
modes of $\dm$ including those from the
decays of $h_1$ and $h_2$ contribute
equally to $Y_{\dm}$. Therefore for
$r\leq10^2$, $Y_{\dm}$ increases
significantly since in this regime
$\dm$ is mainly produced from scalars decay.
Moreover unlike $\zbl$, as the $h_1$ and $h_2$
are in thermal equilibrium, in both panels,
bulk of their contribution to the number density
of $\dm$ occurs when the temperatures of the Universe
are around  $T \sim M_{h_1}$ ($r\sim 1.0$) and
$T \sim M_{h_2}$ ($r\sim 0.03$) respectively.
\begin{figure}[h!]
	\centering
	\includegraphics[height=8.5cm,width=6.5cm,,angle=-90]
	{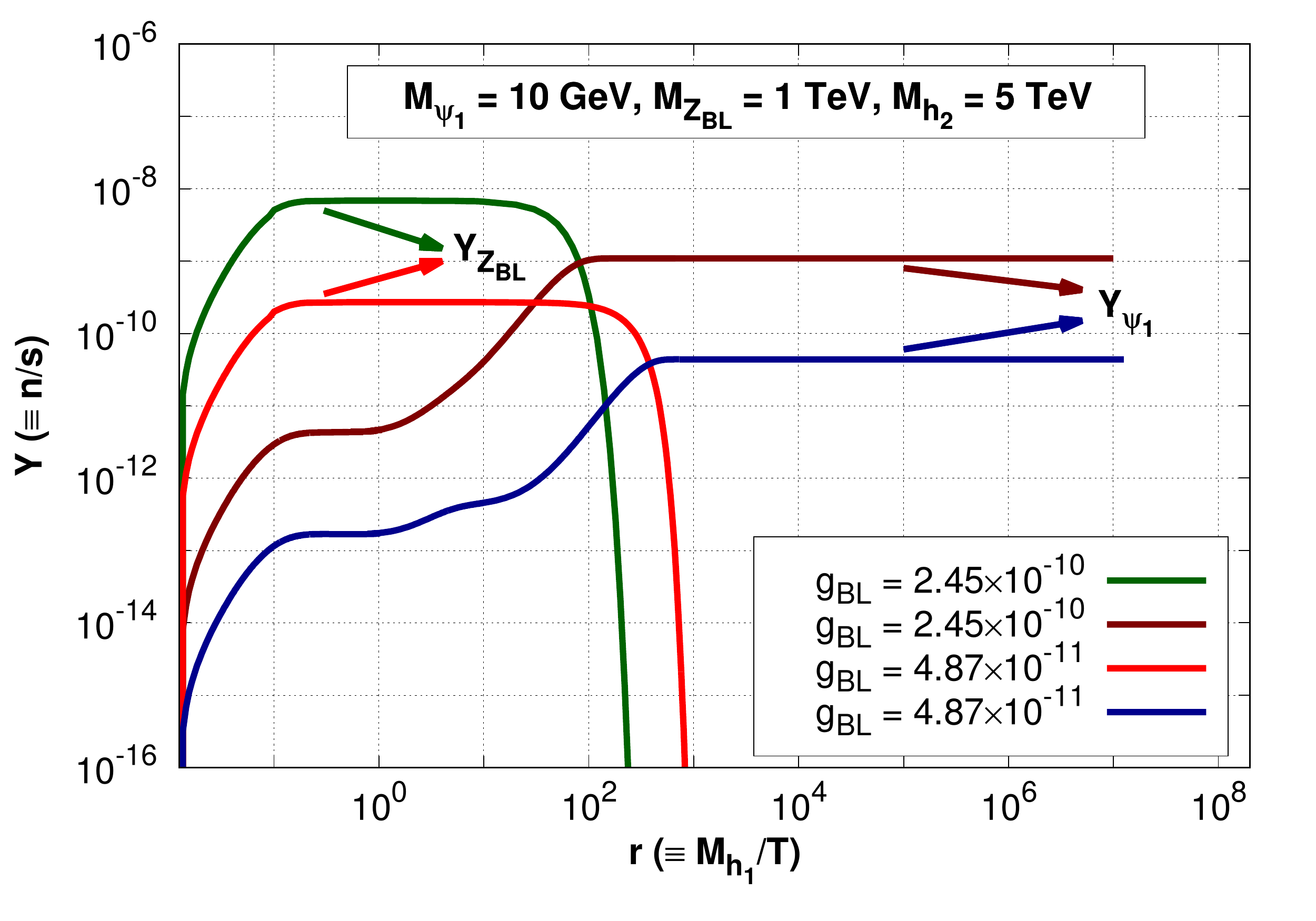}
	\includegraphics[height=8.5cm,width=6.5cm,angle=-90]
	{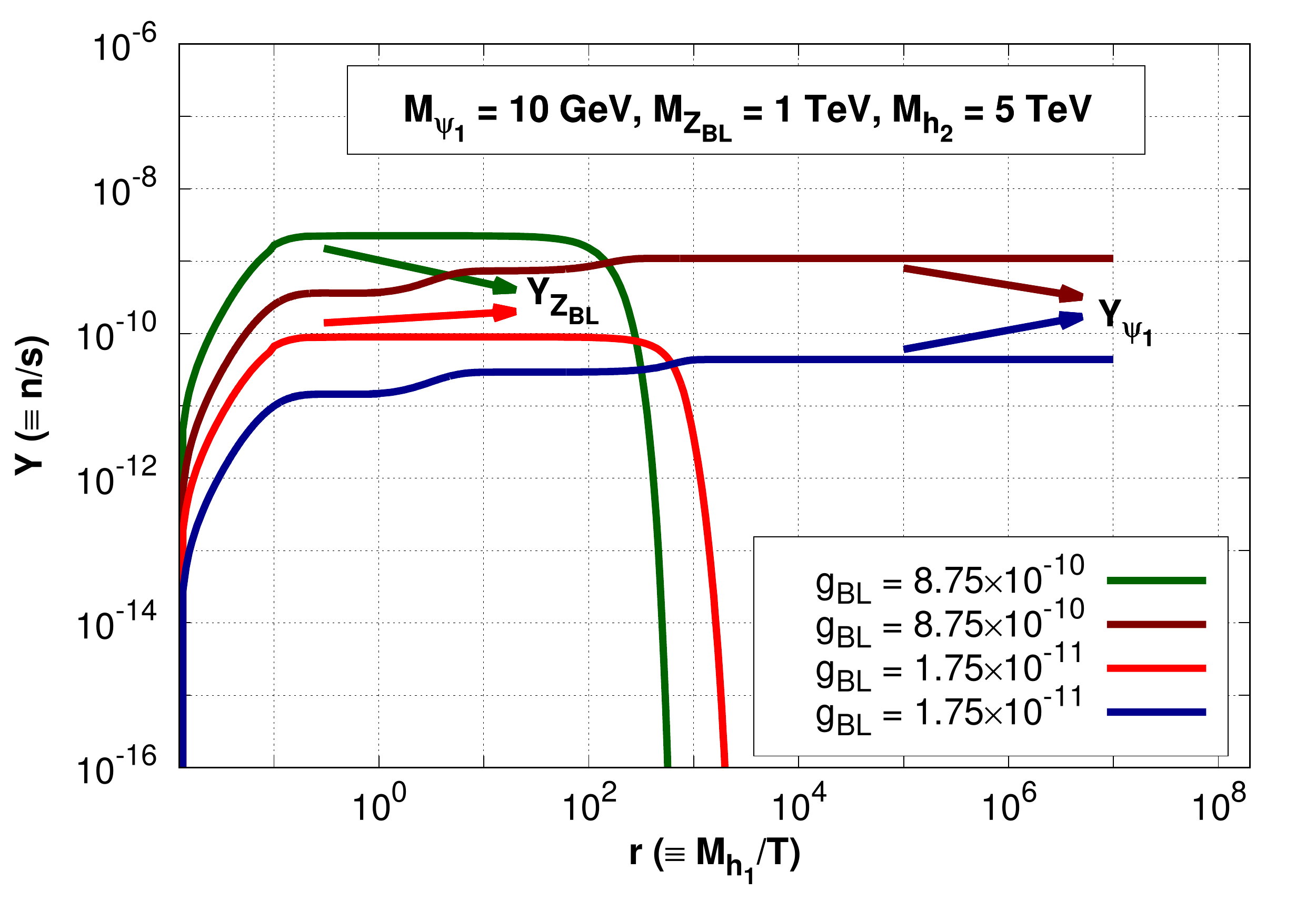}
	\caption{Variation of comoving number density of
	$\zbl$ and $\psi_1$ with $r$ corresponding to
	different values of $\gbl$. {\bf Left}: $\beta=1$ and
	{\bf Right}: $\beta=10^{-3}$.}
	\label{gbl}
\end{figure}

Let us now try to understand how the comoving
number density varies with different model parameters.\,\,Parameters
have varied one at a time, while keeping
the others fixed at their benchmark values. In
Fig. \ref{gbl}, we plot the variation of $Y$ with
varying $\gbl$. Increasing $\gbl$ will result in
an increase in the collision term corresponding
to $h_2 \rightarrow \zbl \zbl$ (since $g_{h_2\zbl\zbl}$
increases, see Appendix \ref{cc}) and hence an enhanced
initial production of $\zbl$. Also, increasing $\gbl$
will enlarge the total decay width of $\zbl$
(see Appendix \ref{width} for the expression of
$\zbl \rightarrow \, all$), and consequently we expect that
the produced $\zbl$ will start to deplete earlier
in the case where $\gbl$ is higher. The curves corresponding
to $Y_{\dm}$ follow the rise of $\zbl$ and in the
case where $\gbl$ is higher, more $\dm$ is produced
in the final state (since there is a corresponding
increment in the production of $\zbl$). As $\zbl$ depletes
off, $\dm$ freezes in to a particular value of
$Y_{\dm}$ as expected. In the right panel of
Fig. \ref{gbl}, the initial sharp rise
of $Y_{\dm}$ for $r\leq 10$ is due to the significant
production of DM from the decaying scalars $h_2$
and $h_1$ respectively. 
\begin{figure}[h!]
	\centering
	\includegraphics[height=8.5cm,width=6.5cm,angle=-90]
	{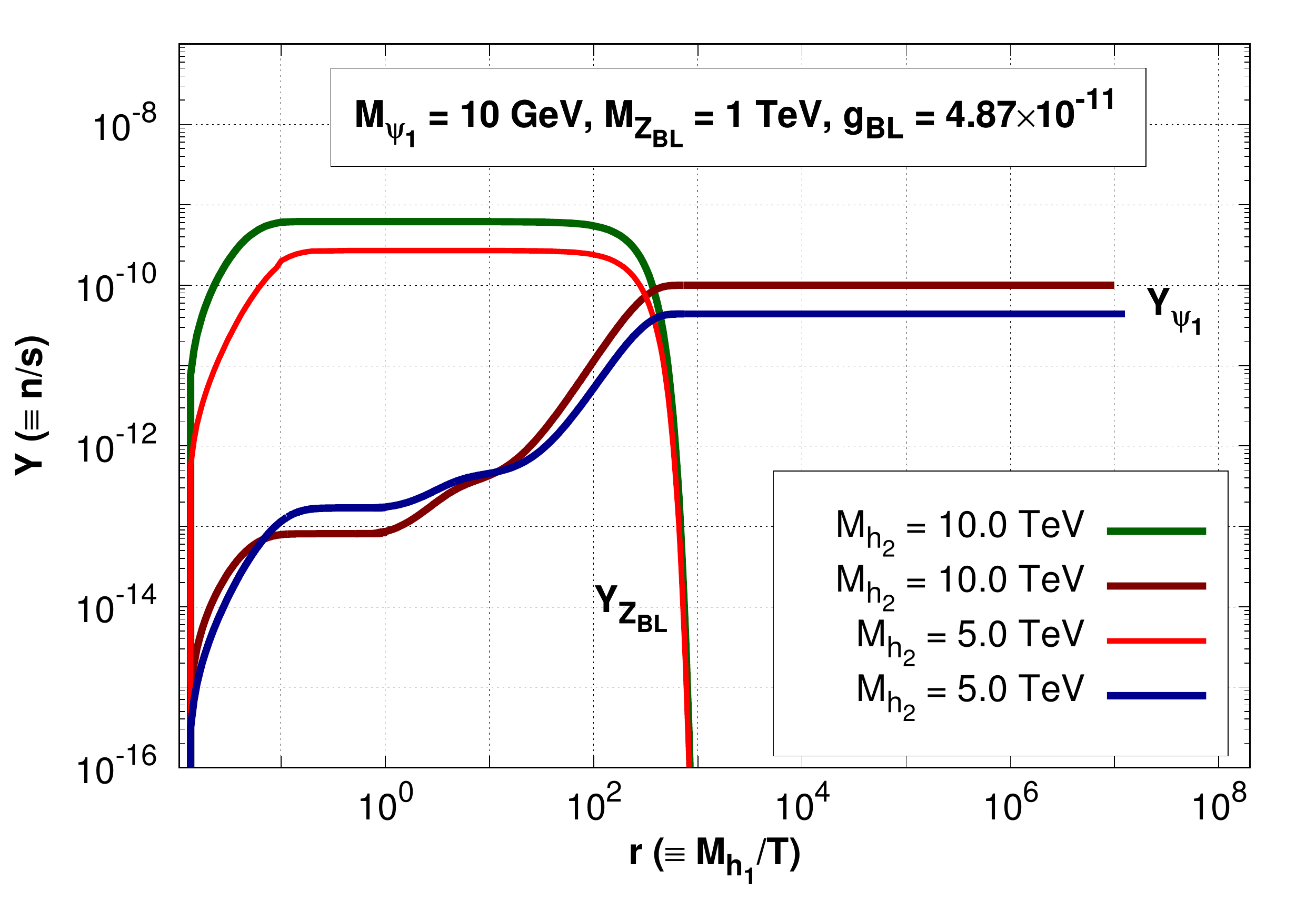}
	\includegraphics[height=8.5cm,width=6.5cm,angle=-90]
	{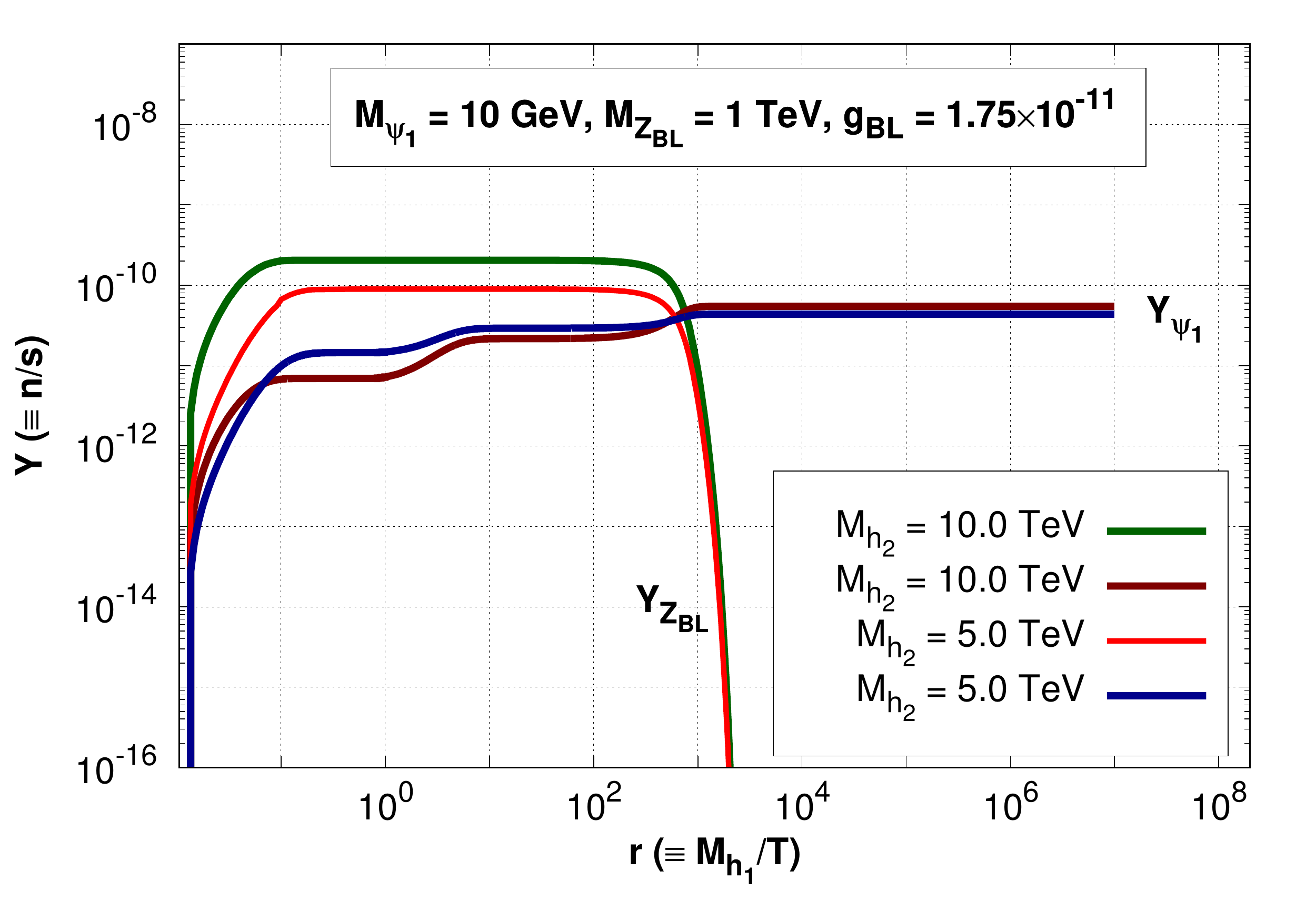}
	\caption{Variation of comoving number density of $\zbl$
	and $\psi_1$ with $r$ corresponding to different values of
	$M_{h_2}$. {\bf Left}: $\beta=1$ and {\bf Right}: $\beta=10^{-3}$.}
	\label{mh2}
\end{figure}
\paragraph{}
In Fig. \ref{mh2}, we have plotted the variations in
$Y$ by changing $M_{h_2}$. Increasing $M_{h_2}$ will
again increase $\mathcal{C}^{h_2 \rightarrow \zbl\,\zbl}$
like the previous case. But unlike before, $Y_{\zbl}$
curves corresponding to the two $M_{h_2}$ values start falling
around the same epoch. This is because, changing
the mass of $h_2$ has no bearing upon the
total decay width of $\zbl \rightarrow \, all$,
while the latter process is responsible for the fall off.
Since more $\zbl$ is produced initially when $M_{h_2}$
is increased, the yield of $\dm$ in this case is also higher,
as is evident from the figure. A qualitative difference between
the right and left panel of Fig. \ref{mh2} is that,
the final abundances of $\dm$ in $\beta=1$ case are
quite different from each other for different values of
$M_{h_2}$, while in $\beta=10^{-3}$ case, we see that
they are almost identical. This is because, in the $\beta=1$ case,
the contribution of the scalars are sub-dominant compared to
$\zbl$, while abundance of the
latter and consequently that of $\dm$ increases with
increasing $M_{h_2}$. Hence the amount of splitting in the
two $Y_{\dm}$ curves (left panel) is almost same as
the difference observed in the corresponding $Y_{\zbl}$
curves. But in the $\beta=10^{-3}$ scenario, things
are a little different. Here, both the scalars as
well as $\zbl$ contribute substantially to the final
abundance of $\dm$. The contribution to the final
abundance from the decays of the two scalars compensates
to reduce the splitting amongst the $Y_{\dm}$ curves
arising from the increment of $Y_{\zbl}$.
\begin{figure}[h!]
	\centering
	\includegraphics[height=8.5cm,width=6.5cm,angle=-90]
	{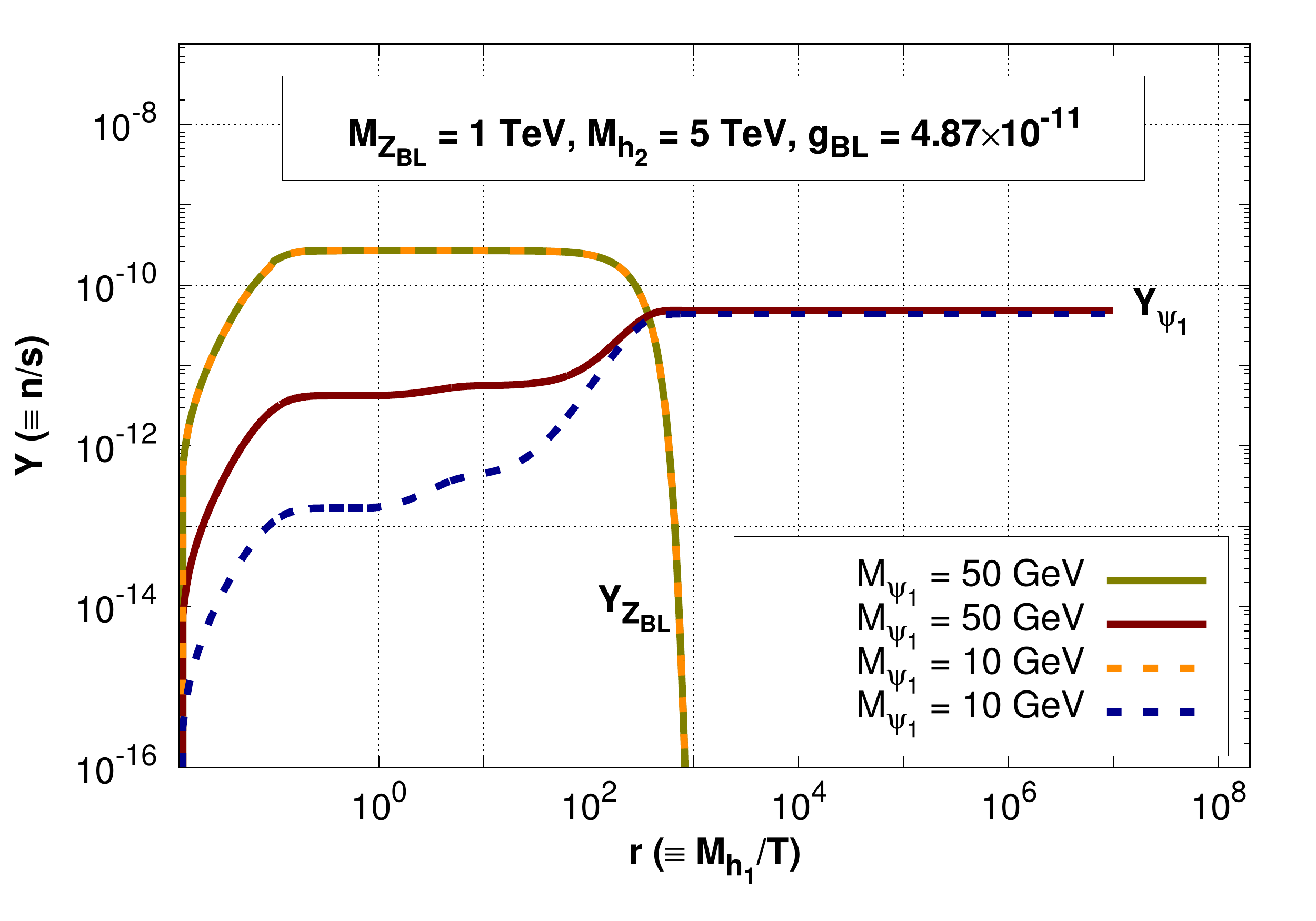}
	\includegraphics[height=8.5cm,width=6.5cm,angle=-90]
	{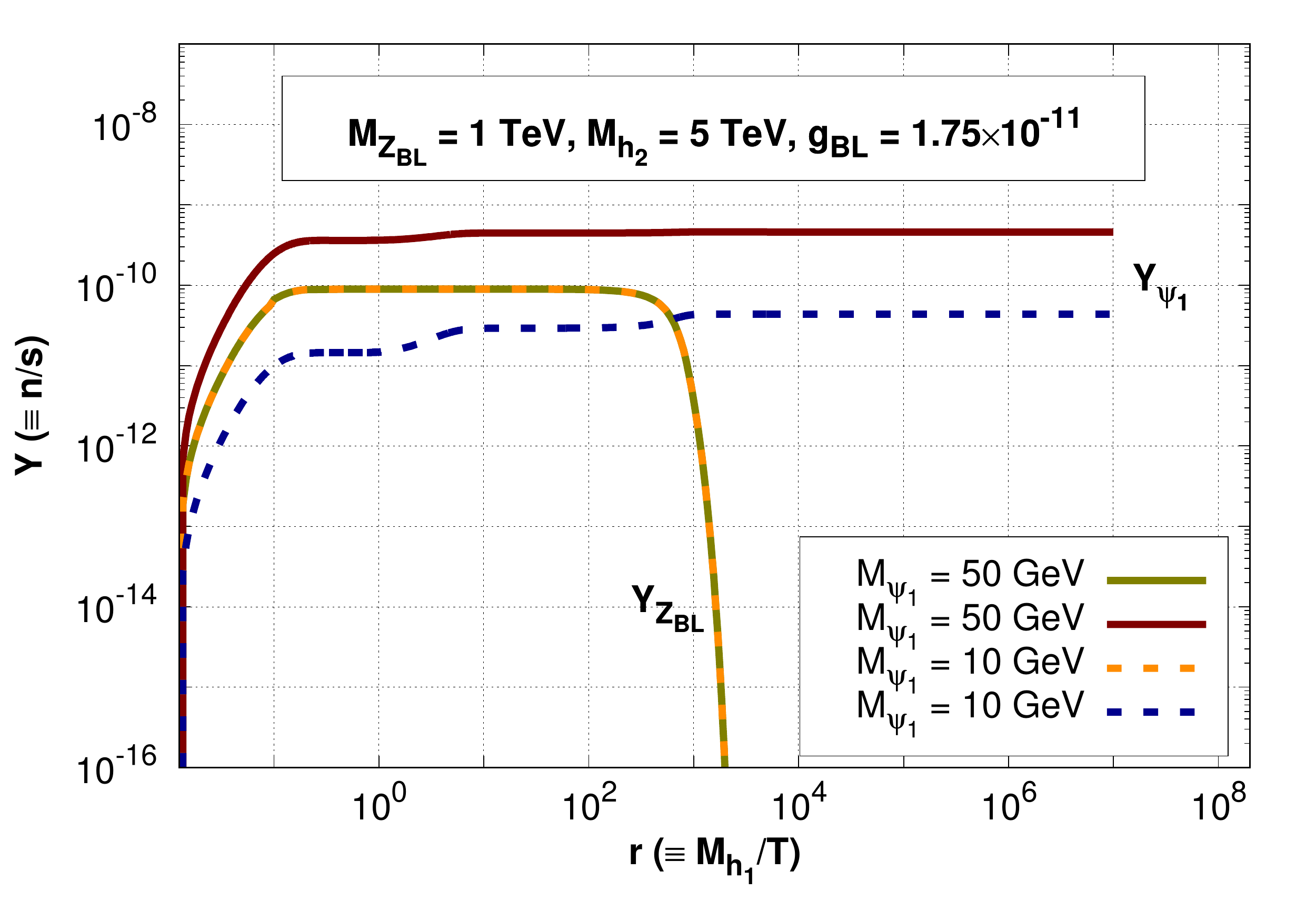}
	\caption{Variation of comoving number density of
	$\zbl$ and $\psi_1$ with $r$ corresponding to different
	values of $M_{\dm}$. {\bf Left}: $\beta=1$ and {\bf Right}: $\beta=10^{-3}$.}
	\label{mdm}
\end{figure}

In Fig. \ref{mdm}, the variation of $Y_{\zbl}$ and $Y_{\dm}$
have been studied by changing the mass of dark matter
itself i.e. $\mdm$. Now in the present scenario
with $\mzbl\gg M_{\dm}$,
any change in the mass of dark matter will in no
way affect $Y_{\zbl}$, since $M_{\dm}$ neither affects the
$\zbl$ total decay width nor does it change
$h_2 \rightarrow \zbl \zbl$ collision term. But the production
of $\dm$ from the scalars decay is however affected. It is
clear from the expression of the scalar--dark matter couplings
given in Appendix \ref{cc}, that with increase in $\mdm$,
the value of the coupling increases and there by
yielding more $\dm$. This observation is corroborated
if we look at the blue dashed line (corresponding to
$\mdm = 10 \rm \,GeV$) and the solid grey line
(corresponding to $\mdm = 50 \rm \, GeV$) in the
left panel of Fig.\,\,\ref{mdm}. For more massive dark matter,
the yield of $\dm$ is higher from the scalars decay.
But there is no effect on the production of $\dm$ from
$\zbl$ decay, which is expected, since the couplings between
$\zbl$ and $\dm$ do not depend on the mass of the latter and
also here $\mzbl\gg M_{\dm}$ . Let us now contrast this case
with the right panel of Fig. \ref{mdm}. Here again
as before the contribution of the scalars become
important. As already mentioned, the benchmark for
$\beta=10^{-3}$ scenario is chosen in such a way,
so that $h_1$, $h_2$ and $\zbl$ contribute equally
in the final relic abundance. Increase in scalar--dark matter
couplings (due to an increase in $\mdm$), hence makes
the contribution from the scalars decay, larger
than that from the $\zbl$.
\begin{figure}[h!]
	\centering
	\includegraphics[height=8.5cm,width=6.5cm,angle=-90]
	{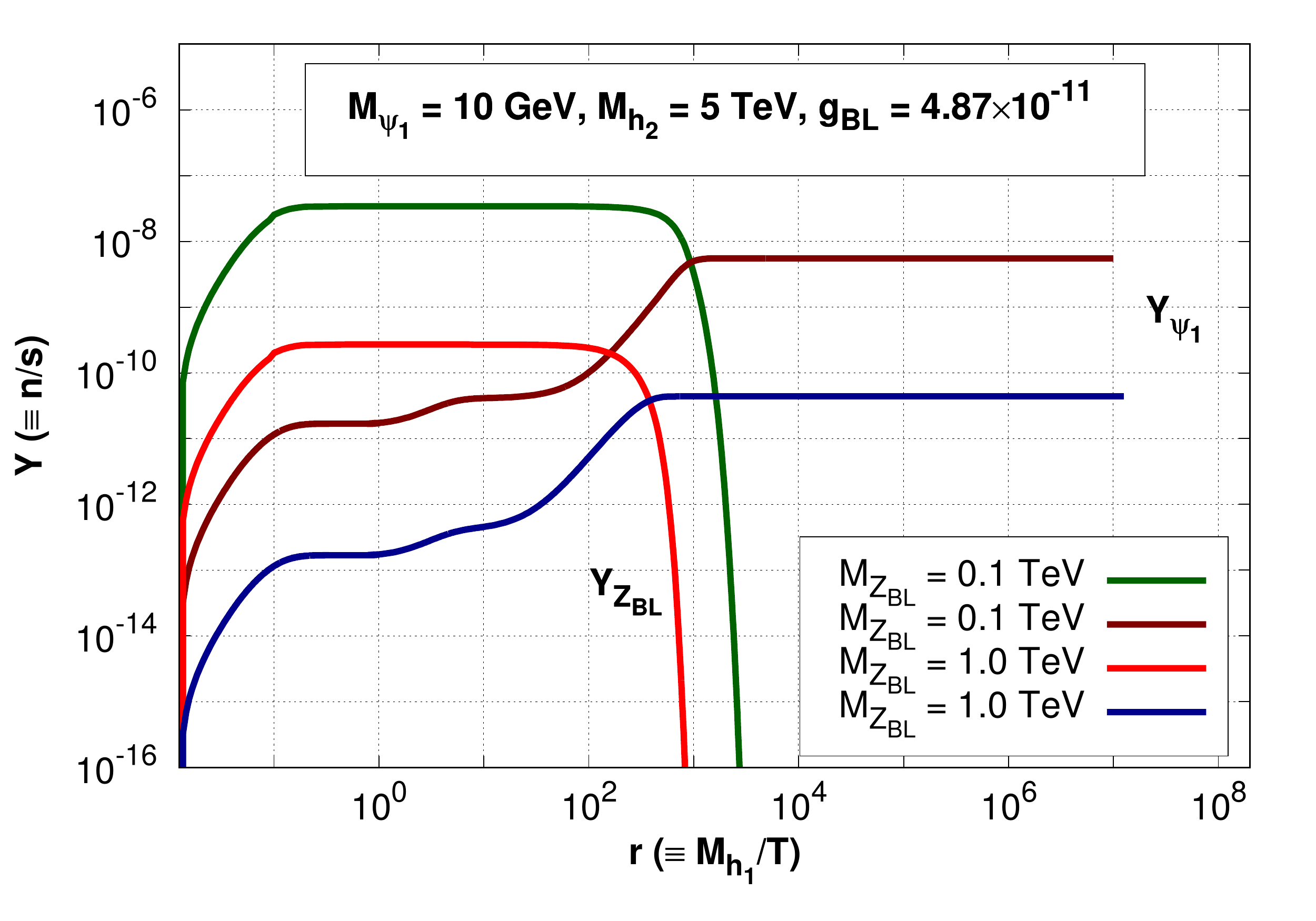}
	\includegraphics[height=8.5cm,width=6.5cm,angle=-90]
	{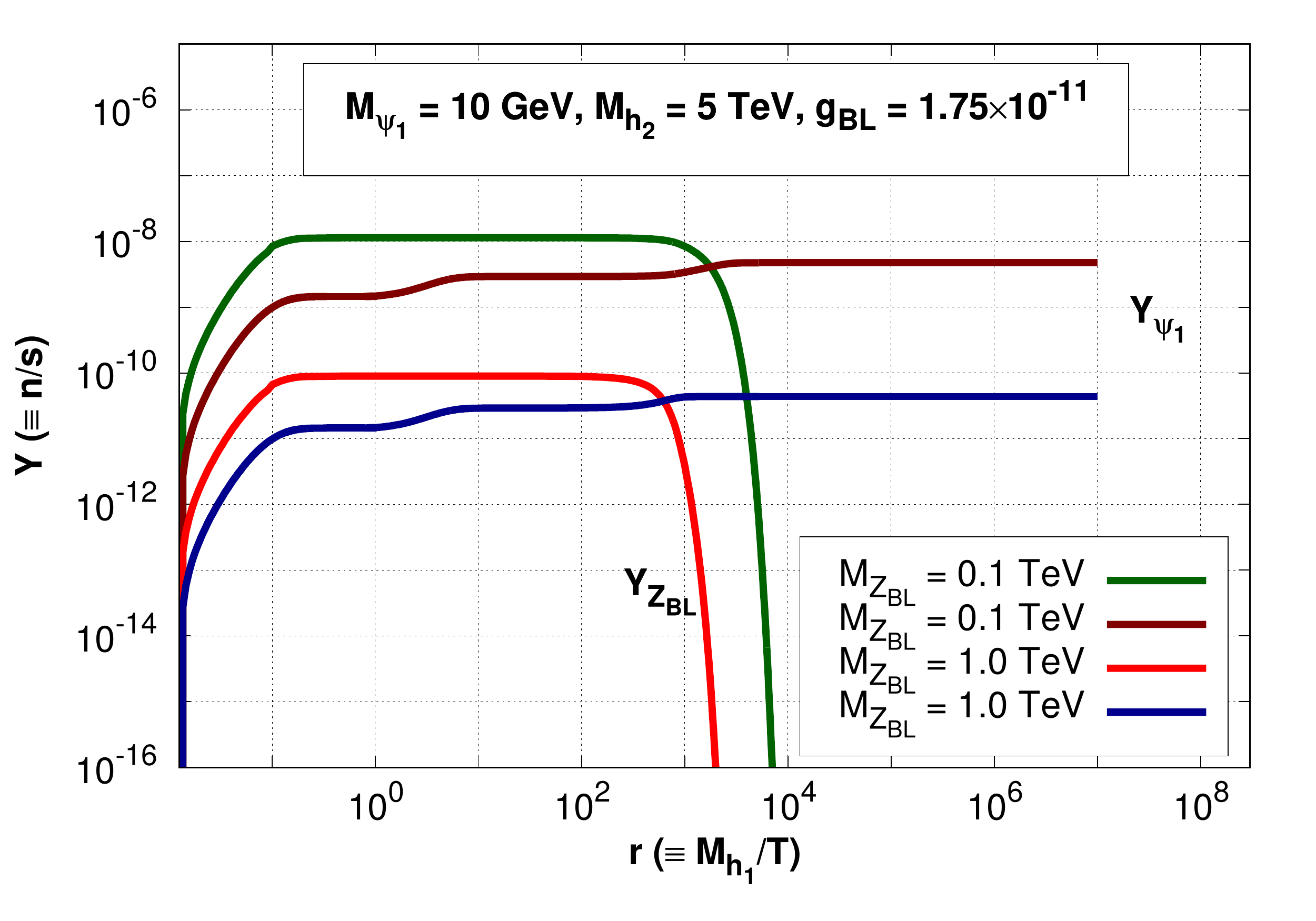}
	\caption{Variation of comoving number density of
	$\zbl$ and $\psi_1$ with $r$ corresponding to
	different values of $M_{\zbl}$. {\bf Left}:
	$\beta=1$ and {\bf Right}: $\beta=10^{-3}$.}
	\label{mzbl}
\end{figure}

In Fig. \ref{mzbl}, variation with respect to $\mzbl$ is
demonstrated. With an increase in $\mzbl$, the total decay width
$\zbl \rightarrow \, all$ increases leading to an earlier
fall in the comoving number density of $\zbl$. Also increasing
$\mzbl$ suppresses the production of $\zbl$ via $h_2$ decay.
$Y_{\dm}$, on the other hand tracks the rise and fall of
$Y_{\zbl}$ (since $\zbl$ is the main production channel
of $\dm$ in the left panel with $\beta =1$ case). For $\beta=10^{-3}$
(right panel), $Y_{\zbl}$ exhibits similar features. The scalar--dark matter
couplings on the other hand increases with an decrease in
$\mzbl$. This leads to higher yield of $\dm$ from scalars
decay. The yield from $\zbl$ decay, for reasons discussed
before, also increases due to a decrease in $\mzbl$.
All these are shown in the right panel of Fig. \ref{mzbl}.

\begin{figure}[h!]
	\captionsetup[subfigure]{justification=centering}
	\centering
	\subfloat[Variation with different $\theta_{13}$
	values for $\beta=1$][Variation with different $\theta_{13}$
	values for \\ $\beta=1$]
	{%
		\includegraphics[height=8.5cm,width=6.5cm,angle=-90]
		{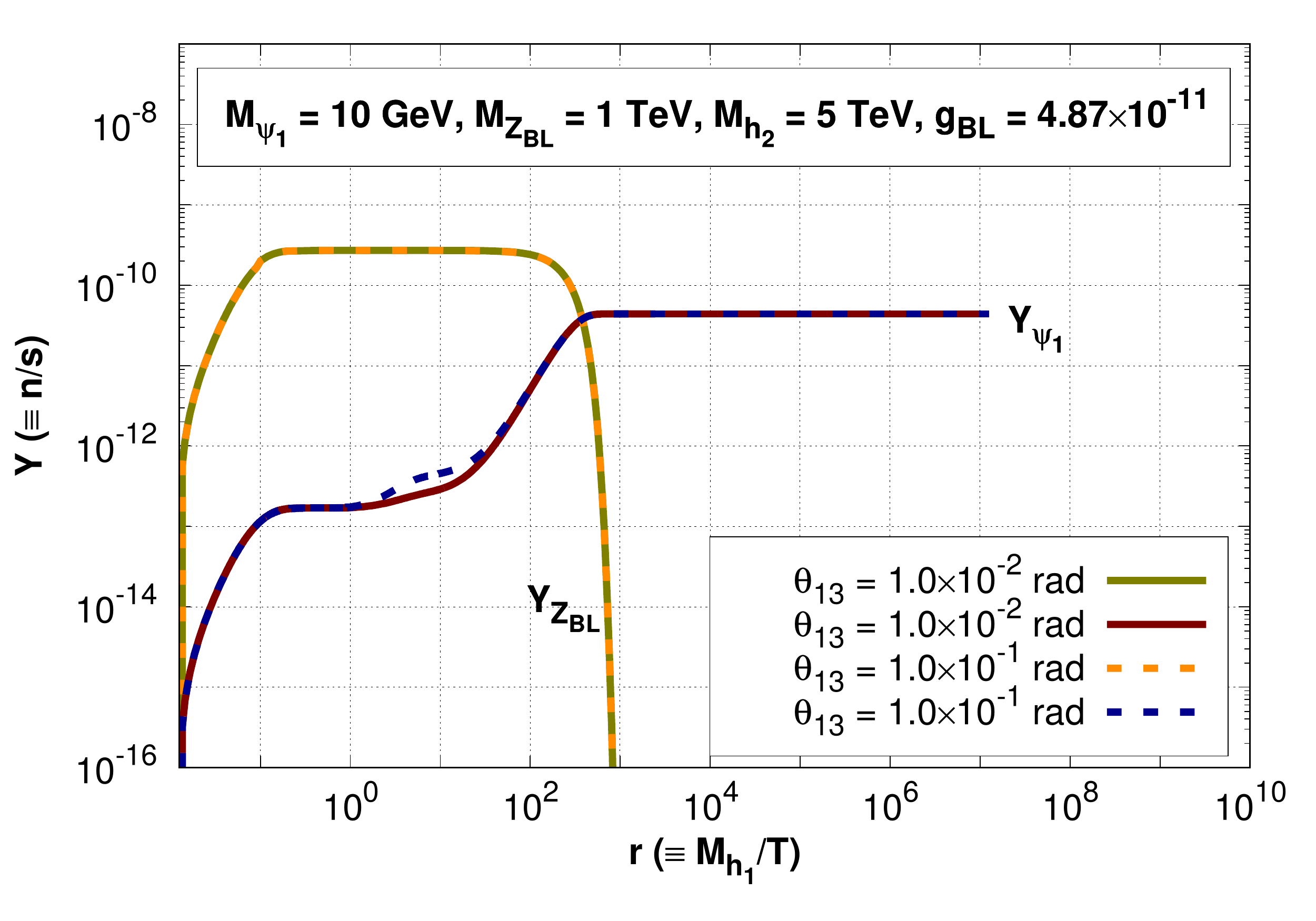}%
	}%
	\subfloat[Variation with different $\theta_{13}$
	values for $\beta=10^{-3}$][Variation with different
	$\theta_{13}$ values for \\ $\beta=10^{-3}$]
	{%
		\includegraphics[height=8.5cm,width=6.5cm,angle=-90]
		{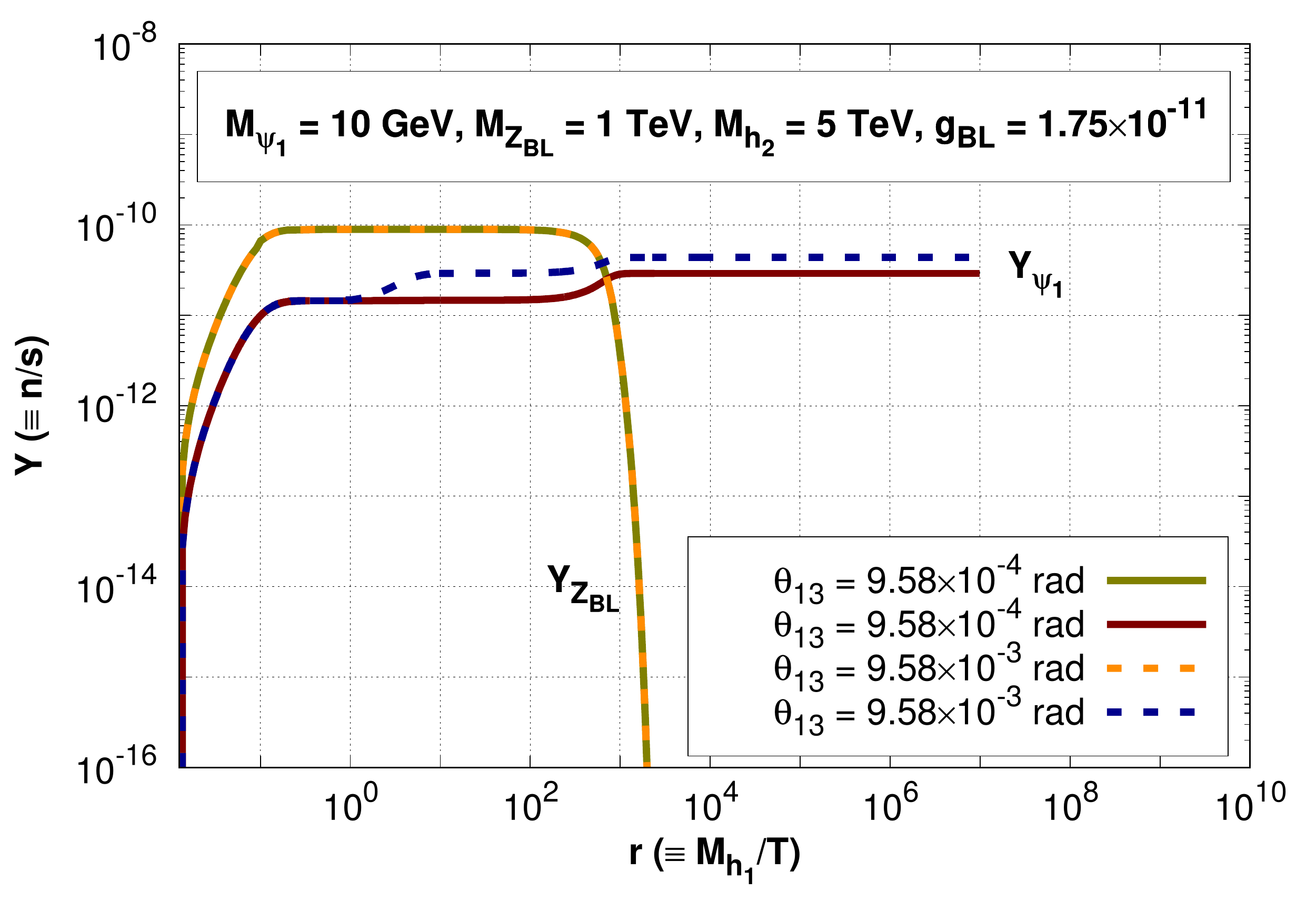}%
	} \\
	\subfloat[Variation with different $\theta_{13}$
	values for $\beta=1$][Variation with different
	$\theta_{23}$ values for \\ $\beta=1$]
	{%
		\includegraphics[height=8.5cm,width=6.5cm,angle=-90]
		{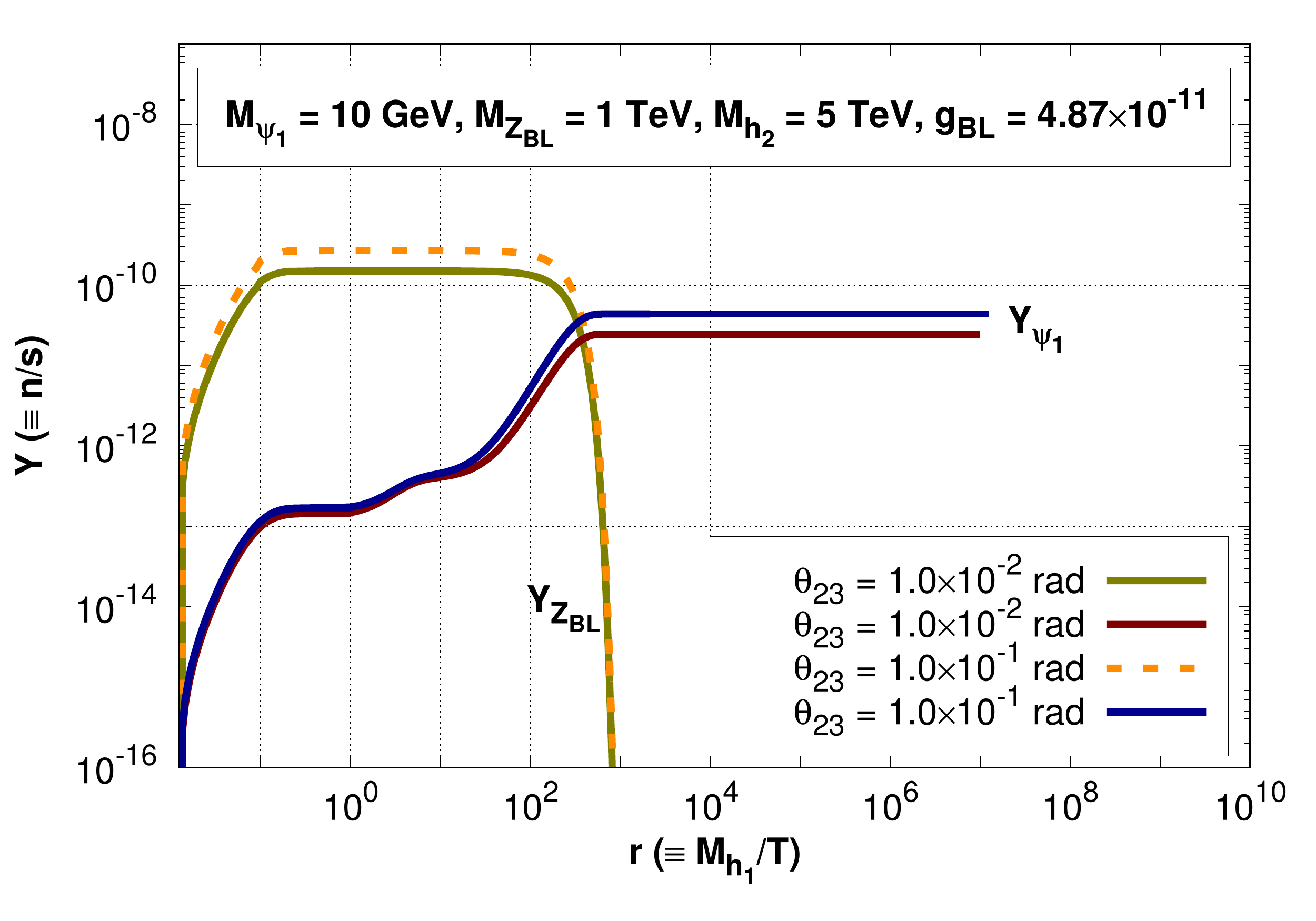}%
	}%
	\subfloat[Variation with different $\theta_{13}$
	values for $\beta=10^{-3}$][Variation with different
	$\theta_{23}$ values for \\ $\beta=10^{-3}$]
	{%
		\includegraphics[height=8.5cm,width=6.5cm,angle=-90]
		{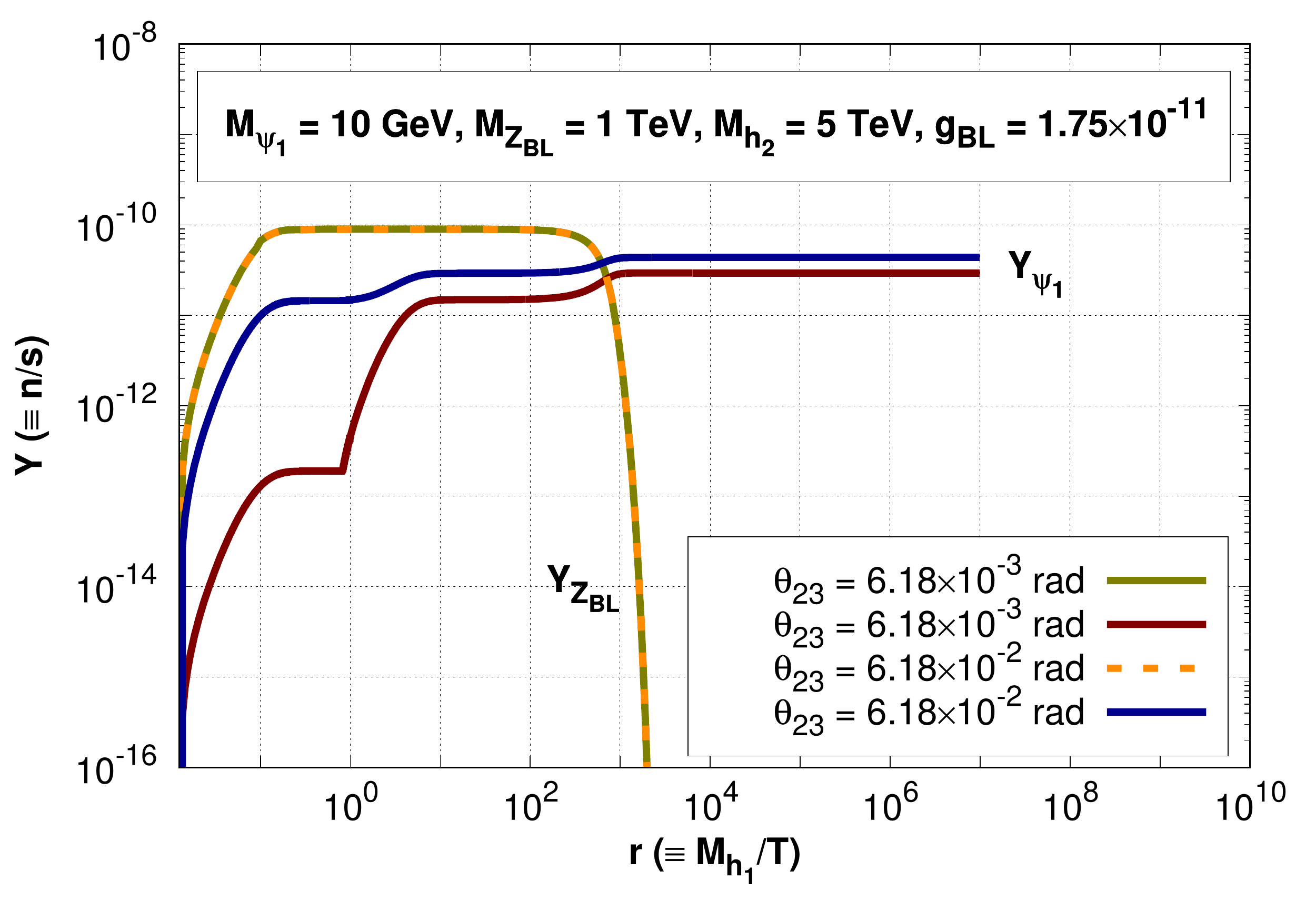}%
	}
	\caption{Comparison of comoving number densities of
	$Z_{\rm BL}$ and $\psi_1$ with respect to mixing angles
	$\theta_{13}$ and $\theta_{23}$.}
	\label{mixing}
\end{figure}

Let us now discuss the variation of $Y$ with respect
to mixing angles. These are shown in Fig.\,\,\ref{mixing}
(a)--\ref{mixing} (d). As mentioned earlier, when $\beta=1$,
the mixing angles have very little effect on the comoving number
density of $\zbl$ and $\dm$. In Fig. \ref{mixing} (a), 
on increasing $\theta_{13}$, we find that there is only a
small increase in the production of $\dm$ from the SM Higgs
($h_1$) due to an increase in $g_{h_1\overline{\psi_1}\dm}$ coupling.
On the other hand, $g_{h_2\zbl\zbl}$ is however insensitive
to variations in $\theta_{13}$
and hence $Y_{\zbl}$ remains unchanged. The $g_{h_2\zbl\zbl}$ coupling,
however, is sensitive to $\theta_{23}$ (because of the presence of
the term like $\cos\theta_{13}\sin{\theta}_{23}$).
So we find a corresponding increase in $\zbl$
yield on increasing $\theta_{23}$ in
Fig. \ref{mixing} (c). Consequently, an increase in
$Y_{\dm}$ is also noted. 

Variation in the yield of $\zbl$ and $\dm$ with
these mixing angles is more pronounced in the $\beta=10^{-3}$ scenario.
Increasing $\theta_{13}$ in Fig. \ref{mixing} (b),
we find that the production of dark matter from SM Higgs decay
is enhanced while the production from $h_2$ is almost unaffected.
This happens since the coupling $g_{h_1\overline{\psi_1} \dm}$ is more
sensitive to changes in $\theta_{13}$ as is seen from its expression
in the Appendix \ref{cc}. On the contrary, the coupling
$g_{h_2\overline{\psi_1} \dm}$ is sensitive to $\theta_{23}$.
So production of $\dm$ from $h_2$ is enhanced in the case
where $\theta_{23}$ is increased (Fig. \ref{mixing} (d)).
Yield of $\zbl$ in this case ($\beta=10^{-3}$),
however remains unaffected 
because the effect of the mixing angle
$\theta_{13}$ on the coupling $g_{h_2\zbl \zbl}$
is always suppressed due to its nature of occurrence within
the trigonometric functions while the low value of $\beta$ makes
the $h_2\zbl \zbl$ coupling insensitive to the
other mixing angle $\theta_{23}$.  

\begin{figure}[h!]
	\centering
	\includegraphics[height=8.5cm,width=10.5cm,angle=0]{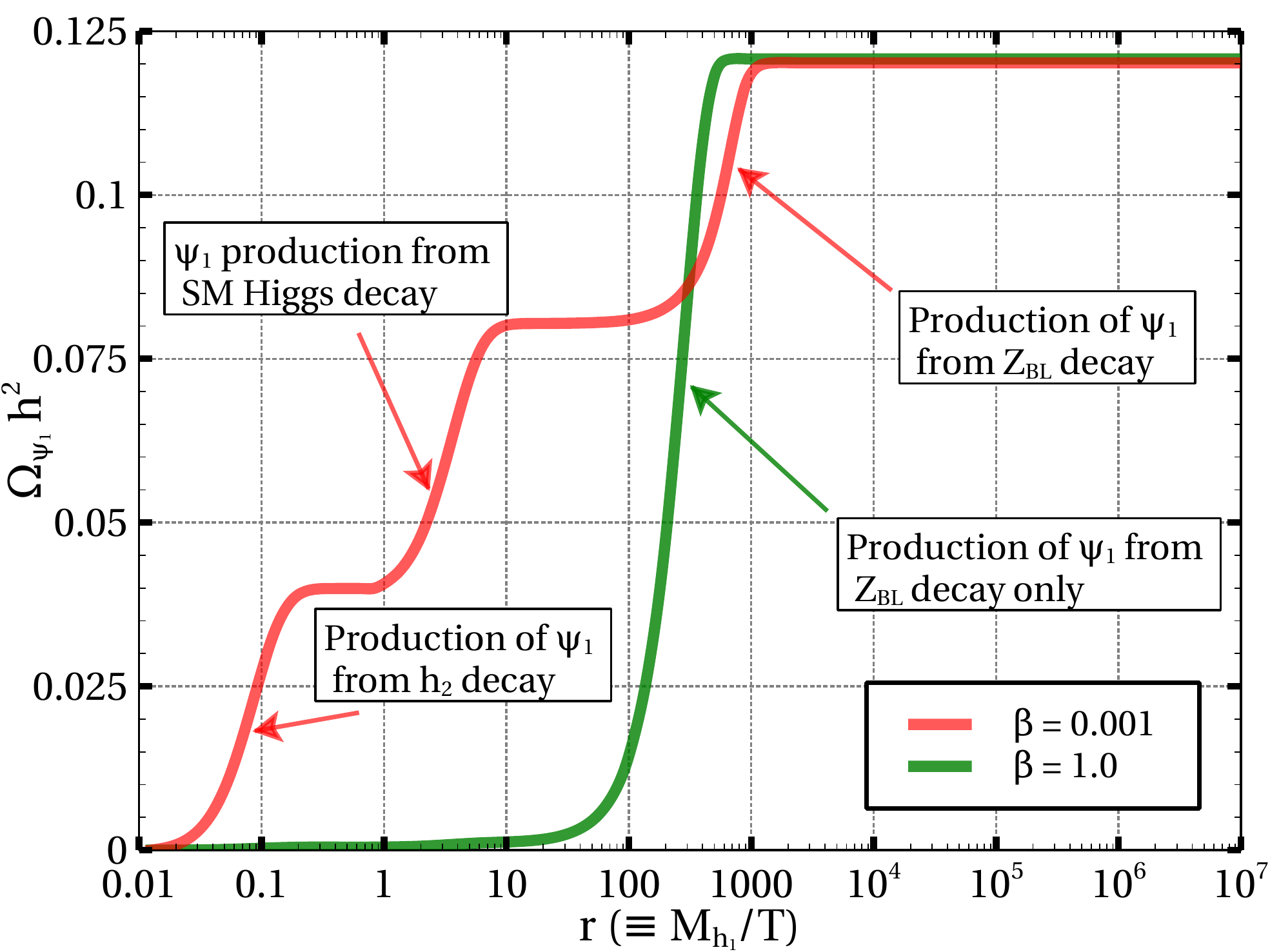}
	\caption{Variation of total relic density of
	$\psi_1$ with $r$ corresponding to $\beta=1$ and $\beta=10^{-3}$.}
	\label{omega}
\end{figure}

Finally, to contrast the two scenarios, we have plotted
the relic density corresponding to the two benchmarks given
in Table \ref{tab1}. The equal contribution of the
scalars as well as $\zbl$ to the final DM relic abundance is clearly
visible in Fig. \ref{omega} for $\beta=10^{-3}$ case. In the
other scenario, all most all of the contribution to the final
abundance of $\dm$ comes from the decay of $\zbl$.

\begin{figure}[h!]
	\centering
	\includegraphics[height=6.5cm,width=8cm,angle=0]{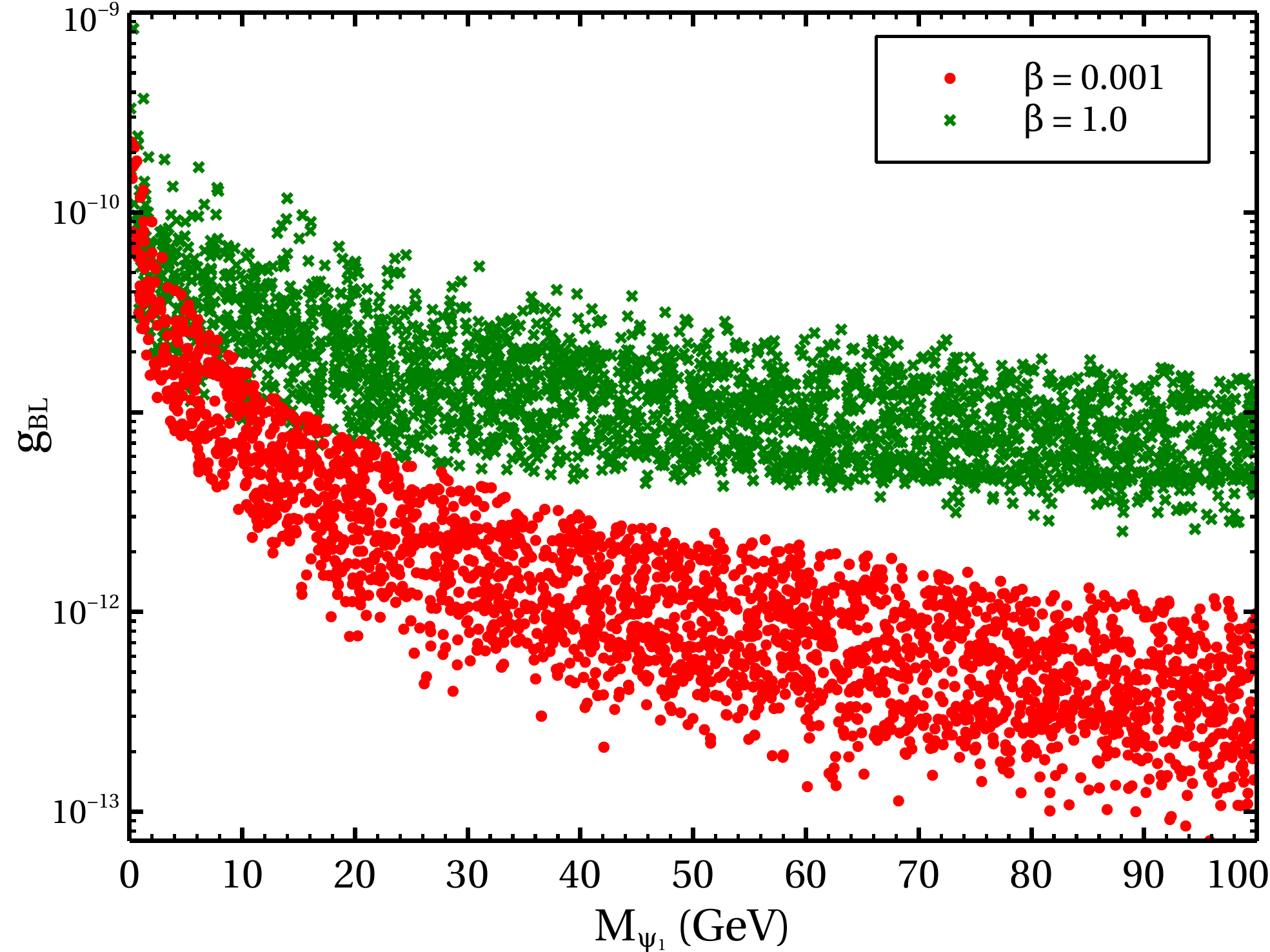}
	\includegraphics[height=6.5cm,width=8cm,angle=0]{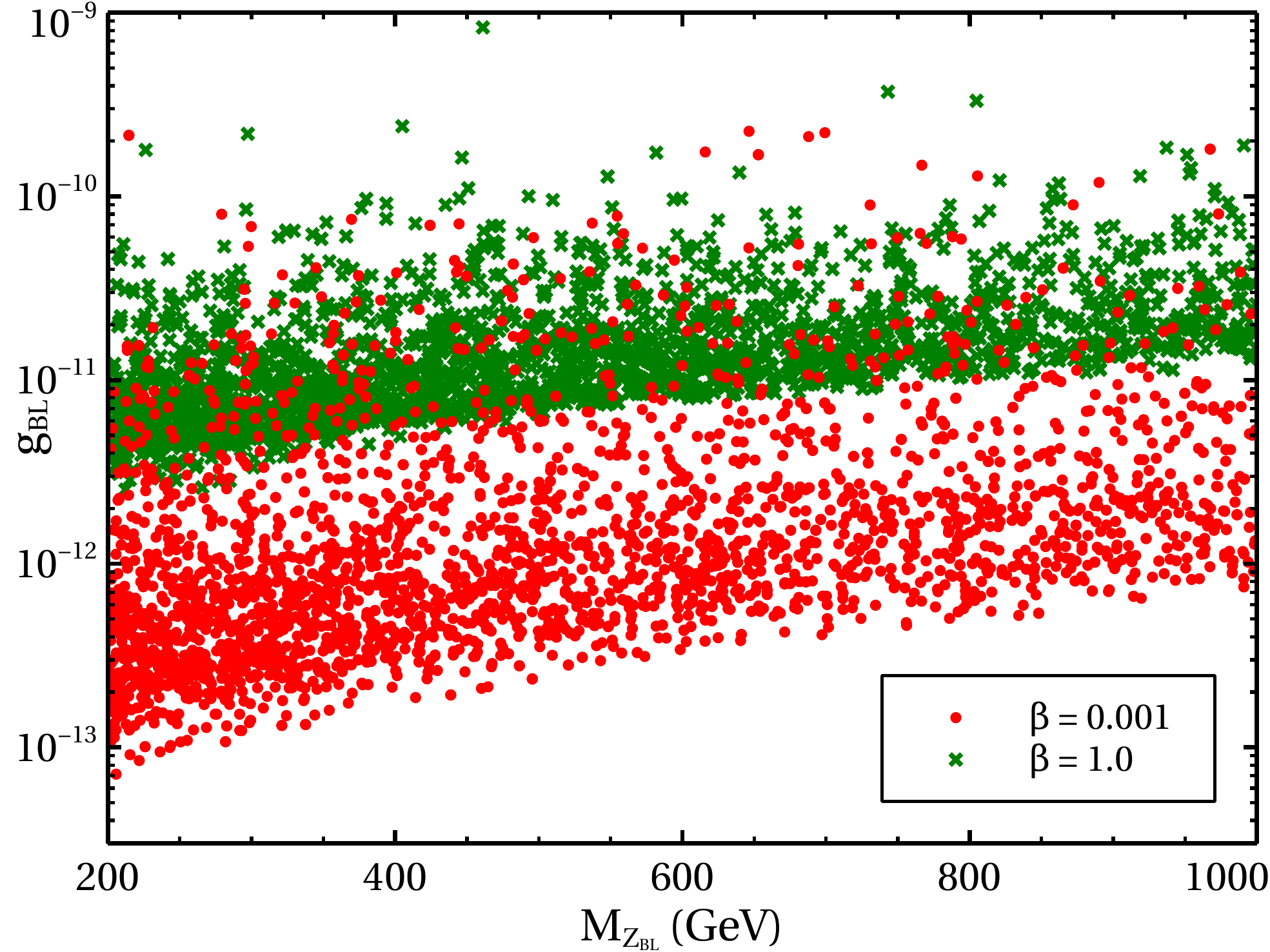}
	\caption{Points allowed by relic density constraint
in $\gbl-\mdm$ (left) and $\gbl-\mzbl$ (right) planes.
The range of variation of other relevant parameters are as
follows: $200\, {\rm GeV} \leq \mzbl \leq 1000$ GeV,
$2\,\mzbl \leq M_{h_2} \leq 7$ TeV, $10^{-14} \leq \gbl \leq 10^{-9}$,
$10^{-1} \,{\rm GeV} \leq \mdm \leq 100$ GeV. Other independent
parameters have been kept fixed to their corresponding benchmark
values.}
	\label{scan}
\end{figure}

To get an overall idea about the allowed parameter range
where our scenario satisfies DM relic density, we next perform
random scans over the appropriate combination of variables.
The results are shown in both panels of Fig. \ref{scan}.
The left panel shows our findings in $\gbl-\mdm$ plane.
Now, if $\gbl$ increases then DM production from $\zbl$
decay will tend to increase its contribution to the relic density.
Hence to satisfy the relic density constraint, DM production
from scalar decay modes should decrease proportionately.
Moreover, from the right panel of Fig. \ref{scan} where we illustrate the allowed
region (which produces correct DM relic density) in $\gbl-\mzbl$ plane, 
one can see that with decreasing $\gbl$, $\mzbl$ also decreases.
Hence, $\mdm$ should decrease with increasing $\gbl$
(see Appendix \ref{scalarCC01}--\ref{scalarCC}) to 
suppress dark matter production from the decaying heavy scalar
bosons ($h_2$, $h_1$).

\begin{figure}[h!]
	\centering
	\includegraphics[height=8.5cm,width=6.5cm,angle=-90]{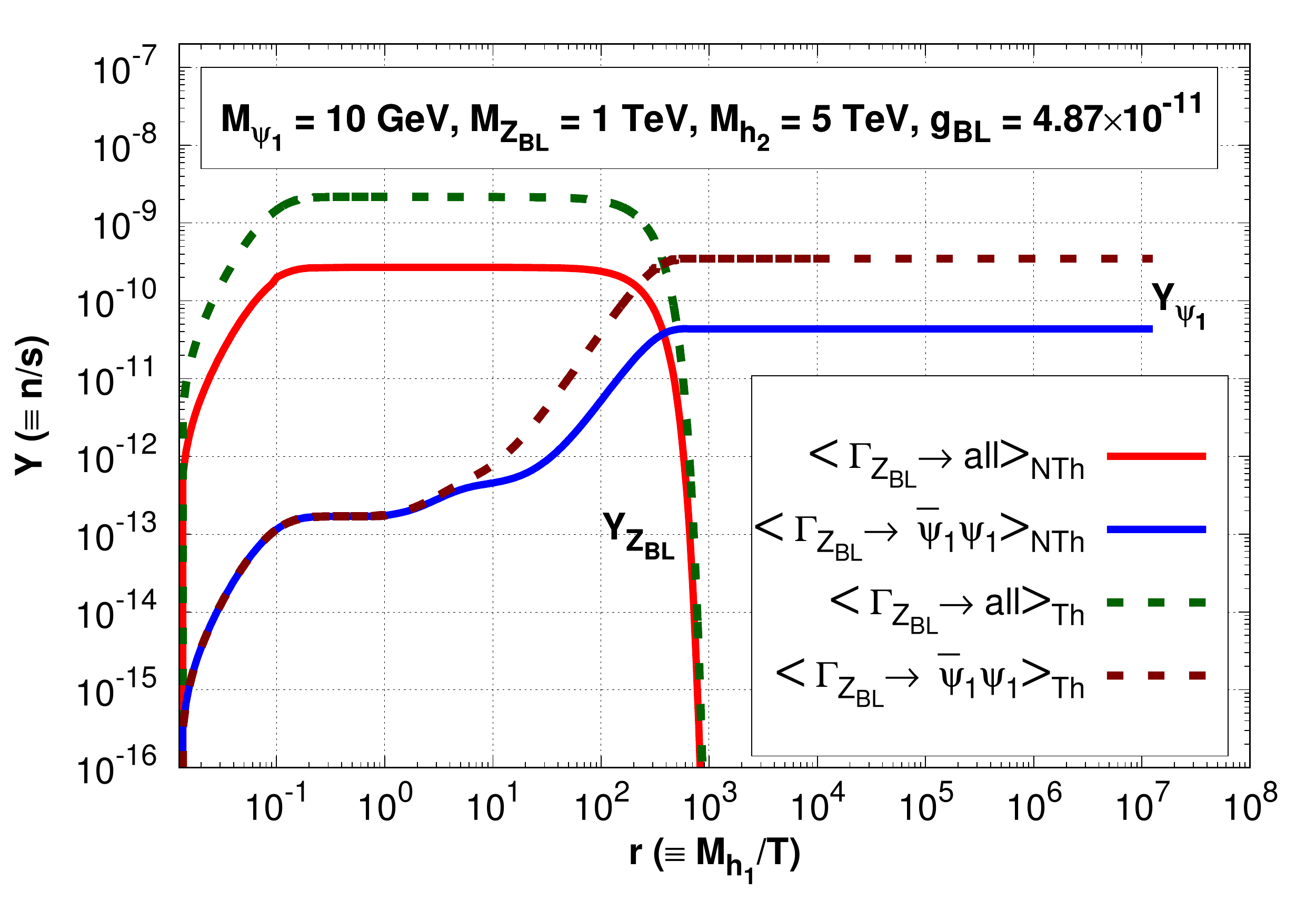}
	\includegraphics[height=8.5cm,width=6.5cm,angle=-90]{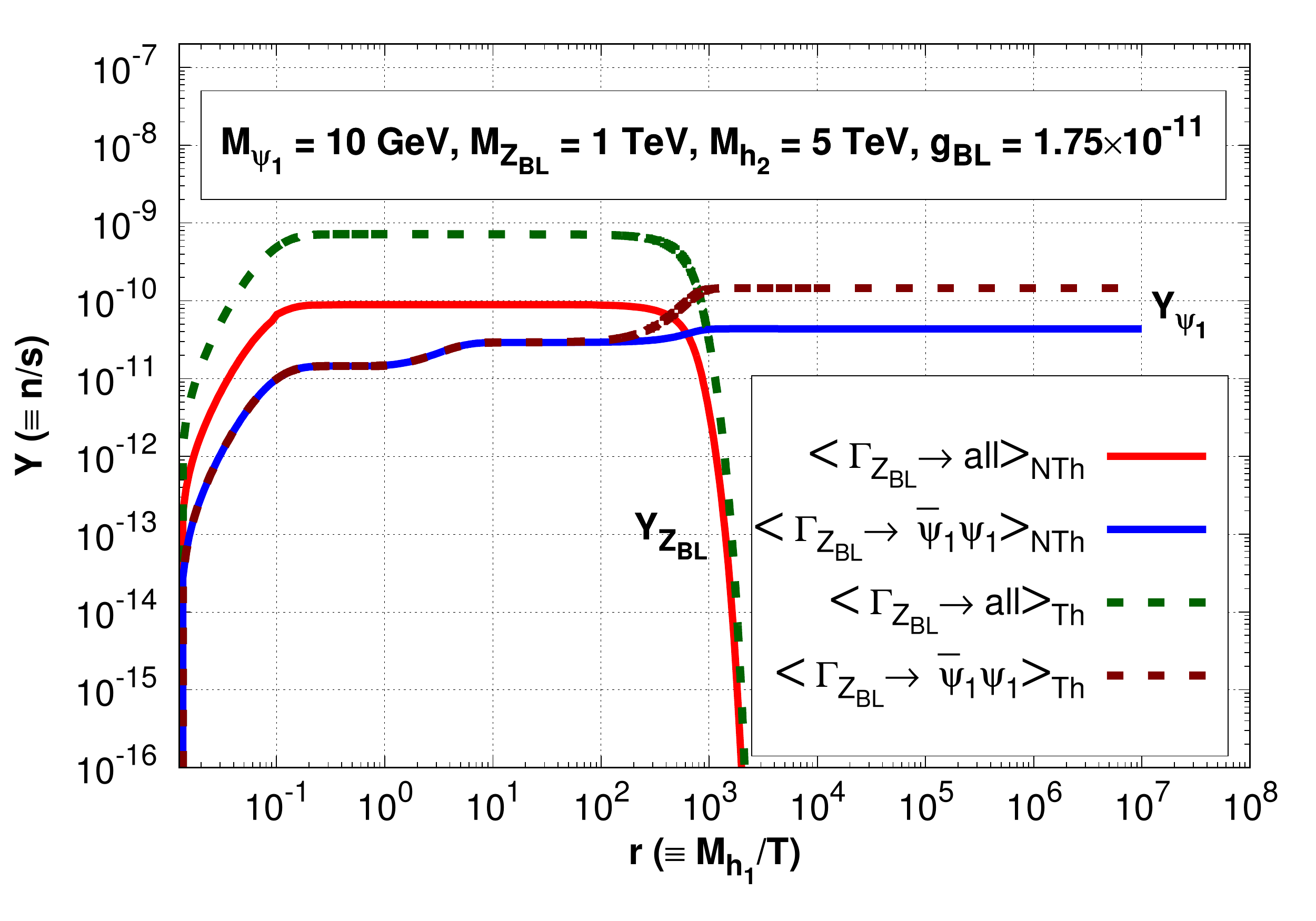}
	\caption{Comparison between the $Y_{\zbl}$ and $Y_{\dm}$
	obtained from the exact calculation using momentum distribution
	approach with that of the approximate method using
	the ${\langle \Gamma \rangle}_{\rm Th}$ for $\beta=1$ (left panel) and
	$\beta=0.001$ (right panel).}
	\label{compare}
\end{figure}

A major portion of this work is focussed on deriving the
distribution function of the dark matter particle $\dm$.
A natural question may hence arise about the need of
following such a procedure. Naively, one may expect to
follow the usual procedure of solving the Boltzmann equation
written in terms of the comoving number density $Y$
\cite{Merle:2013wta,Biswas:2016bfo}. However, if the decaying mother
particle is not in thermal equilibrium, then we need to solve
a separate Boltzmann equation for the comoving number density
of this out of equilibrium mother particle first. Because, the usual
form of the Boltzmann equation in terms of $Y$ depends on the
fact that the species under study is at least close to
thermal equilibrium. For example, in case of DM production
from a decaying species, the thermal average decay
width ${\langle \Gamma \rangle}_{\rm Th}$
appearing in the Boltzmann equation, is usually given by
$\dfrac{K_1(z)}{K_2(z)} \Gamma$, where $K_1$ and $K_2$
are the modified Bessel functions of order 1 and 2 respectively
and $\Gamma$ is the usual decay width in the rest frame of decaying particle.
However, while deriving the above expression of thermally averaged
decay width one assumes that the corresponding decaying particle
is either in thermal equilibrium or at least it is close to
thermal equilibrium such that its obeys Maxwell-Boltzmann
distribution. If this is not the case, such a thermal average
is not guaranteed to give correct results and relic density
should not be computed directly by solving the Boltzmann equation
for $Y$. In such cases average value of the decay width
itself requires the information about the non-equilibrium
momentum distribution function of the decaying mother particle.
Under such circumstances, ${\langle \Gamma \rangle}_{\rm Th}$
should be replaced by non-thermal average, ${\langle \Gamma \rangle}_{\rm NTh}
= m\,\Gamma\,\dfrac{\int\,\frac{\,f_{\rm non-eq}(p)}{\sqrt{p^2+m^2}}\,\,d^3p}
{\int \, f_{\rm non-eq}(p)\,d^3p}$ where m is the mass of the
decaying species and $f_{\rm non-eq}(p)$ is its distribution function.
So we should first solve the distribution function of the
mother particle (here $\zbl$), then use it to calculate
the distribution function of the dark matter directly.
Once this is known, we can
calculate other quantities of interest as we have discussed
elaborately earlier. Thus, finally we make a comparative
study (for both the benchmark points $\beta=1$ and $\beta=0.001$)
of the differences in the results obtained from the
exact calculation and that obtained by assuming the system
to be close to an equilibrium one. The findings are plotted
in both the panels of Fig. \ref{compare} where
left panel is for $\beta=1$ case while the
right one corresponds to $\beta=0.001$.
In both plots, we find considerable
differences in the final abundance of $\dm$ computed using
${\langle \Gamma \rangle}_{\rm NTh}$ (solid lines) and
${\langle \Gamma \rangle}_{\rm Th}$ (dashed lines). 
We also find that the difference in $Y_{\psi_1}$
depends on the contribution of $\zbl$ to comoving
number density of $\dm$. For $\beta =1$ case,
almost all the DM is produced from the decay
of $\zbl$ and hence in this case, $Y_{\dm}$ obtained
from exact calculation is 7.98 times lower than that
obtained from the approximate one. For the other scenario,
with $\beta=0.001$, contribution of $\zbl$ is only 33\%. So 
now, the final value of $Y_{\dm}$ from the exact calculation
using distribution functions is 3.32 times smaller than
the value of $Y_{\dm}$ obtained using the approximate method. 
\section{Constraints}
\label{constraints}
In this section, we will discuss about the relevant constraints
on the model parameters arising from theoretical as
well as experiment and observational results.
To start with, we should first satisfy the nontrivial
theoretical constraints arising from the vacuum stability
criterion. The conditions we need to satisfy are
listed in Eqs. ((\ref{vs1})--(\ref{vs2})) (copositivity
conditions \cite{1205.3781}) For a general
$3\times 3$ scalar mixing matrix it is difficult
to write closed form analytical expressions of
the different $\lambda$'s (Eq. (\ref{lagrang}))
in terms of our chosen set of independent parameters.
We have however checked numerically that for our benchmark
points the conditions are indeed satisfied.

Among the experimental constraints let us first discuss
the implication of the constraint related to the invisible
decay width of SM $Z$ boson. As given in \cite{Agashe:2014kda},
\begin{eqnarray}
\dfrac{\Gamma(\rm Z \rightarrow invisible)}
{\Gamma(\rm Z \rightarrow \nu \bar{\nu})} &=& 2.990 \pm 0.007.
\label{Zbound}
\end{eqnarray}
In our chosen model, in absence of kinetic mixing
terms between the Standard Model $Z$ boson
and the extra gauge boson $\zbl$, the former do not
decay to any BSM particles. Hence Eq. (\ref{Zbound})
is trivially satisfied.

The Standard Model Higgs boson ($h_1$) in the representative
benchmarks is lighter than the other two scalars
($h_2$ and $h_3$) as well as $\zbl$. As already discussed,
the fermion $\psi_2$ and pseudo scalar $A$ are assumed
to be very heavy for simplicity. Hence only allowed
invisible decay mode of $h_1$ is to
$h_1 \rightarrow \bar{\psi_1} \,\dm$. But this channel
is highly suppressed because of the very small value of
the extra gauge coupling $\gbl$ required to satisfy
the non-thermality condition. Thus this decay width
evades the bound from LHC on the invisible decay of
SM Higgs boson \cite{1606.02266}.

The scattering cross section $\psi_1$ with the Standard Model
particles is also very weak in this non-thermal regime
hence the spin independent DM nucleon coherent scattering
cross sections lie much below the present day direct
detection bounds \cite {1608.07648}.

From cosmological point of view, the most important
constraints arise from the structure formation and
somewhat related to it, the bounds from dark radiation.
The presence of highly relativistic particles around
the Big Bang Nucleosynthesis (BBN) epoch can upset
the observed structures of the galaxies. Dark matter
particles if at all produced around that epoch
then it has to be non-relativistic and hence
should not alter the onset of BBN.
Dark matter momentum distribution
functions $f_{\dm}(\xi_p,r)$ can provide us with the vital
information on whether the particle is Hot or Cold.
A suitable quantity to calculate in this regard is
the free-streaming horizon length denoted commonly
by $\lambda_{\rm fs}$ \cite {0812.0010}.
It is defined by:
\begin{eqnarray}
\lambda_{\rm fs} \equiv
\int_{T_{\rm production}}^{T_{\rm now}}
\dfrac{\langle {\rm v}(T)\rangle}{a(T)}\dfrac{dt}{dT}\,dT \,,
\label{lfs}
\end{eqnarray}
where $T_{\rm production}$ is the temperature
when almost all of the dark matter particles have
been produced and the DM comoving number density
already has frozen in. $T_{\rm Now}$ is the present
temperature of the Universe. $\langle {\rm v}(T) \rangle$
is the average velocity of the dark matter particle
and is calculable once its distribution function is known.
The term $\dfrac{dt}{dT}$ can be found from the
time-temperature relationship and
in the radiation dominated era 
$\dfrac{dt}{dT}=-\left(1+\dfrac{1}{4}\dfrac{d\,\ln g_{\rho}(T)}
{d \ln T}\right)\dfrac{1}{H\,T}$, where $H(T)$ is
the Hubble parameter while $g_{\rho}(T)$ is the
number of degrees of freedom related to the
energy density of the Universe. At this point we
transform the variable $T$ to our usual dimensionless
variable $r=\dfrac{M_{sc}}{T}$. With this Eq. (\ref{lfs}) becomes:
\begin{eqnarray}
\lambda_{\rm fs} = \int_{r_{\rm production}}^{r_{\rm now}}
\dfrac{\langle {\rm v}(r)\rangle}{a(r)}\dfrac{\widetilde{g}_{\rho}(r)}
{H(r)}\dfrac{dr}{r}\,,
\label{lfsr}
\end{eqnarray}
where $\widetilde{g}_{\rho}(r)=\left(1-\dfrac{1}
{4}\dfrac{d\,\ln g_{\rho}(r)}{d \ln r}\right)$ and
$r_{\rm production \,(now)}
\equiv \dfrac{M_{sc}}{T_{\rm production \, (now)}}$.
The average velocity $\langle {\rm v}(r) \rangle$ is defined as:
\begin{eqnarray}
\langle {\rm v}(r) \rangle = \mathcal{B}(r)
\dfrac{\bigint_{0}^{\infty}d\xi
\frac{\xi^3f_{\dm}(\xi,r)}{\sqrt{\mathcal{B}(r)^2
\xi^2+r^2\frac{M_{\dm}^2}{M_{sc}^2}}}}
{\bigint_{0}^{\infty}d\xi \,\xi^2f_{\dm}(\xi,r)}\,.
\end{eqnarray}
The scale factor $a(r)$ is obtained by using the conservation
of total entropy of the Universe in a comoving volume.
It is given by:
\begin{eqnarray}
a(r)=\left(\dfrac{43}{11\,g_s(r)}\right)^{1/3}
\dfrac{r}{r_{\rm now}}\,.
\end{eqnarray}
The Hubble parameter in terms of $r$
is written as:
\begin{eqnarray}
H(r)=\dfrac{M_{sc}^2}{M_0(r)\,r^2}\,,
\end{eqnarray}
with $M_0(r)=\left(\dfrac{45M_{Pl}^2}{4\pi^3 g_\rho(r)}\right)^{1/2}$
Using all these in Eq. (\ref{lfsr})
we calculated the free streaming horizon length. According to
Ref. \cite {1306.3996}, if $\lambda_{fs} < 0.01$ Mpc,
then we can attribute the dark matter as ``Cold".
In our case we have found out that $\lambda_{\rm fs} \ll 0.01$ Mpc
for all the benchmark points, and hence respects
the structure formation constraints. 

Another cosmological quantity of interest
that measures the amount of relativistic particles
that can be injected without disrupting the precise
experimental observations around BBN and CMB is the
effective number of neutrinos, denoted by $N_{eff}$.
Its standard value is given by 3.046 \cite {1311.2600}.
This number will change if the highly relativistic particles
are introduced at around the time of BBN and CMB. Stringent
bounds on the amount of {\it extra} relativistic degree
of freedom that can be added is given by $\Delta N_{eff}$.
The present experimental constraints on this quantity
are $\Delta N_{eff} (T_{\rm BBN})< 0.85$ \cite {1308.3240}
and $\Delta N_{eff}(T_{CMB}) < 0.32$ \cite {1502.01589}.
This quantity can be also calculated using the
knowledge of momentum distribution function of $\dm$
following \cite {1502.01011}. The expression of
$\Delta N_{eff}$ is given by :
\begin{eqnarray}
\Delta N_{eff}(r)=\frac{60}{7\pi ^4}
\left(\dfrac{r_{\nu}}{r}\right)^4
\dfrac{M_{\dm}\,r}{M_{sc}}
\mathcal{B}(r)^3\int_{0}^{\infty}
d\xi_p \,\xi_p^2\left(\sqrt{1+
\left(\dfrac{\mathcal{B}(r)\,
\xi_p\, M_{sc}}{M_{\dm}\,r}\right)^2}-1\right)
\,f_{\dm}(\xi_p,r)\,.
\end{eqnarray}
The factor $\left(\dfrac{r_{\nu}}{r}\right)^4
=\left(\dfrac{T}{T_\nu}\right)^4$
is neglected for $T \ga 1$ MeV since the neutrinos
had the same temperature with the background photon
bath during that epoch.
For our benchmarks, the calculated value of
this $\Delta N_{eff}$ (at both the epochs of BBN and CMB)
lies well below the existing upper bounds. It is expected
that our scenario will not disturb the evolution
of Universe during BBN and CMB. This is because in
our chosen benchmarks, the mass of dark matter is
$\mathcal{O}$(Gev) and most of it are produced
at around a temperature of $\mathcal{O}$(100 MeV).
Hence by the time the Universe is cooled to lower
temperatures most of these particles will become
non-relativistic and hence wont affect either
structure formation or CMB.
\section{Conclusion}
\label{conclu}
In this work, we have calculated the momentum distribution
function of a non-thermal fermionic dark matter. Calculation
of momentum distribution function is a general feature of
any non-thermal dark matter scenario if the dark matter
particle under study originates from a parent particle
that itself is outside the thermal soup. On the other hand,
the momentum distribution function of DM is a key quantity for
the computations of all the relevant thermodynamic quantities.
We have demonstrated its use in the calculations of cosmological
constraints, which, though weak in our case, can become
important for other different combination of model parameters.
The model chosen here is also well motivated, since it
is anomaly free and also explains the genesis of neutrino mass,
besides accommodating a non-thermal fermionic dark matter
candidate as well. For the two chosen benchmark scenarios
there are noticeable structural differences in the plots.
In one scenario (with $\beta=1$), the dominant production
of dark matter is seen to be pronounced from $\zbl$
decay and hence the final abundances is not much sensitive
to the scalars mixing angles. In the other scenario
(with $\beta (\ll 1)=10^{-3}$), however all decay modes can 
contribute substantially, resulting in a characteristic
multi-plateau feature in the variation of comoving number
density ($Y_{\dm}$) with $r$. Finally, we have also checked
that our non-thermal dark matter scenario does not
violate any experimental or theoretical constraints.
\section{Acknowledgement}
Authors would like to acknowledge Department of Atomic Energy
(DAE), Govt. of INDIA for financial assistance.
\appendix
\section{Appendix}
\subsection{Relevant Vertex factors}
\label{cc}
We denote the vertex factor by $g_{abc}$ for a vertex containing
fields $a,\,b,\,c$. Vertex factors for the interactions of
$\psi_1$ with CP-even scalars are given below 
\begin{eqnarray}
g_{h_1\overline{\psi_1}\psi_1}&=&
2\sqrt{2}\,\dfrac{\gbl\,\sqrt{1+4\,\beta^2}}{\beta\,\mzbl}
\left(\beta\,\sin\theta_{12}\cos\theta_{13}\cos^2\theta_{\rm L}
M_{\psi_1}+\sin\theta_{13}\sin^2\theta_{\rm L}M_{\psi_1}\right)
\label{scalarCC01}\\
g_{h_2\overline{\psi_1}\psi_1}&=&
2\sqrt{2}\,\dfrac{\gbl\,\sqrt{1+4\,\beta^2}}{\beta\,\mzbl}
\left(\beta\,\left(\cos\theta_{12}\cos\theta_{23}
-\sin\theta_{12}\sin\theta_{23}
\sin\theta_{13}\right)
\cos^2\theta_{\rm L}M_{\psi_1}
\right.\nonumber\\&&\left.
+\sin\theta_{23}\cos\theta_{13}\sin^2\theta_{\rm L} M_{\psi_1}
\right)
\label{scalarCC0}\\
g_{h_3\overline{\psi_1}\psi_1}&=&
2\sqrt{2}\,\dfrac{\gbl\,\sqrt{1+4\,\beta^2}}{\beta\,\mzbl}
\left(-\beta\,\left(\cos\theta_{12}\sin\theta_{23}
+\sin\theta_{12}\cos\theta_{23}\sin\theta_{13}\right)
\cos^2\theta_{\rm L}M_{\psi_1}
\right.\nonumber\\&&\left.
+\cos\theta_{23}\cos\theta_{13}
\sin^2\theta_{\rm L} M_{\psi_1}\right)
\label{scalarCC}
\end{eqnarray}

Vertex factors for the interactions between CP-even scalars
and ${\rm B-L}$ gauge boson ($\zbl$):
\begin{eqnarray}
g_{h_1\zbl\zbl}&=&2\,\dfrac{\gbl\,\mzbl}{\sqrt{1+4\beta^2}}
\left(\sin\theta_{12}\cos\theta_{13}+4\beta\sin\theta_{13}\right)\\
g_{h_2\zbl\zbl}&=&2\,\dfrac{\gbl\,\mzbl}{\sqrt{1+4\beta^2}}
\left(\cos\theta_{12}\cos\theta_{23}-\sin\theta_{12}
\sin\theta_{23}\sin\theta_{13}+4\beta\cos\theta_{13}
\sin\theta_{23}\right)\\
g_{h_3\zbl\zbl}&=&2\,\dfrac{\gbl\,\mzbl}{\sqrt{1+4\beta^2}}
\left(-\left(\cos\theta_{12}\sin\theta_{23}+\sin\theta_{12}
\cos\theta_{23}\sin\theta_{13}\right)+4\beta\cos\theta_{13}
\cos\theta_{23}\right)
\end{eqnarray}

The interaction vertex of the dark matter
$\psi_1$ with the new gauge boson can be written as: 
$\frac{\gbl}{6}\,\overline{\psi}_1
\gamma^\mu\left(a-b\,\gamma_5\right)\psi_1$, where
$a=(1-3\sin^2\theta_{\rm L})$ and $b=-3\,(1+\sin^2\theta_{\rm L})$.

The Yukawa couplings of the fermions i.e. the ${y_{\xi}}_i$'s and
${y_{\eta}}_i$'s are also listed below for completeness.
\begin{eqnarray}
{y_{\xi}}_1  &=& \sqrt 2\,\dfrac{\gbl\sqrt{1+4\beta^2}}{\beta\,\mzbl}
\left(\cos\theta_{\rm L}\cos\theta_{\rm R} \,M_{\psi_2}
+\sin\theta_{\rm L}\sin\theta_{\rm R} \,M_{\psi_1}\right),\\
{y_{\xi}}_2  &=& \sqrt 2\,\dfrac{\gbl\sqrt{1+4\beta^2}}{\beta\,\mzbl}
\left(-\cos\theta_{\rm L}\sin\theta_{\rm R}\,M_{\psi_2}
+\sin\theta_{\rm L}\cos\theta_{\rm R} \,M_{\psi_1}\right),\\
{y_{\eta}}_1  &=& \sqrt 2\,\dfrac{\gbl\sqrt{1+4\beta^2}}{\mzbl}
\left(-\sin\theta_{\rm L}\cos\theta_{\rm R}\,M_{\psi_2}
+\cos\theta_{\rm L}\sin\theta_{\rm R} \,M_{\psi_1}\right),\\
{y_{\eta}}_2  &=& \sqrt 2\,\dfrac{\gbl\sqrt{1+4\beta^2}}{\mzbl}
\left(\sin\theta_{\rm L}\sin\theta_{\rm R}\,M_{\psi_2}
+\cos\theta_{\rm L}\cos\theta_{\rm R} \,M_{\psi_1}\right).
\end{eqnarray}
\subsection{Relevant Decay Widths}
\label{width}
The only relevant decay widths that we need during the
computation of dark matter momentum distribution functions 
are those corresponding to the decay of the extra gauge boson
$\zbl$ to fermions.
\begin{eqnarray}
\Gamma_{\zbl \rightarrow f \bar{f}}&=&
\dfrac{\mzbl}{16\pi}\dfrac{4}{3}
\left(a_f^2+b_f^2\right)\Bigg(1+\dfrac{2\left(a_f^2-2b_f^2\right)}
{\left(a_f^2+b_f^2\right)}\dfrac{M_f^2}{\mzbl^2}\Bigg)
\sqrt{1-\dfrac{4M_f^2}{\mzbl^2}}\,.
\end{eqnarray}
If $f$ is a Standard Model fermion then
$b_f=0$ and $a_f=\gbl \,Q_{\rm BL}(f)$,
where $Q_{\rm BL}(f)$ is the $B-L$ charge corresponding
to the fermion $f$ (see Table \ref{tabcharge}). If, on the
other hand $f$ is a beyond Standard Model particle
(say, $f=\psi_1$), we have $a_{\psi_1}=
\frac{\gbl}{6}\left(1-3\sin^2\theta_{\rm L}\right)$
and $b_{\psi_1}=\frac{\gbl}{2}\left(1+\sin^2\theta_{\rm L}\right)$.
Hence the total decay width (assuming other BSM particles
such as $\psi_2$, $A$ and $h_3$ are heavier than $\zbl$)
is given as the sum of the individual decay widths
to all these individual SM and BSM channels i.e. 
\begin{equation}
\Gamma_{\zbl \rightarrow all}=
\left(\sum\limits_{\rm SM \, fermions}
\Gamma_{\zbl \rightarrow f \bar{f}}\right)
+\Gamma_{\zbl \rightarrow \psi_1 \overline{\psi_1}}\,.
\label{zdecay}
\end{equation} 
Note, that this expression of decay width is valid only
in the rest frame of the decaying particle (here $\zbl$).
In a reference frame where $\zbl$ is not at rest but moving
with an energy $E_{\zbl}$, the total
decay width is given by:
\begin{eqnarray}
\Gamma_{\zbl \rightarrow all}^{\prime}
= \Gamma_{\zbl \rightarrow all}\,\frac{\mzbl}{E_{\zbl}}
\label{decywdth}
\end{eqnarray}
\subsection{Collision terms}
\label{app:colterm}
The first step while solving the Boltzmann equation is
to derive the collision terms ($\mathcal{C}[f]$).
The generic form of the collision term in
case of $1 \rightarrow 2$ decay process
(say, $\chi \rightarrow a\,b$) is given by
\cite{Kolb:1990vq, Gondolo:1990dk}:
\begin{eqnarray}
\mathcal{C}[f_\chi(p)]&=&\dfrac{1}{2\,E_p}\bigintsss
\dfrac{g_a\,d^{3} p^{\prime}}{(2\pi)^{3} \,2E_{p^{\prime}}}
\dfrac{g_b\,d^{3} q^{\prime}}{(2\pi)^{3} \,2E_{q^{\prime}}}
(2\pi)^4\,\delta^4(\tilde{p}-\tilde{p}^{\prime}-\tilde{q}^{\prime})
\times\overline{\,\lvert \mathcal{M}\rvert^2}\nonumber \\
&&~~~~~~~~~~~~~~\times\,[f_a\,f_b\,\left(1 \pm f_\chi\right)
-f_\chi\left(1 \pm f_a\right)\left(1 \pm f_b\right)]\,.
\label{colision1}
\end{eqnarray}
In this expression $p,p^\prime,q^\prime$ are the absolute
values of three momenta of $\chi,\, a$ and $b$ respectively.
The corresponding four momenta are given by $\tilde{p},\,\tilde{p}^\prime$
and $\tilde{q}^\prime$ while $E_{p},\,E_{p^\prime}$ and
$E_{q^\prime}$ are the energies of $\chi,\,a$ and $b$ respectively.
These energies are of course related to the absolute
value of the corresponding three momenta by the
usual relativistic dispersion relation. For e.g.
$E_p=\sqrt{p^2+m_\chi^2}$ and so on. The internal degrees
of freedom corresponding to the particles $a$ and $b$
are indicated by $g_a$ and $g_b$ respectively.
The matrix element squared denoted by
$\overline{\,\lvert \mathcal{M}\rvert^2}$ for the
corresponding process (here, $\chi \rightarrow a\,b$)
is averaged over the spins of both the initial as well as
final state particles. The distribution function
corresponding to the particle $x$ is denoted by $f_{x}$
and $(1 \pm f_x)$ are the Pauli blocking and the
stimulated emission factors respectively. These factors can be
approximated $\sim 1$ in absence of Bose condensation
and Fermi degeneracy. If any of the particles $a$ or $b$
is in thermal equilibrium then the corresponding $f$
can be approximated by the Maxwell-Boltzmann distribution
function i.e. $f \sim e^{-\frac{E}{T}}$, where $E$ is the energy
of the particle and $T$ is the temperature of the Universe.
\subsubsection{\Large$\boldsymbol{\mathcal{C}^{\zbl \rightarrow all}}$}
\label{app:zbl-all}
As a concrete example let us try to calculate
${\mathcal{C}^{\zbl \rightarrow all}}$
(the second collision term in Eq. (\ref{fzbl})).
Let us first calculate the collision term for a specific channel,
say $\mathcal{C}^{\zbl \rightarrow f \bar{f}}$, where $f$ is any
fermion. In Eq. (\ref{fzbl}), we are interested in solving
the non-equilibrium distribution function for $\zbl$.
Hence this collision term denotes the depletion of
the particle under study. If we neglect the back reactions
i.e. the inverse decay processes (which is a legitimate
approximation for a particle in non-thermal regime \cite{1404.2220,
1502.01011}) and approximate the Pauli blocking factors
and stimulated emission terms to be $\sim$ 1,
then from Eq. (\ref{colision1}) we have:
\begin{eqnarray}
\mathcal{C}^{\zbl \rightarrow f \bar{f}}[f_{\zbl}(p)]&=&
\dfrac{1}{2\,E_p}\bigintsss \dfrac{g_f\,d^{3} p^{\prime}}
{(2\pi)^{3} \,2E_{p^{\prime}}}\dfrac{g_f\,d^{3} q^{\prime}}
{(2\pi)^{3} \,2E_{q^{\prime}}}(2\pi)^4\,\delta^4
(\tilde{p}-\tilde{p}^{\prime}-\tilde{q}^{\prime})
\times\overline{\,\lvert \mathcal{M}\rvert^2} \nonumber \\
&&~~~~~~~~~~~~~~\times[-f_{\zbl}(p)]\,, \nonumber  \\
&=& -f_{\zbl}(p)\times\dfrac{1}{2\,E_p}
\bigintsss \dfrac{g_f\,d^{3} p^{\prime}}{(2\pi)^{3} \,2E_{p^{\prime}}}
\dfrac{g_f\,d^{3} q^{\prime}}{(2\pi)^{3} \,2E_{q^{\prime}}}
(2\pi)^4\,\delta^4(\tilde{p}-\tilde{p}^{\prime}
-\tilde{q}^{\prime})\times\overline
{\,\lvert \mathcal{M}\rvert^2}\,. \nonumber \\
\label{colision2}
\end{eqnarray}
But we know that the decay width (in an arbitrary frame)
for the process $\zbl \rightarrow f \bar{f}$ 
is given by the expression:
\begin{eqnarray}
\Gamma_{\zbl \rightarrow f\bar{f}}^{\prime}&=&
\dfrac{1}{2\,E_p}\bigintsss \dfrac{g_f\,d^{3} p^{\prime}}
{(2\pi)^{3} \,2E_{p^{\prime}}}\dfrac{g_f\,d^{3} q^{\prime}}
{(2\pi)^{3} \,2E_{q^{\prime}}}(2\pi)^4\,
\delta^4(\tilde{p}-\tilde{p}^{\prime}-\tilde{q}^{\prime})
\times\overline{\,\lvert \mathcal{M}\rvert^2}
\Bigg\lvert_{\zbl \rightarrow f\bar{f}}\,,
\label{decay}
\end{eqnarray}
where, as discussed before, $p,\,p^{\prime}$ and $q^{\prime}$
are the three momenta corresponding to $\zbl,\,f$ and $\bar{f}$
respectively. Now using Eq. (\ref{decay}) in Eq. (\ref{colision1})
and making the change of variables
$\xi_p \equiv \dfrac{1}{\mathcal{B}(r)}\,\dfrac{p}{T}$
and $r_{\zbl} \equiv \dfrac{M_{\zbl}}{T}$, we get
\begin{eqnarray}
\mathcal{C}^{\zbl \rightarrow f\bar{f}}[f_{\zbl}(\xi_p)]&=&
-f_{\zbl}(\xi_p)\times\Gamma_{\zbl \rightarrow f\bar{f}}
\times\dfrac{\mzbl}{E_{\zbl}}\,, \nonumber \\
\nonumber \\
&=& -f_{\zbl}(\xi_p)\times\Gamma_{\zbl \rightarrow f\bar{f}}
\times\dfrac{r_{\zbl}}{\sqrt{\xi_p^2\,\mathcal{B}(r)^2+
r_{\zbl}^2}}\,,
\end{eqnarray}
where we have used Eq. (\ref{decywdth}). Hence the collision
term $\mathcal{C}^{\zbl \rightarrow all}$ is now simply given by:
\begin{eqnarray}
\mathcal{C}^{\zbl \rightarrow all}&=&-f_{\zbl}(\xi_p)\times
\Gamma_{\zbl \rightarrow all}\times
\dfrac{r_{\zbl}}{\sqrt{\xi_p^2\,\mathcal{B}(r)^2+r_{\zbl}^2}}\,.
\end{eqnarray}
We can easily rewrite the above equation in terms of
$r\equiv \frac{M_{sc}}{T}$ by writing
$r_{\zbl}=\frac{\mzbl}{M_{sc}}\,r$.

The derivation of this collision term is greatly simplified
by the use of the expression of the decay width (Eq. (\ref{decay})).
This simplification is possible because the distribution
function of the particle we are interested in (i.e. $\zbl$)
is itself the decaying particle.

However, the situation may be such that the particle whose
non-equilibrium momentum distribution function we are interested
in, is the daughter particle produced from the decay of another
mother particle (where it is assumed that the distribution
function of the latter is already known). In that case, the
final expression for the collision term will not be so simple.
We will illustrate such a case now with a definite example.
Let us hence derive the first collision term in Eq. (\ref{fzbl}) i.e.
${\mathcal{C}^{h_2 \rightarrow \zbl \zbl}}$.
\subsubsection{\Large$\boldsymbol{\mathcal{C}^{h_2
\rightarrow \zbl \zbl}}$}
\label{app:h2-zblzbl}
The starting point is again Eq. (\ref{colision1}). This is actually
the first collision term in Eq. (\ref{fzbl}). Proceeding as before
we now have:
\begin{eqnarray}
\mathcal{C}^{h_2 \rightarrow \zbl\zbl}[f_{\zbl}(p)]&=&
2\times\dfrac{1}{2\,E_p}\bigintsss \dfrac{g_{h_2}\,d^{3} k}{(2\pi)^{3}
\,2E_{k}}\dfrac{g_{\zbl}\,d^{3} q^{\prime}}{(2\pi)^{3} \,2E_{q^{\prime}}}
(2\pi)^4\,\delta^4(\tilde{k}-\tilde{p}-\tilde{q}^{\prime})
\times\overline{\,\lvert \mathcal{M}\rvert^2} \bigg 
\lvert _{h_2 \rightarrow \zbl \zbl}\nonumber \\
&&~~~~~~~~~~~~~~\times\,[f_{h_2}\left(1 \pm f_{\zbl}\right)
\left(1 \pm f_{\zbl}\right)-f_{\zbl}\,f_{\zbl}
\,\left(1 \pm f_{h_2}\right)] \,.
\end{eqnarray}
Here $k$ is the three momentum of the decaying particle ($h_2$)
while $p$ and $q^{\prime}$ are the three momenta of the final state
particles ($\zbl$). The factor of $2$ in front is due the production
of two $\zbl$ in the final state from $h_2$ decay. $g_{\zbl}$ and
$g_{h_2}$ are the internal degrees of freedom for the extra gauge
boson and extra scalar respectively. Hence, $g_{\zbl}=3$ and $g_{h_2}=1$.

Using the usual approximations of neglecting the back
reactions as well as the Pauli blocking and stimulated
emission factors, we finally get:
\begin{eqnarray}
\mathcal{C}^{h_2 \rightarrow \zbl\zbl}[f_{\zbl}(p)]&=&
2\times\dfrac{1}{2\,E_p}\bigintsss \dfrac{g_{h_2}\,d^{3} k}{(2\pi)^{3}
\,2E_{k}}\dfrac{g_{\zbl}\,d^{3} q^{\prime}}{(2\pi)^{3}
\,2E_{q^{\prime}}}(2\pi)^4\,\delta^4(\tilde{k}-\tilde{p}
-\tilde{q}^{\prime})\times\overline{\,\lvert \mathcal{M}\rvert^2}
\bigg \lvert _{h_2 \rightarrow \zbl \zbl}\nonumber \\
&&~~~~~~~~~~~~~~\times\,[f_{h_2}(k)] \,.
\label{collision3}
\end{eqnarray}
The matrix element squared average for
the decay process $h_2 \rightarrow \zbl \zbl$
is given by:
\begin{eqnarray}
\overline{\,\lvert \mathcal{M}\rvert^2}
\bigg \lvert _{h_2 \rightarrow \zbl \zbl}&=&
\dfrac{g_{h_2 \zbl \zbl}^2}{2\times 9}\Bigg(2+
\dfrac{\left(E_{p}E_{q^{\prime}}-
\vec{p}\,.\,\vec{q^{\prime}}\,\right)^2}
{\mzbl^4}\,\Bigg)\,.
\end{eqnarray}
In Eq. (\ref{collision3}), $\delta^{(4)}
(\tilde{k}-\tilde{p}-\tilde{q}^{\prime})$ can be
written as $\delta^{(3)}(\vec{k}-\vec{p}-
\vec{q}^{\prime})\,\delta(E_k-E_p-E_{q^{\prime}})$.
We can then do the integral over $q^{\prime}$. So we should
replace every occurrence of $\vec{q^{\prime}}$ with $\vec{k}-\vec{p}$.
As already stated earlier that, to simplify notations
we will write $\lvert \vec{k}\rvert = k$ and so on. Hence
now $E_{q^{\prime}}$ has become a function of
$p$ and $k$ (and of the masses of the corresponding particles
which have three momenta $\vec{p}$ and $\vec{k}$ respectively),
i.e. $E_{q^{\prime}}=E_{q{\prime}}\,(p,k)$.
Therefore Eq. (\ref{collision3}) becomes:
\begin{eqnarray}
\mathcal{C}^{h_2 \rightarrow \zbl\zbl}[f_{\zbl}(p)]&=&
\dfrac{g_{h_2\zbl \zbl}^2}{6\,(4\pi)^2}\dfrac{1}{E_p}
\bigintsss \dfrac{d^{3} k}{E_{k}\,E_{q^{\prime}}(p,k)}
\,\delta\left(E_k-E_p-E_{q^{\prime}}(p,k)\right)\times \nonumber \\
&&~~~~~~~\Bigg(2+\dfrac{\left(E_{p}E_{q^{\prime}}\,
(k,p)+p^2-p\, k\cos\theta\right)^2}{\mzbl^4}\,\Bigg)
\times\,[f_{h_2}(k)]\,,
\label{collision3.1}
\end{eqnarray}
where $\theta$ is the angle between the $\vec{k}$ and $\vec{p}$.
Also, we have, $E_{q^{\prime}} = \sqrt{k^2+p^2+\mzbl^2-2p\,
k\cos\theta}$. At this point let us transform variables
to $\xi_k =\dfrac{1}{\mathcal{B}(r)}\, \dfrac{k}{T},
\,\xi_p=\dfrac{1}{\mathcal{B}(r)}\,\dfrac{p}{T}$
and $\cos\theta=y$, where $\mathcal{B}(r)$ is defined by
Eq. (\ref{Br}). Also $r_{\zbl}=\dfrac{\mzbl}{T}$
and $r_{h_2}=\dfrac{M_{h_2}}{T}$. Hence, 
\begin{eqnarray}
E_{q^{\prime}}&=&T\sqrt{\xi_k^2\,\mathcal{B}(r)^2+
\xi_p^2\,\mathcal{B}(r)^2+r_{\zbl}^2-2\mathcal{B}(r)^2\,
\xi_k\,\xi_p\,y}\equiv T\,H_1\,(\xi_k,\xi_p,y).
\end{eqnarray}
From here onwards, for notational fluidity, we will suppress
the explicit dependence on $r\equiv \frac{M_{sc}}{T}$. Every
occurrence of $r_{\zbl}$ and/or $r_{h_2}$ should be replaced
by $r_{\zbl}=\frac{\mzbl}{M_{sc}}\,r$ and $r_{h_2}=\frac{M_{h_2}}
{M_{sc}}\,r$. Hence it is easy to identify the functional
dependence on $r$. Eq. (\ref{collision3.1}) now simplifies to:
\begin{eqnarray}
\mathcal{C}^{h_2 \rightarrow \zbl\zbl}[f_{\zbl}(\xi_p)]=
\dfrac{g_{h_2\zbl \zbl}^2}{48\pi\,T}\dfrac{\mathcal{B}(r)^3}
{\sqrt{\xi_p^2\,\mathcal{B}(r)^2+r_{\zbl}^2}}
\bigintsss \dfrac{\xi_k^2\,d\xi_k\,dy}
{\sqrt{\xi_k^2\,\mathcal{B}(r)^2+r_{h_2}^2}\,H_1(\xi_k,\xi_p,y)}
\, \delta\left(\mathcal{F}(\xi_k,\xi_p,y)\right) \nonumber \\
\times\Bigg(2+\dfrac{\left(\sqrt{\xi_p^2\,
\mathcal{B}(r)^2+r_{\zbl}^2}\,H_1(\xi_k,\xi_p,y)+
\xi_p^2\mathcal{B}(r)^2-\mathcal{B}(r)^2\,\xi_p\,
\xi_k\,y\right)^2}{r_{\zbl}^4}\,\Bigg)
\times[f_{h_2}(\xi_k)]\,. \nonumber \\
\label{collision3.2}
\end{eqnarray}
For convenience we have defined:
\begin{eqnarray} \mathcal{F}(\xi_k,\xi_p,y)&\equiv&
\sqrt{\xi_k^2+r_{h_2}^2}-\sqrt{\xi_p^2\,
\mathcal{B}(r)^2+r_{\zbl}^2}-H_1(\xi_k,\xi_p,y)\,.
\end{eqnarray}
Also let,
\begin{eqnarray}
H_2(\xi_k,\xi_p,y)&\equiv& \dfrac{\left(\sqrt{\xi_p^2\,
\mathcal{B}(r)^2+r_{\zbl}^2}\,H_1(\xi_k,\xi_p,y)+\xi_p^2\,
\mathcal{B}(r)^2 - \mathcal{B}(r)^2\,\xi_p\,
\xi_k\,y\right)^2}{r_{\zbl}^4} \,.
\end{eqnarray}
The $y$ integral in Eq. (\ref{collision3.2}) can easily be done.
For this, we have used the well known property of $\delta$ function
which is
$\delta(\mathcal{F}(\xi_k,\xi_p,y))=
\dfrac{\delta(y-y_0)}{\lvert \mathcal{F}^
\prime(\xi_k,\xi_p,y_0)\rvert}$, $\mathcal{F}^\prime(\xi_k,\xi_p,y_0)$
denotes differentiation of $\mathcal{F}(\xi_k,\xi_p,y)$ with respect to $y$
at $y=y_0$ where $y_0$ is the root of the equation
$\mathcal{F}(\xi_k,\xi_p,y)=0$. The expression
of $y_0$ is given by:
\begin{eqnarray}
y_0 (\xi_k,\xi_p)&=& \dfrac{1}{2\,\mathcal{B}(r)^2\,\xi_k\,\xi_p}
\left(2\sqrt{\xi_k^2\,\mathcal{B}(r)^2+r_{h_2}^2}
\sqrt{\xi_p^2\,\mathcal{B}(r)^2+r_{\zbl}^2}-r_{h_2}^2\right)\,.
\end{eqnarray} 
Using this we find that
\begin{eqnarray}
\mathcal{F}^{\prime}(\xi_k,\xi_p)&=&
\dfrac{\mathcal{B}(r)^2\,\xi_k\,\xi_p}
{\sqrt{\xi_k^2\,\mathcal{B}(r)^2+r_{h_2}^2}
-\sqrt{\xi_p^2\,\mathcal{B}(r)^2+r_{\zbl}^2}}
\equiv F(\xi_k,\xi_p)\,.
\end{eqnarray}
But since $y_0$ is a function of $\xi_k$ itself,
integration over $y$ puts a limit on the $\xi_k$
integral as well. The limit(s) can be derived by
remembering that $y_0$ is actually $\cos\theta_0$,
and hence $\lvert y_0 \rvert \leq 1$.
The minimum and maximum limits on $\xi_k$
turn out to be:
\begin{eqnarray}
\xi_k^{\rm min} (\xi_p)&=&
\dfrac{1}{2\,\mathcal{B}(r)\,r_{\zbl}}\bigg\lvert \,\eta (\xi_p)
-\mathcal{B}(r)\dfrac{\xi_p\,r_{h_2}^2}{r_{\zbl}}
\bigg \rvert\,, \\
\xi_k^{\rm max} (\xi_p)&=&\dfrac{1}{2\,\mathcal{B}(r)\,r_{\zbl}}
\bigg( \,\eta (\xi_p)+\mathcal{B}(r)
\dfrac{\xi_p\,r_{h_2}^2}{r_{\zbl}} \bigg)\,,
\end{eqnarray}
where
\begin{eqnarray*}
\eta(\xi_p)&\equiv& r_{h_2}\,\sqrt{\left(\dfrac{M_{h_2}^2}
{\mzbl^2}-4\right)}\,\sqrt{\bigg(\xi_p^2\,\mathcal{B}(r)^2
+r_{\zbl}^2\bigg)}\,.
\end{eqnarray*}
So, finally when the smoke clears, Eq. (\ref{collision3.2})
reduces to:
\begin{eqnarray}
&&\mathcal{C}^{h_2 \rightarrow \zbl\zbl}[f_{\zbl}(\xi_p)]
=\dfrac{g_{h_2\zbl \zbl}^2}{48\pi\,M_{sc}}
\dfrac{r\,\mathcal{B}(r)^3}
{\sqrt{\xi_p^2\,\mathcal{B}(r)^2+\left(\frac{\mzbl \,r}
{M_{sc}}\right)^2}} \times \nonumber \\
&&\bigintsss_{\xi_k^{min}}^{\xi_k^{max}}
\dfrac{\xi_k^2\,f_{h_2}(\xi_k)\,d\xi_k}
{\sqrt{\xi_k^2\mathcal{B}(r)^2+\left(\frac{M_{h_2} \,r}
{M_{sc}}\right)^2}\,H_1(\xi_k,\xi_p,y_0(\xi_k,\xi_p))}
\times\,\dfrac{1}{F(\xi_k,\xi_p)}\,
\Bigg(2+ H_2\left(\xi_k,\xi_p,y_0(\xi_k,\xi_p)
\right)\,\Bigg)\,. \nonumber \\
\label{collision3.3}
\end{eqnarray}
For completeness, let us now
plug back in the explicit dependence of
the functions in Eq. (\ref{collision3.3}) on $r$ and list them below:
\begin{eqnarray}
F(\xi_k,\xi_p,r)&=&\dfrac{\mathcal{B}(r)^2\,\xi_k\,\xi_p}
{\sqrt{\xi_k^2\,\mathcal{B}(r)^2+\left(\frac{M_{h_2}\,r}
{M_{sc}}\right)^2}-\sqrt{\xi_p^2\,\mathcal{B}(r)^2+
\left(\frac{M_{\zbl}\,r}{M_{sc}}\right)^2}}\,,
\label{F} \\
y_0 (\xi_k,\xi_p,r)&=& \dfrac{1}{2\mathcal{B}(r)^2
\,\xi_k\,\xi_p}\left(2\sqrt{\xi_k^2\,\mathcal{B}(r)^2+
\left(\frac{M_{h_2}\,r}{M_{sc}}\right)^2}
\sqrt{\xi_p^2\,\mathcal{B}(r)^2+\left(\frac{\mzbl\,r}
{M_{sc}}\right)^2}-\left(\frac{M_{h_2}\,r}{M_{sc}}
\right)^2\right)\,,
\nonumber \\
\label{y0} \\
H_1(\xi_k,\xi_p,r)&=&\sqrt{\xi_k^2\,\mathcal{B}(r)^2+
\left(\frac{M_{h_2}\,r}{M_{sc}}\right)^2}-
\sqrt{\xi_p^2\,\mathcal{B}(r)^2+
\left(\frac{\mzbl\,r}{M_{sc}}\right)^2}\,,
\label{H1} \\
H_2(\xi_k,\xi_p,r)&=& \dfrac{\left(\sqrt{\xi_p^2\,
\mathcal{B}(r)^2+\left(\frac{\mzbl\,r}{M_{sc}}\right)^2}
\,H_1(\xi_k,\xi_p,r)+\xi_p^2\,\mathcal{B}(r)^2-
\mathcal{B}(r)^2\,\xi_p\, \xi_k\,y_0(\xi_k,\xi_p,r)\right)^2}
{\left(\frac{\mzbl\,r}{M_{sc}}\right)^4}\,.
\label{H2}
\end{eqnarray}
The limits of the integration are as follows:
\begin{eqnarray}
\xi_k^{\rm min} (\xi_p,r)&=&\dfrac{M_{sc}}{2\,\mathcal{B}(r)\,r\,\mzbl}
\bigg\lvert \,\eta (\xi_p,r)-\dfrac{\mathcal{B}(r)
\times M_{h_2}^2}{\mzbl \times M_{sc}}\,\xi_p\,r
\,\bigg \rvert\,, \\
\xi_k^{\rm min} (\xi_p,r)&=&\dfrac{M_{sc}}{2\,\mathcal{B}(r)\,r\,\mzbl}
\bigg( \,\eta (\xi_p,r)+\dfrac{\mathcal{B}(r)
\times M_{h_2}^2}{\mzbl\times M_{sc}}
\,\xi_p\,r \,\bigg)\,,
\end{eqnarray}
where
\begin{eqnarray}
\eta(\xi_p,r)&=& \left(\frac{M_{h_2}\,r}{M_{sc}}\right)
\,\sqrt{\dfrac{M_{h_2}^2}{\mzbl^2}-4}\,\,
\sqrt{\xi_p^2\,\mathcal{B}(r)^2+
\left(\frac{\mzbl\,r}{M_{sc}}\right)^2}\,.
\label{etap}
\end{eqnarray}
With the explicit forms of the functions at hand
(Eqs. ((\ref{F})--(\ref{etap}))) and remembering
that $f_{h_2}(\xi_k)$ is the equilibrium distribution function
(here Maxwell-Boltzmann distribution function), Eq. (\ref{collision3.3})
can be greatly simplified. The final form of the collision term
after performing the integral over $\xi_k$ thus turns out to be:
\begin{eqnarray}
\mathcal{C}^{h_2 \rightarrow \zbl\zbl} &=&
\dfrac{r}{8\pi M_{sc}}\dfrac{\mathcal{B}^{-1}(r)}
{\xi_p \sqrt{\xi_p^2\mathcal{B}(r)^2+
\left(\dfrac{M_{\zbl}\,r}{M_{sc}}\right)^2}}
\dfrac{g_{h_2\zbl\zbl}^2}{6}
\left(2+\dfrac{(M_{h_2}^2-2M_{\zbl}^2)^2}{4M_{\zbl}^4}\right) \\ \nonumber
&&\times \left(e^{-\sqrt{\left(\xi_{k}^{\rm min}\right)^2
\mathcal{B}(r)^2+\left(\frac{M_{h_2}\,r}{M_{sc}}\right)^2}}
\,-\,e^{-\sqrt{\left(\xi_{k}^{\rm max}\right)^2
\mathcal{B}(r)^2+\left(\frac{M_{h_2}\,r}{M_{sc}}\right)^2}}
\right) \,.
\label{ch2zblzbl-final}
\end{eqnarray}

Having derived in detail all the collision terms in Eq. (\ref{fzbl}),
it is now a straight forward exercise to derive the expressions
for the other collision terms appearing in Eq. (\ref{fpsi}).
Hence, for rest of the collision terms, we will simply write
the analytical expressions for the different functions
analogous to those in Eqs. ((\ref{F})--(\ref{H2}))
without going into the detailed derivations.
Finally, we will provide the most simplified forms
of the corresponding collision terms (where ever possible).
\subsubsection{\Large$\boldsymbol{\mathcal{C}^{s
\rightarrow \overline{\psi_1} \psi_1}}$}
\label{app:spsi1psi1}
In this case, the matrix element squared average is given by:
\begin{eqnarray}
\overline{\,\lvert \mathcal{M}\rvert^2} \bigg \lvert
_{s \rightarrow \overline{\psi_1} \psi_1}&=&
(g_{s\overline{\psi_1}{\psi_1}})^2
\left(E_p\,E_q-\vec{p}\, .\,\vec{q}-
m_{\psi_1}^2\right)\,,
\end{eqnarray}
where $\vec{p},\,\vec{q}$ are the three momenta of
the final state particles and $E_p,\,E_q$ are the
corresponding energies. The generic form of this collision
term is given as:
\begin{eqnarray}
\mathcal{C}^{s\rightarrow \overline{\psi_1}\psi_1}&=&
\dfrac{(g_{s\overline{\psi_1}\psi_1}^2)}{8\,\pi\,r}
\dfrac{M_{sc}}{\sqrt{\xi_p^2\,\mathcal{B}(r)^2+
\left(\frac{M_{\psi_1}\,r}{M_{sc}}\right)^2}}
\times [\mathcal{B}(r)]^3 \times g_{s}\,g_{\dm}\times \nonumber \\
&&~~~~~\bigintss_{\widehat{\xi_k}^{min}}^{\widehat{\xi_k}^{max}}
\dfrac{\xi_k^2\, f_{s}(\xi_k)\,\widehat{H}^s_2(\xi_k,\xi_p,r)\,d\xi_k}
{\sqrt{\xi_k^2\,\mathcal{B}(r)^2+\left(\frac{M_{s}\,r}
{M_{sc}}\right)^2}\widehat{H}^s_1(\xi_k,\xi_p,r)
\widehat{F}(\xi_k,\xi_p,r)}\,, \nonumber \\
\label{spsi2}
\end{eqnarray}
where $g_{\psi_1}$, $g_s$ are the internal degrees of freedom
of scalar ($s=h_1$, $h_2$) and fermion ($\psi_1$) respectively.
Below we list the expressions
of all the relevant functions which have appeared in Eq.~(\ref{spsi2}).
\begin{eqnarray}
\widehat{F}(\xi_k,\xi_p,r)=\dfrac{\mathcal{B}(r)^2
\,\xi_k\,\xi_p}{\sqrt{\xi_k^2\mathcal{B}(r)^2+
\left(\frac{M_{s}\,r}{M_{sc}}\right)^2}-
\sqrt{\xi_p^2\,\mathcal{B}(r)^2+
\left(\frac{M_{\psi_1}\,r}{M_{sc}}\right)^2}}\,,
\label{F1} 
\end{eqnarray}
\begin{eqnarray}
\widehat{y}_0 (\xi_k,\xi_p,r)=\dfrac{1}{2\,
\mathcal{B}(r)^2\,\xi_k\,\xi_p}
\left(2\sqrt{\xi_k^2\,\mathcal{B}(r)^2+
\left(\frac{M_{s}\,r}{M_{sc}}\right)^2}
\sqrt{\xi_p^2\,\mathcal{B}(r)^2+
\left(\frac{M_{\psi_1}\,r}{M_{sc}}\right)^2}-
\left(\frac{M_{s}\,r}{M_{sc}}\right)^2\right)\,, \nonumber\\
\label{y01} 
\end{eqnarray}
\begin{eqnarray}
\widehat{H}^{s}_1(\xi_k,\xi_p,r)=
\sqrt{\xi_k^2\,\mathcal{B}(r)^2+
\left(\frac{M_{s}\,r}{M_{sc}}\right)^2}
-\sqrt{\xi_p^2\,\mathcal{B}(r)^2+
\left(\frac{M_{\psi_1}\,r}{M_{sc}}\right)^2}\,,
\label{H11}
\end{eqnarray}
\begin{eqnarray}
\widehat{H}^s_2(\xi_k,\xi_p,r)&=&
\left(\sqrt{\xi_p^2\,\mathcal{B}(r)^2+
\left(\frac{M_{\psi_1}\,r}{M_{sc}}\right)^2}\,
\widehat{H}^s_1(\xi_k,\xi_p,r)+\xi_p^2\,
\mathcal{B}(r)^2 
\right.
\nonumber \\
&&\left.
-\mathcal{B}(r)^2\,\xi_p\, \xi_k\,
\widehat{y}_0(\xi_k,\xi_p,r)-
\left(\frac{M_{\psi_1}\,r}
{M_{sc}}\right)^2\right)\,.
\label{H21}
\end{eqnarray}
Here, $s = h_1,\,h_2$ and $M_s$ is the mass of the
of the scalar under consideration. All of these functions
(except $\widehat{H}_2^s$) have the same structural
form as those in the expression of ${\mathcal{C}^{h_2
\rightarrow \zbl \zbl}}$ (i.e. Eqs. ((\ref{F})--(\ref{H1}))).
The only difference is that the masses of the particles
have been modified accordingly. This is because, if we look
into the derivation of collision term as presented in the above section,
we will see that these functions are mostly derived
from kinematical conditions. The functions $\widehat{H}^s_2$
(Eq. (\ref{H21})) and $H_2$ (Eq. (\ref{H2})) are however
different since they depend on the dynamics of the processes
concerned (i.e. the type of the interaction involved). 

The limits of the integration are given by:
\begin{eqnarray}
\widehat{\xi_k}^{\rm min} (\xi_p,r)&=&
\dfrac{M_{sc}}{2\,\mathcal{B}(r)\,r\,M_{\psi_1}}\bigg\lvert
\,\widehat{\eta} (\xi_p,r)-\dfrac{M_{s}^2
\times\mathcal{B}(r)}{M_{\psi_1}\times M_{sc}}\,
\xi_p\,r \,\bigg \rvert \,,\\
\widehat{\xi_{k}}^{\rm max} (\xi_p,r)&=&
\dfrac{M_{sc}}{2\,\mathcal{B}(r)\,r\,M_{\psi_1}}\bigg( \,\widehat{\eta}
(\xi_p,r)+\dfrac{M_{s}^2 \times
\mathcal{B}(r)}{M_{\psi_1}\times M_{sc}}\,\xi_p\,r
\,\bigg)
\end{eqnarray}
where
\begin{eqnarray}
\widehat{\eta}(\xi_p,r)&=& \left(\frac{M_{s}\,r}{M_{sc}}\right)\,
\sqrt{\dfrac{M_{s}^2}{M_{\psi_1}^2}-4}
\,\,
\sqrt{\xi_p^2\,\mathcal{B}(r)^2+
\left(\frac{M_{\psi_1}\,r}{M_{sc}}\right)^2}\,.
\end{eqnarray}
Like the previous case, here also using
Eqs. ((\ref{F1})--(\ref{H21})) we can simplify Eq. (\ref{spsi2}).
The final expression (after putting in the numerical
values of the internal degrees of freedom) for the
collision term hence turns out to be:
\begin{eqnarray}
\mathcal{C}^{s \rightarrow \overline{\psi_1} \psi_1}
&=& \dfrac{r}{8\pi M_{sc}}\dfrac{\mathcal{B}^{-1}(r)}
{\xi_p \,\sqrt{\xi_p^2\mathcal{B}(r)^2
+\left(\dfrac{M_{\dm}\,r}{M_{sc}}\right)^2}}
\,\,g_{s\overline{\psi_1}\dm}^2\,
\left(M_{s}^2-4M_{\dm}^2\right)\\ \nonumber
&&\times \left(e^{-\sqrt{\left(\widehat{\xi_{k}}^{\rm min}\right)^2
\mathcal{B}(r)^2+\left(\frac{M_{s}\,r}{M_{sc}}\right)^2}}\,
-\,e^{-\sqrt{\left(\widehat{\xi_{k}}^{\rm max}\right)^2
\mathcal{B}(r)^2+\left(\frac{M_{s}\,r}{M_{sc}}\right)^2}}
\right)\,.
\label{cspsi1psi1-final}
\end{eqnarray}
\subsubsection{\Large$\boldsymbol{\mathcal{C}^{\zbl
\rightarrow \overline{\psi_1} \psi_1}}$}
\label{app:zblpsi1psi1}
The matrix element squared average for the decay mode 
$\zbl\rightarrow \overline{\psi_1} \psi_1$ is given by:
\begin{eqnarray}
\overline{\,\lvert \mathcal{M}\rvert^2}
\bigg \lvert _{\zbl \rightarrow \overline{\psi_1} \psi_1}&=&
\dfrac{1}{3}\left((a_{\psi_1}^2+b_{\psi_1}^2)
(\tilde{p}_1.\tilde{p}_2)+3M_{\psi_1}^2
(a_{\psi_1}^2-b_{\psi_1}^2)+2
\dfrac{(a_{\psi_1}^2+b_{\psi_1}^2)}
{\mzbl^2}(\tilde{p}_1.\tilde{k})
(\tilde{p}_2.\tilde{k})\right)\,,
\nonumber \\
\end{eqnarray}
where $\tilde{p}_1$ and $\tilde{p}_2$ are the four momenta
of the final state particles while $\tilde{k}$ is the
corresponding four momenta for the mother particle ($\zbl$)
and $\tilde{k}=\tilde{p}_1+\tilde{p}_2$.
The couplings $a_{\psi_1}=\frac{\gbl}{6}
\left(1-3\sin^2\theta_{\rm L}\right)$
and $b_{\psi_1}=\frac{\gbl}{2}
\left(1+\sin^2\theta_{\rm L}\right)$. 
The masses of $\psi_1$ and $\zbl$ are $M_{\psi_1}$ and $\mzbl$
respectively.

As before, the collision term has the following form:
\begin{eqnarray}
\mathcal{C}^{\zbl \rightarrow \overline{\psi_1}\psi_1}&=&
\dfrac{1}{24\,\pi\,r}\dfrac{M_{sc}}
{\sqrt{\xi_p^2\,\mathcal{B}(r)^2+\left(\frac{M_{\psi_1}\,r}
{M_{sc}}\right)^2}} \times [\mathcal{B}(r)]^3
\times g_{\zbl}\,g_{\dm} \times \nonumber \\
&&~~~~~\bigintss_{\widetilde{\xi_k}^{min}}^{\widetilde{\xi_k}^{max}}
\dfrac{\xi_k^2\, f_{\zbl}(\xi_k,\,r)\,
\mathcal{H}_2(\xi_k,\xi_p,r)\,d\xi_k}
{\sqrt{\xi_k^2\,\mathcal{B}(r)^2+\left(\frac{M_{\zbl}\,r}{M_{sc}}
\right)^2}\mathcal{H}_1(\xi_k,\xi_p,r)
\,\mathscr{F}(\xi_k,\xi_p,r)}\,. \nonumber \\
\label{zblpsi2}
\end{eqnarray}

As expected, the functions $\mathcal{H}_1$ and
$\mathscr{F}$ in Eq. (\ref{zblpsi2}) are structurally
quite similar to those in Eqs. ((\ref{F})--(\ref{H1}))
and Eqs. ((\ref{F1})--(\ref{H11})), since they arise
from kinematical considerations. Only the masses will
change in accordance with the particles involved. 
Thus we have:
\begin{eqnarray}
\mathscr{F}(\xi_k,\xi_p,r)=
\dfrac{\mathcal{B}(r)^2\,\xi_k\,\xi_p}
{\sqrt{\xi_k^2\,\mathcal{B}(r)^2+\left(\frac{\mzbl\,r}
{M_{sc}}\right)^2}-\sqrt{\xi_p^2\,\mathcal{B}(r)^2+
\left(\frac{M_{\psi_1}\,r}
{M_{sc}}\right)^2}} \,,
\label{F2}
\end{eqnarray}
\begin{eqnarray}
\mathscr{Y}_0 (\xi_k,\xi_p,r)=
\dfrac{1}{2\,\mathcal{B}(r)^2\,\xi_k\,\xi_p}
\left(2\sqrt{\xi_k^2\,\mathcal{B}(r)^2+
\left(\frac{\mzbl\,r}{M_{sc}}\right)^2}
\sqrt{\xi_p^2\,\mathcal{B}(r)^2+
\left(\frac{M_{\psi_1}\,r}{M_{sc}}\right)^2}
-\left(\frac{\mzbl\,r}{M_{sc}}
\right)^2\right)\,,
\nonumber \\
\label{y02}
\end{eqnarray}
\begin{eqnarray}
\mathcal{H}_1(\xi_k,\xi_p,r)=
\sqrt{\xi_k^2\,\mathcal{B}(r)^2+
\left(\frac{\mzbl\,r}{M_{sc}}\right)^2}
-\sqrt{\xi_p^2\,\mathcal{B}(r)^2+
\left(\frac{M_{\psi_1}\,r}{M_{sc}}\right)^2}\,.
\label{H12}
\end{eqnarray}
To write down the exact analytical form of
$\mathcal{H}_2$ defined in Eq. (\ref{zblpsi2}) in a compact way,
it is useful to define some auxiliary functions first.
They are:
\begin{eqnarray}
{G}_1(\xi_k,\xi_p,r)&=&
\left(\sqrt{\xi_p^2\,\mathcal{B}(r)^2+
\left(\frac{M_{\psi_1}\,r}{M_{sc}}\right)^2}
\,\mathcal{H}_1(\xi_k,\xi_p,r)+
\xi_p^2\,\mathcal{B}(r)^2 
\right.
\nonumber \\
&&\left.
-\,\mathcal{B}(r)^2\,\xi_p\,
\xi_k\,\mathscr{Y}_0(\xi_k,\xi_p,r)-
\left(\frac{M_{\psi_1}\,r}
{M_{sc}}\right)^2\right)\,,
\label{G1}
\end{eqnarray}
\begin{eqnarray}
{G}_2(\xi_k,\xi_p,r)&=&
\sqrt{\xi_k^2\,\mathcal{B}(r)^2+
\left(\frac{\mzbl\,r}{M_{sc}}\right)^2}
\sqrt{\xi_p^2\,\mathcal{B}(r)^2+
\left(\frac{M_{\psi_1}\,r}
{M_{sc}}\right)^2} 
-\mathcal{B}(r)^2\,\xi_p\,\xi_k\,
\mathscr{Y}_0(\xi_k,\xi_p,r)\,,\nonumber
\\
\label{G2}\\
{G}_3(\xi_k,\xi_p,r)&=&
\sqrt{\xi_k^2\,\mathcal{B}(r)^2+
\left(\frac{\mzbl\,r}{M_{sc}}\right)^2}
\,\mathcal{H}_1(\xi_k,\xi_p,r)-\xi_k^2\,
\mathcal{B}(r)^2+\mathcal{B}(r)^2\,
\xi_p\,\xi_k\,\mathscr{Y}_0
(\xi_k,\xi_p,r)\,. \nonumber
\label{G3}\\
\end{eqnarray}
Therefore using Eqs. ((\ref{G1})--(\ref{G3})) we have:
\begin{eqnarray}
\mathcal{H}_2(\xi_k,\xi_p,r)&=&
(a_{\psi_1}^2+b_{\psi_1}^2)\,{G_1}(\xi_k,\xi_p,r)
+2\left(\dfrac{M_{\psi_1}\,r}{M_{sc}}\right)^2
(2\,a_{\psi_1}^2-b_{\psi_1}^2) \nonumber \\
&&+\,2\,\dfrac{(a_{\psi_1}^2+b_{\psi_1}^2)}
{\left(\dfrac{\mzbl\,r}{M_{sc}}\right)^2}\,
{G_2}(\xi_k,\xi_p,r)\,{G_3}(\xi_k,\xi_p,r)\,.
\end{eqnarray}
However, using Eqs. ((\ref{F2})--(\ref{H12})),
$G_1,\,G_2$ and $G_3$ are greatly simplified :
\begin{eqnarray}
G_1 &=& \dfrac{1}{2}\left(\dfrac{M_{\zbl}\,r}{M_{sc}}
\right)^2-2\left(\dfrac{M_{\dm}\,r}{M_{sc}}\right)^2\,, \\
G_2&=&G_3 = \dfrac{1}{2}\left(\dfrac{M_{\zbl}\,r}{M_{sc}}\right)^2.
\end{eqnarray}
Consequently, $\mathcal{H}_2$ is also simplified to:
\begin{eqnarray}
\mathcal{H}_2 &=& \left(\dfrac{M_{\zbl}\,r}
{M_{sc}}\right)^2 \left(a_{\dm}^2+b_{\dm}^2\right)+
2\left(\dfrac{M_{\dm}\,r}{M_{sc}}\right)^2
\left(a_{\dm}^2-2\,b_{\dm}^2\right)\,.
\label{H12sim}
\end{eqnarray}
The limits of the integration are given by:
\begin{eqnarray}
\tilde{\xi_k}^{\rm min} (\xi_p,r)&=&
\dfrac{M_{sc}}{2\,\mathcal{B}(r)\,r\,M_{\psi_1}}
\bigg\lvert \,\tilde{\eta} (\xi_p,r)-
\dfrac{M_{\zbl}^2\times\mathcal{B}(r)}
{M_{\psi_1}\times M_{sc}}\,
\xi_p\,r \,\bigg \rvert \,,\\
\tilde{\xi_{k}}^{\rm max} (\xi_p,r)&=&
\dfrac{M_{sc}}{2\,\mathcal{B}(r)\,r\,M_{\psi_1}}
\bigg( \,\tilde{\eta} (\xi_p,r)+
\dfrac{M_{\zbl}^2 \times \mathcal{B}(r)}{
M_{\psi_1}\times M_{sc}}\,\xi_p\,r \,\bigg)\,.
\end{eqnarray}
where,
\begin{eqnarray}
\tilde{\eta}(\xi_p,r)&=& \left(\frac{\mzbl\,r}{M_{sc}}\right)\,
\sqrt{\dfrac{\mzbl^2}{M_{\psi_1}^2}-4}\,\,
\sqrt{\xi_p^2\,\mathcal{B}(r)^2+
\left(\frac{M_{\psi_1}\,r}{M_{sc}}\right)^2}\,.
\end{eqnarray}

Finally, using Eqs. ((\ref{F2})--(\ref{H12}))
and Eq. (\ref{H12sim}) in Eq. (\ref{zblpsi2}), we get:
\begin{eqnarray}
\mathcal{C}^{\zbl \rightarrow \overline{\psi_1}\psi_1}&=&
\dfrac{r}{4\,\pi\,M_{sc}}\dfrac{\mathcal{B}(r)}
{\xi_p\,\sqrt{\xi_p^2\,\mathcal{B}(r)^2+\left(\frac{M_{\psi_1}\,r}
{M_{sc}}\right)^2}} \times \left(M_{\zbl}^2
\left(a_{\dm}^2+b_{\dm}^2\right)+2 M_{\dm}^2
\left(a_{\dm}^2-2\,b_{\dm}^2\right)\right)\nonumber \\
&&~~~~~\times \bigintss_{\widetilde{\xi_k}
^{\rm min}}^{\widetilde{\xi_k}^{\rm max}}
\dfrac{\xi_k \, f_{\zbl}(\xi_k,\,r)\,d\xi_k}
{\sqrt{\xi_k^2\,\mathcal{B}(r)^2+
\left(\frac{M_{\zbl}\,r}{M_{sc}}
\right)^2}}\,.
\label{czblpsi1psi1-final}
\end{eqnarray}
Unlike the previous cases, here the integration over
$\xi_k$ can not be analytically performed since, we do
not apriori know the distribution function of $\zbl$.
The Boltzmann equation (Eq.~(\ref{fzbl})) has been solved
for finding this $f_{\zbl}$ and hence the integration has
been done numerically.
\bibliographystyle{jhep}
\bibliography{sterile_new}
\end{document}